\documentclass[10pt]{article}

\usepackage{latexsym}
\usepackage{amsfonts}
\usepackage{amsmath}
\usepackage{mathrsfs}  
\usepackage{dsfont}    
\usepackage{bbold}     
\usepackage[english]{babel}
\usepackage{caption}
\usepackage{epsfig}
\usepackage{float}
\usepackage{isolatin1}

\DeclareMathOperator{\Val}{Val}

\newtheorem{theorem}{Theorem}

\newtheorem{defi}[theorem]{Definition}
\newtheorem{lema}[theorem]{Lemma}
\newtheorem{prop}[theorem]{Proposition}
\newtheorem{rmk}[theorem]{Remark}

\newcommand{\zerarcounters}{\setcounter{equation}{0}\setcounter{theorem}{0}}

\newcommand{\Z}{\mathds{Z}}
\newcommand{\C}{\mathds{C}}
\newcommand{\N}{\mathds{N}}

\newcommand{\R}{\mathds{R}}
\newcommand{\T}{\mathds{T}}

\newcommand{\calA}{{\mathcal A}}

\newcommand{\calE}{{\mathcal E}}
\newcommand{\calF}{{\mathcal F}}

\newcommand{\calI}{{\mathcal I}}

\newcommand{\calL}{{\mathcal L}}
\newcommand{\calM}{{\mathcal M}}
\newcommand{\calN}{{\mathcal N}}

\newcommand{\calQ}{{\mathcal Q}}
\newcommand{\calR}{{\mathcal R}}
\newcommand{\calS}{{\mathcal S}}
\newcommand{\calT}{{\mathcal T}}

\newcommand{\calV}{{\mathcal V}}

\newcommand{\calmT}{{\mathscr T}}

\newcommand{\gotm}{{\mathfrak m}}

\newcommand{\gotD}{{\mathfrak D}}

\newcommand{\matT}{{\mathscr T}}

\newcommand{\undm}{\underline{m}}

\newcommand{\undzero}{\underline{0}}
\newcommand{\undomega}{\underline{\omega}}

\newcommand{\Fullbox}{{\rule{2.0mm}{2.0mm}}}
\newcommand{\ES}{\hfill\Box}
\newcommand{\EP}{\hfill\Fullbox}
\newcommand{\prova}{\noindent{\it Proof. }}
\newcommand{\eps}{\varepsilon}

\newcommand{\laa}{\langle}
\newcommand{\raa}{\rangle}
\newcommand{\Me}[1]{\laa#1\raa}

\newcommand{\vnu}{\boldsymbol{\nu}}

\newcommand{\vomega}{\boldsymbol{\omega}}
\newcommand{\vz}{\boldsymbol{0}}
\newcommand{\ov}{\overline}
\newcommand{\olu}{\ov{u}}
\newcommand{\der}{{\rm d}}

\begin{document}

\title{\bf Stability for quasi-periodically perturbed \\ Hill's equations}

\author
{\bf Guido Gentile$^\dagger$, Daniel A. Cortez$^\ast$ and 
João C. A. Barata$^\ast$ 
\vspace{2mm} \\ \small 
$^\dagger$Dipartimento di Matematica, Università di Roma Tre, Roma,
I-00146, Italy.
\\ \small 
E-mail: gentile@mat.unirom3.it
\\ \small
$^\ast$Instituto de Física, Universidade de São Paulo, 
Caixa Postal 66 318, 
\\ \small
São Paulo, 05315 970 SP, Brasil.
\\ \small 
E-mails: dacortez, jbarata@fma.if.usp.br}

\date{}

\maketitle

\begin{abstract}
  We consider a perturbed Hill's equation of the form $\ddot \phi +
  \left( p_{0}(t) + \eps p_{1}(t) \right) \phi = 0$, where $p_0$ is
  real analytic and periodic, $p_1$ is real analytic and
  quasi-periodic and $\eps\in \R$ is ``small''.
  Assuming Diophantine conditions on the frequencies of the
  decoupled system, i.e.\ the frequencies of the external potentials
  $p_{0}$ and $p_{1}$ and the proper frequency of the unperturbed
  ($\eps=0$) Hill's equation, but without making
  non-degeneracy assumptions on the perturbing potential $p_{1}$,
  we prove that quasi-periodic solutions
  of the unperturbed equation can be continued into quasi-periodic
  solutions if $\eps$ lies in a Cantor set of relatively large measure
  in $[-\eps_0,\eps_0] \subset \R$, where $\eps_0$ is small enough.
  Our method is based on a resummation procedure of a formal Lindstedt
  series obtained as a solution of a generalized Riccati equation
  associated to Hill's problem.
\end{abstract}

%\newpage

%\tableofcontents

%\newpage

%%%%%%%%%%%%%%%%%%%%%%%%%%%%%%%%%%%%%%%%%%%%%%%%%%%%%%%%%%%%%%%%%%%%%%%%%
%%%%%%%%%%%%%%%%%%%%%%%%%%%%%%%%%%%%%%%%%%%%%%%%%%%%%%%%%%%%%%%%%%%%%%%%%
\zerarcounters
\section{Introduction}
%%%%%%%%%%%%%%%%%%%%%%%%%%%%%%%%%%%%%%%%%%%%%%%%%%%%%%%%%%%%%%%%%%%%%%%%%
%%%%%%%%%%%%%%%%%%%%%%%%%%%%%%%%%%%%%%%%%%%%%%%%%%%%%%%%%%%%%%%%%%%%%%%%%

In the present work we will consider the one-dimensional Hill's
equation (for a standard reference, see~\cite{Magnus_Winkler}) with a
quasi-periodic perturbation
\begin{equation} \label{eq:hill}
\ddot \phi + \left( p_{0}(t) + \eps p_{1}(t) \right) \phi\; =\; 0\; ,
\end{equation}
where $p_{0}$ and $p_{1}$ are two real analytic functions, the first
periodic with frequency $\omega_{0}$ and the latter quasi-periodic
with frequency vector $\undomega_{1}\in\R^{A}$, for an integer $A\geq
1$ (for notational details see Section~\ref{ssec:notation}).  No
further assumption is made on the equation, besides requiring that the
real parameter $\eps$ is small and that the unperturbed equation
(i.e. for $\eps\equiv 0$) has a fundamental system of real
quasi-periodic solutions.

For $p_{0}$ constant such an equation has been extensively studied,
also in connection with the spectrum of the corresponding Schrödinger
equation $\ddot \phi + \eps V(\undomega_{1}t) \phi = E \phi$, with $V$
analytic and periodic in its arguments; see for instance
Refs.~\cite{Dinaburg_Sinai,Russmann,Eliasson,Johnson_Moser,
Sorets_Spencer,Moser_Poschel}.
We also mention the recent Ref.~\cite{Broer_Puig_Simo} and also
\cite{Broer_Simo}, where some properties of the gaps and of the
instability tongues have been investigated.
Different perturbations of Hill's equation, with a $L^{1}$
perturbing potential, have been considered for instance
in Refs.~\cite{RofeBeketov,Zeludev1,Zeludev2,Hinton_Shaw}.

We are interested in the problem of studying conservation of
quasi-periodic motions for $\eps$ different from zero but small enough.
Of course, equation (\ref{eq:hill}) can be considered as arising
from an autonomous Hamiltonian system with $d=A+2$
degrees of freedom, described by the Hamiltonian
\begin{equation} \label{eq:hamiltonian}
H\; =\; \Omega_{0} A + \omega_{0} A_{0} + \undomega_{1}\cdot \underline 
A_{1} + \eps p_{1}(\underline \alpha_{1})\,
f(A,\, A_{0},\, \alpha,\, \alpha_{0}) ,
\end{equation}
where $(A,\, A_{0},\, \underline A_{1},\, \alpha,\, \alpha_{0},\, \underline 
\alpha_{1}) \in \R\times\R\times\R^{A}\times\T\times\T\times\T^{A}$
are action-angle variables, and $f$ and $\Omega_{0}$ depend on the
periodic potential $p_{0}$. For instance if $p_{0}$ is a constant,
say $p_{0}=1$, then the variables $(A_{0},\, \alpha_{0})$ disappear,
$\Omega_{0}=1$ and $f(A,\, \alpha)=2A\cos^{2}\alpha$.
In general the change of variables leading to (\ref{eq:hamiltonian})
is slightly more complicated, but it can be easily worked out;
we refer for instance to Refs.~\cite{Chierchia1,Chierchia2}.
Also in such a case the function $f$ is linear in the action variables.
Hence systems like (\ref{eq:hamiltonian}) are not typical
in KAM theory, because the perturbation does not remove
isochrony. What one usually does is to study the behavior of the
solutions, in particular to understand if they are
bounded (quasi-periodic) or unbounded (linearly or exponentially 
growing),
when varying the parameters characterizing the external potential.
In the case of the Schrödinger equation this can be done
for a fixed potential, by varying the energy, which represents an
extra free parameter, and information can be obtained about the 
spectrum. In Ref.~\cite{Chierchia2} this is done for bounded solutions,
so that conditions on $E$ are obtained characterizing the
spectrum of the corresponding Schrödinger operator.

Here we are interested in the case in which the potential is fixed,
and the parameters of $p_{0}$ are such that the fundamental solutions of the
corresponding Hill's equation $\ddot\phi+p_{0}(t)\,\phi=0$ are
quasi-periodic (this means that we are inside the stability regions).
Hence for $\eps=0$ we have $d=A+2$ fundamental frequencies
$\undomega_{1},\,\omega_{0},\,\Omega_{0}$, where $\Omega_{0}$ is the
proper frequency of the unperturbed Hill's equation.  Then we want to
study if the solutions remain quasi-periodic when the perturbation is
switched on. Even when this occurs, one expects that the proper frequency
of the system is changed as an effect of the perturbation.
Since the system is in fact a perturbation of an isochronous one,
and we have no free parameter to adjust,
either the proper frequency is changed to some perturbation
order or it is never changed (if disposing of the extra parameter $E$
the frequency changes to first order up to a zero-measure set).  But
to follow all the possibilities requires some careful analysis, which
one can avoid by assuming some non-degeneracy condition on the
perturbation in order to control the change of the frequencies.  On
the contrary we do not want to impose any condition on the
perturbation.

Degeneracy problems of this kind are known to be not easy to handle.
An example is given by Herman's conjecture in the case in which
one has a system of $N$ harmonic oscillators where no assumption
is made on the coupling terms of order higher than two:
in such a case the conservation of a large measure of invariant tori
has been be proved only for $N=2$ \cite{Herman}.
We can mention also Cheng's results on the conservation
of lower $(N-1)$-dimensional tori for systems with $N$ degrees
of freedom \cite{Cheng1,Cheng2}.

To come back to our problem, we fix the unperturbed torus and study
for which values of $\eps$ (small enough) such a torus is conserved.
In particular we are interested in the dependence on $\eps$ of the
conserved torus: we shall find that the torus will be defined for
$\eps$ in a Cantor set of large relative measure, and for such values
of $\eps$ the system turns out to be reducible.  We shall see also
that one can give a meaning to the perturbation series, through a
suitable resummation, in an analogous way to what was done in similar
contexts in
Refs.~\cite{Gentile,Gallavotti_Gentile,Gentile_Gallavotti}.

We do not study directly the equation (\ref{eq:hill}). Rather, we
shall write $\phi$ in terms of a suitable function $u$, for which
a very simple-looking equation can be derived. Indeed by setting
\begin{equation*}
\phi_{0}(t) \;=\; {\rm const.}
\exp \left( i \int_{0}^{t} g_{0}(t') \, {\rm d}t' \right) , \qquad
Q(t) \;= \;\exp \left( -2 i \int_{0}^{t} g_{0}(t') \, {\rm d}t' \right) ,
\end{equation*}
where $\phi_{0}$ is a quasi-periodic solution of (\ref{eq:hill})
for $\eps=0$, with rotation vector $(\omega_{0},\Omega_{0})$, where
the proper frequency $\Omega_{0}$ is the average of $g_{0}$, and 
defining
\begin{equation} \label{eq:g}
\phi(t) \;=\; \phi_{0}(t)
\exp \left( i \int_{0}^{t} g(t') \, {\rm d}t' \right) ,
\qquad g(t) \;=\; i\eps Q(t) u(t) ,
\end{equation}
one finds that $u$ has to solve the equation (see
Section~\ref{ssec:ansatz} for details)
\begin{equation} \label{eq:u}
\dot u \; = \;R + \eps Q u^{2} , \qquad R\; = \;p_{1}Q^{-1} ,
\end{equation}
which is an ordinary differential equation which
could be of interest by its own.

The advantage of this procedure is that we can look
for a solution of (\ref{eq:u}) with the same rotation vector
$\vomega=(\undomega_{1},\, \omega_{0},\, \Omega_{0})$ of the
unperturbed system, something which cannot be done
for the full unperturbed system, as the proper frequency
$\Omega_{0}$ is expected to change (as usually happens
when perturbing an isochronous system).

That such a solution $u(t)$ exists can be shown, and this is the core
of the paper, provided one assumes, besides an obvious Diophantine
condition on $\vomega$, that $\eps$ is small enough, say
$|\eps| \leq \eps_{0}$, and belongs to a suitable Cantor set $\calE$ of
large relative measure in $[-\eps_{0},\eps_{0}]$. By the latter we mean
that one has $\lim_{\eps\to0^{+}} {\rm meas}(\calE\cap[-\eps,\eps])/
2\eps=1$, with ${\rm meas}$ denoting Lebesgue measure.

To recover the solution $\phi(t)$ we have to express it in terms of
$u$. By using the relations given in (\ref{eq:g}) one realizes that,
first, the solution could be unbounded (if the imaginary part of the
average $\Me{g}$ of $g$ did not vanish), and, second, even if this did
not occur, an extra frequency $\Omega_{\eps}= \Omega_{0} + \Me{g}$
would appear in addition to the $d$ frequencies already
characterizing the model, which would sound strange.
But one can check that both problems are spurious,
as $\Me{g}$ turns out to be real and dependence
on time of the function $\phi(t)$, which, in principle, could be
through the variables $\undomega_{1}t,\, \omega_{0}t,\, \Omega_{0}t,\,
\Omega_{\eps}t$ (by construction), is indeed only through the
variables $\undomega_{1}t,\, \omega_{0}t,\, \Omega_{\eps}t$, as
formally noticed in the case treated in \cite{Barata}. In other
words, the dependence on $\Omega_{0}t$ disappears, and this means that
the maximal torus, which in absence of perturbation has rotation
vector $(\undomega_{1},\, \omega_{0},\, \Omega_{0})$, can be continued
for $\eps\in\calE$, and the last component of the rotation vector is
changed into an $\eps$-dependent quantity $\Omega_{\eps}$ (that the
other components cannot change is obvious by the form of the
equations of motion). Hence the solution of (\ref{eq:u}) provides
directly a perturbation expansion for the correction
of the proper frequency of the system: indeed $\Omega_{\eps}-\Omega_{0}=
\Me{g}$, and $g$ is expressed in terms of the solution $u$.

We can now state our results in the following theorem.

\begin{theorem} \label{theo:main}
  Let $p_0:\R\to\R$ be real analytic and periodic with frequency
  $\omega_0$ and such that the fundamental solutions of the corresponding
  Hill's equation $\ddot\phi+p_{0}(t)\,\phi=0$ are quasi-periodic with
  a proper frequency $\Omega_0\in\R$.  Let $p_1:\R\to\R$ be real analytic
  and quasi-periodic with frequency vector $\undomega_1\in\R^A$ for
  some $A\geq 1$.  Define $\vomega := (\undomega_1,\, \omega_0,\,
  \Omega_0) \in \R^d$ with $d = A+2$ and assume that
  $\undm \cdot \undomega_1 + n \omega_0 + 2\Omega_0 \neq 0$, 
  $\forall (\undm, \, n) \in \Z^{A+1} $ and, moreover
\begin{equation*}
|\vomega \cdot \vnu| \;\geq\; \frac{C_0}{|\vnu|^{\tau}} \, ,
\quad \forall \vnu \in \Z^{d} \setminus \{\vz\} \, ,
\end{equation*}
for two fixed positive constants $C_0>0$ and $\tau > d-1$ 
(Diophantine conditions).
  Then, there exists $\eps_0 > 0$ small enough and a Cantor set $\calE
  \subset [-\eps_0, \eps_0]$ of large relative measure in $[-\eps_0,
  \eps_0]$ such that, for all $\eps \in \calE$, (\ref{eq:u}) admits a
  quasi-periodic solution of the form 
\begin{equation*}
\ov{u}(t)\; =\; U(\vomega t;\; \eps)\; =\; 
\sum_{\vnu \in \Z^d} \tilde{U}_{\vnu}(\eps) e^{i \vomega\cdot\vnu t} \, ,
\end{equation*} 
  where the sum above is absolutely and uniformly convergent for all $t \in
  \R$ and all $\eps \in \calE$. Moreover, for all $\eps \in \calE$,
  the system (\ref{eq:hill}) is reducible and it has a
  quasi-periodic solution of the form   
\begin{equation*}
\phi(t) \;=\; \Phi(\Omega_\eps t, \; \undomega_1 t,\;\omega_0t;\;\eps)
\; =\; 
e^{i\Omega_\eps t} 
\left(
\sum_{(\undm, n) \in \Z^{A+1}} \tilde{\Phi}_{\undm,
n}(\eps) e^{i(\undm\cdot\undomega_1 + n \omega_0) t} 
\right)
\, ,
\end{equation*} 
  where, by denoting with $\Me{\cdot}$ the average of a quasi-periodic
  function (that is the constant term in its Fourier expansion), one has
  $\Omega_\eps := \Omega_0 + \Me{g} = \Omega_0 +i\eps\Me{Qu}$ is real, 
  and the sum above is absolutely and uniformly convergent for all 
  $t \in \R$ and all $\eps \in \calE$. Finally, if $\Me{g} = 0$ then $\calE =
  [-\eps_0, \eps_0]$ and $\Omega_\eps$ reduces to $\Omega_0$. $\ES$
\end{theorem}

In particular the proof of the result will imply that the equation is
reducible for $\eps\in\calE$. It would be interesting to study what
happens for $\eps$ outside the set $\calE$ (cf. the results
proved for the case of the Schrödinger equation with $p_{0}=0$
and other related models \cite{Eliasson,Krikorian1,Krikorian2}).

The rest of this paper is devoted to the proof of the above theorem.

We organize this work as follows: in Section~\ref{sec:perturbation} we
motivate and discuss the Ansatz used to solve (\ref{eq:hill}) and
introduce the tree representation of the perturbative coefficients
obtained, which is the basis for the forthcoming analysis.
Section~\ref{sec:fixing} is devoted to the solution of the ``zero
mode'' problem, which is essential for constructing a consistent
quasi-periodic solution for (\ref{eq:u}). Section~\ref{sec:formal}
shows that our naive perturbative solution is merely formal, i.e. not
convergent as a power series in $\eps$. This is related to small
divisors problems. Next, Section~\ref{sec:renormalization} brings the
core idea of this paper: the renormalization of the formal solution.
This process is implemented through a multiscale decomposition of
propagators and a suitable resummation technique. As described in
Theorem~\ref{theo:main}, the result is a convergent quasi-periodic
solution for (\ref{eq:u}), well defined in a Cantor set $\calE$ of
relatively large measure in $[-\eps_0, \eps_0]$.
Section~\ref{sec:properties} is devoted to the proof of some technical
lemmas which are related to estimates on the so called ``self-energy
values''. This lemmas are crucial in the proof of
Theorem~\ref{theo:main}, which is essentially performed in
Section~\ref{sec:convergence}, where convergence of the renormalized
expansion is shown. Next, in Section~\ref{sec:measure} we provide
estimates on the measure of the set $\calE$ where the renormalized
solution exists. It is shown that $\calE$ is of relatively large
measure in a compact set $[-\eps_0, \eps_0]$. Finally,
Section~\ref{sec:prenormalized} completes the proof of
Theorem~\ref{theo:main} by analyzing properties of the renormalized
expansion. Section~\ref{sec:null} closes the paper by discussing the
rather trivial situation where we cannot fix the zero modes as in
Section~\ref{sec:fixing}. This is the situation where the proper
frequency of the unperturbed Hill's equation in unchanged when the
perturbation is switched on, i.e. $\Omega_\eps = \Omega_0$.
    
%%%%%%%%%%%%%%%%%%%%%%%%%%%%%%%%%%%%%%%%%%%%%%%%%%%%%%%%%%%%%%%%%%%%%%%%%
\subsection{Basic notations} 
\label{ssec:notation}
%%%%%%%%%%%%%%%%%%%%%%%%%%%%%%%%%%%%%%%%%%%%%%%%%%%%%%%%%%%%%%%%%%%%%%%%%

In this paper $\N$ will denote the set of positive integers, $\Z$ the set
of all integers and $\R$ the set of real numbers. Note that $0 \notin \N$. 
For any $n \in \N$, $\Z^n$ (or $\R^n$) is the Cartesian product of $\Z$ (or
$\R$) $n$ times. The set $\T$ denotes the one-dimensional torus, i.e. $\T =
\R/2\pi\Z$. $\T^n$ is the $n$-dimensional torus. 

Vectors in $\Z^n$ (or $\R^n$) will be denoted either by boldface or underline
characters. Boldface characters will be used to denote vector in a certain
dimension $d$, i.e. $\vomega \in \R^d$, $\vnu \in \Z^d$. Underline
characters will be used to denote vector in a certain dimension $A < d$, 
i.e. $\undomega_1 \in \R^A$, $\undm \in \Z^A$. 

For any $n \in \N$, $\Z^n_\ast$ is defined as $\Z^n \setminus \{\vz\}$,
i.e. $\Z^n_\ast$ is $\Z^n$ with the exception of the zero. The same applies
to $\R^n$.  

The scalar product in $\R^n$ will be denoted as usual by a dot: $\mathbf{v}
\cdot \mathbf{w} := v_1 w_1 + \cdots + v_n w_n$, for $\mathbf{v}, \mathbf{w}
\in \R^n$. The $\ell^1$-norm of a vector $\mathbf{v}=(v_1, \, \ldots ,
\, v_n) \in \R^n$ is
$|\mathbf{v}| := |v_1|+\cdots+|v_n|$, where in the r.h.s. $|\cdot|$ denotes
the usual absolute value in $\R$ (or $\C$).
The complex conjugate of  $z \in \C$ will be denoted by $z^\ast$. 

Given a periodic or, more generally, a quasi-periodic function $f$
we denote by $\Me{f}$ the average of $f$, 
\begin{equation*}
\Me{f} \; := \; \lim_{T\to\infty}\frac{1}{2T}\int_{-T}^{T} 
\der t \, f(t) \; = \; f_{\vz} \, , 
\end{equation*}  
where $f_{\vz}$ is the constant term of the Fourier expansion of $f$
\cite{Katznelson}.

The symbol $\Box$ will be used at the end of the statement of a
theorem, lemma or proposition and $\Fullbox$ will be used at the end
of a proof.
  
%%%%%%%%%%%%%%%%%%%%%%%%%%%%%%%%%%%%%%%%%%%%%%%%%%%%%%%%%%%%%%%%%%%%%%%%%
%%%%%%%%%%%%%%%%%%%%%%%%%%%%%%%%%%%%%%%%%%%%%%%%%%%%%%%%%%%%%%%%%%%%%%%%%
\zerarcounters
\section{Perturbative analysis}
\label{sec:perturbation}
%%%%%%%%%%%%%%%%%%%%%%%%%%%%%%%%%%%%%%%%%%%%%%%%%%%%%%%%%%%%%%%%%%%%%%%%%
%%%%%%%%%%%%%%%%%%%%%%%%%%%%%%%%%%%%%%%%%%%%%%%%%%%%%%%%%%%%%%%%%%%%%%%%%

In this Section we will begin our perturbative analysis. We start from
a given complex quasi-periodic solution for the unperturbed version of
(\ref{eq:hill}), i.e. for $\eps = 0$, and search for a perturbative
solution for the full equation that formally tends to this
unperturbed solution as $\eps \to 0$. For this, we apply an
exponential Ansatz, whose geometrical motivation we briefly discuss
below, leading to a generalized Riccati equation (equation (\ref{eq:equ}),
ahead). In the core of this paper we prove that this generalized
Riccati equation admits a quasi-periodic solution under suitable
conditions on the frequencies and on the coupling parameter $\eps$ and,
as we prove below, this implies quasi-periodicity of the perturbed
solution of (\ref{eq:hill}). In Section~\ref{ssec:TreeExpansion} we
present a formal tree expansion for the solution of (\ref{eq:equ})
that will be the starting point of our renormalization analysis.

However, as we shall see, boundness on the solutions of (\ref{eq:equ})
will automatically imply stability on the associate solutions of
Hill's equation. This will become more clear with
Proposition~\ref{proposition:average}. 
   
%%%%%%%%%%%%%%%%%%%%%%%%%%%%%%%%%%%%%%%%%%%%%%%%%%%%%%%%%%%%%%%%%%%%%%%%%
\subsection{Unperturbed equation}
%%%%%%%%%%%%%%%%%%%%%%%%%%%%%%%%%%%%%%%%%%%%%%%%%%%%%%%%%%%%%%%%%%%%%%%%%

The following elementary result presents some basic properties of
complex quasi-periodic solutions of the unperturbed Hill's equation
that partially motivates the approach of Section~\ref{ssec:ansatz}.

%%%%%%%%%%%%%%%%%%%%%%%%%%%%%%%%%%%%%%%%%%%%%%%%%%%%%%%%%%%%%%%%%%%%%%%%%
\begin{prop}
Let
$p_0: \R \to \R$ be an analytic periodic function
with period $T_0 = 2\pi/\omega_0$, such that the equation
\begin{equation}
\ddot{\phi}(t) + p_0(t) \, \phi(t) \; =\; 0 
\label{equacaonaoperturbada}
\end{equation} 
has two non-trivial, real, analytic, quasi-periodic and independent
solutions $\phi_a $ and $\phi_b$.  Then, the complex quasi-periodic
solution $\phi_0(t)=\phi_a(t)+i\phi_b(t)$ can be expressed in the form
\begin{equation*}
\phi_0(t)\; =\; \exp\left( i\Omega_0 t + i\psi_0(t) \right) \, ,
\end{equation*}
where $\Omega_0 \in \R$ and $\psi_0: \R \to \C$ is an analytic
periodic function with frequency $\omega_0$.$\ES$
\end{prop} 
%%%%%%%%%%%%%%%%%%%%%%%%%%%%%%%%%%%%%%%%%%%%%%%%%%%%%%%%%%%%%%%%%%%%%%%%%

\prova Since the Wronskian 
$W(t)=\phi_a(t)\dot{\phi_b}(t)- \phi_b(t)\dot{\phi_a}(t)$ is a 
non-vanishing constant, $W(t)=W_0\neq 0, \; \forall t\in\R$, one has
\begin{equation*}
|W_0| \; \leq \; |\phi_a(t)|\,
|\dot{\phi_b}(t)|+|\phi_b(t)|\,|\dot{\phi_a}(t)|
\; \leq \; D (|\phi_a(t)|+|\phi_b(t)|), 
\end{equation*}
where $D := \max \{ \sup_{t\in \R}|\dot{\phi_a}(t)| , \; 
\sup_{t\in \R}|\dot{\phi_b}(t)|\} < \infty$, because $\dot{\phi_a}$
and $\dot{\phi_b}$ are both, by hypothesis, quasi-periodic. Let
$\phi_0:= \phi_a + i\phi_b$. By the equivalence of the $\ell^1$ and
$\ell^2$ norms, there exists a constant $C>0$ such that
\begin{equation}
\label{eq:modphi0}
|\phi_0(t)| \; = \; 
\sqrt{|\phi_a(t)|^2+|\phi_b(t)|^2} \; \geq
\; C (|\phi_a(t)|+|\phi_b(t)|) \; \geq \; \frac{C|W_0|}{D}, \quad
\forall \, t.
\end{equation}
This tells us that the quasi-periodic complex function $\phi_0$
remains outside of a neighborhood of the origin for all times. 
Under these circumstances, a theorem of H.~Bohr~\cite{Bohr}, implies
that we can write
\begin{equation} \label{eq:bohr}
\phi_0(t) \; = \; \exp \Bigl(i\Omega_0 t + i \psi_0(t) \Bigr),
\end{equation}
where $\Omega_0 \in \R$ and $ \psi_0(t): \R \to \C$ is almost
periodic. Floquet's theorem guarantees
that $\psi_0$ is periodic with the same frequency of $p_0$.$\EP$

\vspace{0.4cm}

We clearly see from (\ref{eq:bohr}) that $\Omega_0$ is the rotation
number of $\phi_0$.

Since $\phi_0^\ast$ is also a solution of
(\ref{equacaonaoperturbada}) (because (\ref{equacaonaoperturbada}) is
real), the most general (complex) solution is
\begin{equation} \label{eq:generalsolution}
A_1\exp \Bigl(+i\Omega_0 t + i \psi_0(t) \Bigr) \;+\;
A_2\exp \Bigl(-i\Omega_0 t - i \psi_0(t)^\ast \Bigr) ,
\end{equation}
with $A_1, \; A_2 \in \C$.  

Defining the periodic function $ g_0(t) := \dot{\psi}_0(t) + \Omega_0
$, we can write
\begin{equation}
\label{eq:phi0g0}
\phi_0(t) \;=\; \exp\left(i \int_0^t g_0(t') \, \der t' \right) e^{i
\psi_0(0)} \, .
\end{equation}
Since $\Me{\dot{\psi}_0} = 0$, we have $\Omega_0 = \Me{g_0}$.

%%%%%%%%%%%%%%%%%%%%%%%%%%%%%%%%%%%%%%%%%%%%%%%%%%%%%%%%%%%%%%%%%%%%%%%%%
\subsection{Perturbed equation and the exponential Ansatz} \label{ssec:ansatz}
%%%%%%%%%%%%%%%%%%%%%%%%%%%%%%%%%%%%%%%%%%%%%%%%%%%%%%%%%%%%%%%%%%%%%%%%%

As we mentioned, the representation (\ref{eq:phi0g0}) is possible
because (\ref{eq:modphi0}) tells us that the quasi-periodic complex
function $\phi_0$ runs outside of a neighborhood of the origin for
all times. It is tempting to presume that this sort of stability
property is preserved when the perturbation is switched on and that
the periodic function $g_0$ is replaced by a quasi-periodic one in the
form $g_0+g$, where $g$ vanishes when $\eps\to 0$. This is the
motivation for the steps that follow.

Let us now consider the perturbed equation (\ref{eq:hill}) with
$p_1:\R \to \R$ analytic and quasi-periodic, with 
frequencies in the set $\{\undm \cdot \undomega_1, \, \undm \in
\Z^A\}$ for some $A \geq 1$. The motivations presented above (see
also \cite{Barata}) lead us to search for a solution of
(\ref{eq:hill}) with the following form
\begin{equation} \label{eq:phi}
\phi(t)\; =\; \phi_0(t) \exp\left(i \int_0^t g(t') \, \der t' \right)\; =\;
\exp\left(i\int_0^t [g_0(t') + g(t')] \, \der t'\right)e^{i\psi_0(0)} ,
\end{equation}
with $g$ vanishing identically for $\eps = 0$. It is easily
verifiable that $g$ must satisfy the following generalized Riccati
equation:
\begin{equation} \label{eq:R}
\frac{\der}{\der t}\left(g \phi_0^2\right) + i\left(g \phi_0\right)^2 
- i\eps p_1 \phi_0^2 \;=\; 0 \; ,
\end{equation}
or, in another form,
\begin{equation} \label{eq:R-II}
\dot g + i g^2 + 2ig_0g - i \eps p_1 \; =\; 0\; .
\end{equation}

\begin{rmk} \label{rmk:menoiomega0}
Of course in this way we are considering a solution which reduces
to the first function in (\ref{eq:generalsolution}) for $\eps=0$.
In the following we could also consider solutions continuing
for $\eps\neq0$ the second function in (\ref{eq:generalsolution}),
and the analysis would be the same.
\end{rmk}

The idea now is to search for a quasi-periodic solution $g$ for the
above equation. In this case, 
\begin{equation*}
\phi(t) \; = \; \exp\left(i \Omega_\eps t + i \psi_\eps(t) \right) \, ,
\end{equation*}
where
\begin{equation*}
\Omega_\eps := \Omega_0 + \Me{g} \qquad \text{ and } \qquad 
\psi_\eps(t) :=  \psi_0(t) + \int_0^t \left(g(t') - \Me{g}\right) \, \der t'\, .
\end{equation*}
Note that, if such a $g$ exists, $\psi_\eps$ would be also
quasi-periodic.  However, in order to assure that $\phi$ is
quasi-periodic we have to show that $\Omega_\eps$ is a real number,
which is the case iff $\Me{g} \in\R$. This is established by the
following proposition that shows that if $g$ is quasi-periodic, then
$\phi$ is automatically stable, i.e. the Lyapunov exponent
$\mbox{Im}(\Omega_\eps)$ vanishes.

%%%%%%%%%%%%%%%%%%%%%%%%%%%%%%%%%%%%%%%%%%%%%%%%%%%%%%%%%%%%%%%%%%%%

%%%%%%%%%%%%%%%%%%%%%%%%%%%%%%%%%%%%%%%%%%%%%%%%%%%%%%%%%%%%%%%%%%%%%%%%%
\begin{prop} \label{proposition:average}
  Let us assume that (\ref{eq:R}) has a quasi-periodic solution $g$.
  Then the average of $g$ is real, that is $\Me{g}\in\R$.$\ES$
\end{prop}
%%%%%%%%%%%%%%%%%%%%%%%%%%%%%%%%%%%%%%%%%%%%%%%%%%%%%%%%%%%%%%%%%%%%%%%%%

\prova Write $g_0=x_0+iy_0$ and $g=x+iy$.  Note that
$\Me{g_{0}}=\Omega_{0}\in\R$, hence $\Me{y_{0}}=0$.  One has $i\dot
g_0 - g_0^{2} + p_0 = 0$, whose imaginary part gives $\dot x_0 =
2x_0y_0$.  Moreover, one has $\dot g + i g^2 + 2ig_0g - i \eps p_1 =0$
(equation~(\ref{eq:R-II})), whose real part is $\dot x - 2xy - 2xy_0 - 2y
x_0 = 0$.  Combining the two equations we obtain $\dot x - 2xy -2x_0 y
- 2 x y_0 + ( - 2x_0 y_0 + \dot x_0 ) = 0$, hence $\dot x + \dot x_0 -
2(y + y_0)(x + x_0) = 0$.

By defining $z=x+x_{0}$ the above equation becomes
$\dot z = f(t)\,z$, where the function
$f(t)=2(y(t)+y_{0}(t))$ is bounded (and quasi-periodic), hence,
by explicit integration,
\begin{equation*}
z(t) = \exp \left( 2 \int_{0}^{t} [ y_{0}(t') + y(t') ] \, {\rm d}t'
\right) z(0) ,
\end{equation*}
where $z(0)=x_{0}(0)+x(0)\neq 0$ (if $z(0)=0$ then $z(t)\equiv 0$
for all $t$, hence $x(t)=-x_{0}(t)$ for all $t$, which requires
$x_{0}(t)=x(t)\equiv 0$ for all $t$, and this is not possible
as $\Me{x_{0}}=\Omega_{0}\neq0$, so that $x_{0}(t)$ cannot
vanish identically). On the other hand $z(t)$ has to be a bounded
quasi-periodic function, and this requires $\Me{y_{0}+y}=0$,
so that one has $\Me{y}=0$.$\EP$ 

\vspace{0.4cm}

Therefore, we can establish that $\phi(t)$ given in (\ref{eq:phi}) is
quasi-periodic provided we find a quasi-periodic $g$. Further remarks
on properties of $\phi$ will be discussed in
Section~\ref{sec:prenormalized}.

A slightly simpler version of the generalized Riccati equation 
(\ref{eq:R}) above was studied in \cite{Gentile} by a tree 
expansion method (see, e.g., \cite{Gentile_Mastropietro} and 
references therein). So, the idea now is to try to write the same
expansion of~\cite{Gentile} for a solution of (\ref{eq:R}) and to adapt 
its analysis (and results) to the context of the problem posed here.

First of all, let us rewrite the Riccati equation (\ref{eq:R}) as in
\cite{Gentile}. Since $\phi_0 \neq 0$ for all $t \in \R$, we define
$u(t)$ by
\begin{equation*}
g(t) \; = \; i \eps Q(t) u(t) \, ,
\end{equation*} 
where
\begin{equation*}
Q(t) \; := \; \exp\left(-2i \int_0^t g_0(t') \, \der t'\right) 
\; = \; 
\left( \phi_0(t)\right)^{-2}
\, ,
\end{equation*}
which, by (\ref{eq:bohr}), is also quasi-periodic.
We also define,
\begin{equation*}
R(t) \; := \; p_1(t) Q(t)^{-1} \; = \; p_1(t) \phi_0^2(t) 
\; = \; p_1(t)
\exp\left(2i \int_0^t g_0(t') \, \der t'\right) \, .
\end{equation*}
With the above definitions one trivially checks from (\ref{eq:R}) that
\begin{equation} \label{eq:equ}
\dot{u} \; = \; R + \eps Q u^2 \, ,
\end{equation}
which is very similar to the equation studied in \cite{Gentile}.

%%%%%%%%%%%%%%%%%%%%%%%%%%%%%%%%%%%%%%%%%%%%%%%%%%%%%%%%%%%%%%%%%%%%%%%%%
\subsection{Tree expansion}\label{ssec:TreeExpansion}
%%%%%%%%%%%%%%%%%%%%%%%%%%%%%%%%%%%%%%%%%%%%%%%%%%%%%%%%%%%%%%%%%%%%%%%%%

Now we pass to the perturbative expansions and a graphic
representation that will conduct our analysis. As a first attempt (and
also just to introduce notations) we search for a solution of
(\ref{eq:equ}) as a power series in $\eps$:
\begin{equation*}
u(t)\; =\; \sum_{k=0}^\infty \eps^k u^{(k)}(t) \, .
\end{equation*}
Note that, in principle, $u$ does not vanish identically for $\eps =
0$, but $g$ does, since $g \sim \eps u$. By inserting the above Ansatz
into equation (\ref{eq:equ}), we arrive at
\begin{equation} \label{eq:ukdot}
\begin{array}{rcl}
\dot{u}^{(0)} &=& R \, , \\ \\
\dot{u}^{(k)} &=& Q \displaystyle \sum_{k_1 + k_2 = k - 1} u^{(k_1)}
u^{(k_2)} \, , \;\; \forall k \geq 1 \, .
\end{array}
\end{equation}
Since we search for a quasi-periodic solution $u$ of (\ref{eq:equ}), 
it is natural to introduce the following Fourier decomposition:
\begin{equation} \label{eq:uknu}
u^{(k)}(t) \; =\; \sum_{\vnu \in \Z^d} u_{\vnu}^{(k)} e^{i\vnu \cdot
\vomega t} \, ,
\end{equation}
for some $d \geq 1$ to be conveniently fixed later. Note that with
the above decomposition, we have
\begin{equation} \label{eq:expanalitico}
u(t)\; = \;\sum_{k = 0}^{\infty} \eps^k \sum_{\vnu \in \Z^d} 
u_{\vnu}^{(k)} e^{i \vnu \cdot \vomega t} \, .
\end{equation}
Our goal now is to find a graphical representation in terms of trees 
for
the Fourier coefficients $u_{\vnu}^{(k)}$, as
in \cite{Gentile}. 

We now proceed and write the Fourier decomposition
of the functions $p_0$, $p_1$, $\phi_0$, $Q$ and $R$. 
Since $p_0$ is assumed periodic (with period $T_0 = 2\pi/\omega_0$),
we simply have
\begin{equation*}
p_0(t)\; =\; \sum_{n \in \Z} P_n^{(0)} e^{i n \omega_0 t} \, . 
\end{equation*}

The function $p_1$ is assumed quasi-periodic with spectrum of
frequencies contained in the set $\{\undm \cdot \undomega_1, \; \undm
\in \Z^A\}$.
Hence,
\begin{equation*}
p_1(t) \;=\; \sum_{\undm \in \Z^A} P_{\undm}^{(1)} e^{i \undm \cdot
\undomega_1 t} \, .
\end{equation*}

We write the Fourier decompositions of $\phi_0^2$ and $\phi_0^{-2}$ as
follows:
\begin{equation*} 
\left(\phi_0(t)\right)^2 \; = \; \sum_{n \in \Z} \calF_n^{(2)} e^{i(n\omega_0 +
2\Omega_0)t} \, , \qquad
\left(\phi_0(t)\right)^{-2} \; = \; \sum_{n \in \Z} \calF_n^{(-2)} e^{i(n\omega_0 -
2\Omega_0)t} \, .
\end{equation*}
Therefore, the Fourier decomposition of $R$ is 
\begin{equation*}
R(t)\; =\; \sum_{\undm \in \Z^A} P^{(1)}_{\undm}  e^{i \undm \cdot
\undomega_1 t} \sum_{n \in \Z} \calF_n^{(2)} e^{i(n\omega_0 +
2\Omega_0)t} = \sum_{\vnu \in \Z^d} R_{\vnu} e^{i \vnu \cdot \vomega t} 
\, ,
\end{equation*}
where
\begin{equation} \label{eq:nudomega}
\vnu \;:=\; (\undm,\;  n_1,\; n_2) \, , \quad d :=
A+2 \, , \quad \vomega\; :=\; (\undomega_1, \;\omega_0, \;\Omega_0)
\end{equation}
and
\begin{equation*}
R_{\vnu} \;:=\; P^{(1)}_{\undm} \calF_{n_1}^{(2)} \delta_{n_2, 2} \, .
\end{equation*}

With this notation, the Fourier decomposition of $Q$ is as follows:
\begin{equation*}
Q(t)\; = \;\sum_{n \in \Z} \calF_n^{(-2)} e^{i(n\omega_0 -
2\Omega_0)t} = \sum_{\vnu \in \Z^d} Q_{\vnu} e^{i \vnu \cdot \vomega t} 
\, ,
\end{equation*}
where $\vnu$, $d$ and $\vomega$ are as (\ref{eq:nudomega}) and
\begin{equation*}
Q_{\vnu} \;:= \;\delta_{\undm,\undzero} \calF_{n_1}^{(-2)} 
\delta_{n_2, -2} \, .
\end{equation*}

\begin{rmk} \label{rmk:Diophantine}
We assume the following
non-resonant condition on the frequency vector $\vomega$:
\begin{equation*}
\undm \cdot \undomega_1 + n \omega_0 + 2\Omega_0 \;\neq\; 0
\qquad \forall (\undm,\, n) \in \Z^{A+1} .
\end{equation*}
We also impose a Diophantine condition on $\vomega$, namely:
\begin{equation} \label{eq:Diophantine}
|\vomega \cdot \vnu| \;\geq\; \frac{C_0}{|\vnu|^{\tau}}
\qquad \forall \vnu \in \Z^{d}_{*} \; ,
\end{equation}
with $\Z^{d}_{*} := \Z^{d} \setminus \{ \vz \}$,
for two fixed positive constants $C_0$ and $\tau > d-1$.
\end{rmk}

\begin{rmk} \label{rmk:decQR}
By the analyticity assumption on $p_0$ and $p_1$ one obtain 
the following decay for the Fourier coefficients of $Q$ and $R$:
\begin{equation} \label{eq:decQR}
|R_{\vnu}| \;\leq\; \calQ e^{-\kappa|\vnu|} \, , \quad |Q_{\vnu}|\; \leq\;
\calQ e^{-\kappa|\vnu|} \, ,
\end{equation}
for some positive constants $\calQ$ and $\kappa$. 
This will be essential in our forthcoming analysis.
\end{rmk}

We now proceed and insert the decomposition (\ref{eq:uknu}) into 
(\ref{eq:ukdot}). The result is the following recursive relations 
for the coefficients $u^{(k)}_{\vnu}$, $\vnu \neq \vz$:
\begin{equation} \label{eq:recursive}
\begin{array}{rcl}
(i \vomega \cdot \vnu) u^{(0)}_{\vnu} &=& R_{\vnu} \, , \\ \\
(i \vomega \cdot \vnu ) u^{(k)}_{\vnu} &=& \displaystyle
\sum_{k_1 + k_2 = k - 1} \sum_{\vnu_0 + \vnu_1 + \vnu_2 = \vnu}
Q_{\vnu_0} u^{(k_1)}_{\vnu_1} u^{(k_2)}_{\vnu_2} \, ,
\;\; \forall k \geq 1 \, ,
\end{array}
\end{equation}
for all $\vnu \neq \vz$. Since the l.h.s. of (\ref{eq:ukdot}) has zero
average, one must also impose
\begin{equation} \label{eq:masteralphak}
\begin{array}{rcl}
0 &=& R_{\vz} \, , \\ \\
0 &=& \displaystyle \sum_{k_1 + k_2 = k - 1} \sum_{\vnu_0 + \vnu_1 +
\vnu_2 = \vz} Q_{\vnu_0} u^{(k_1)}_{\vnu_1} u^{(k_2)}_{\vnu_2} =:
\left\laa [Qu^2]^{(k-1)}\right\raa \, ,
\quad \forall k \geq 1 \, .
\end{array}
\end{equation}
We note that $R_{\vz} = P^{(1)}_{\undzero} \calF_{0}^{(2)} \delta_{0,
2} = 0$ so there is no problem with the requirement $0 = R_{\vz}$.

%%%%%%%%%%%%%%%%%%%%%%%%%%%%%%%%%%%%%%%%%%%%%%%%%%%%%%%%%%%%%%%%%%%%%%%%%
%\subsubsection{Graphical Representation}
%%%%%%%%%%%%%%%%%%%%%%%%%%%%%%%%%%%%%%%%%%%%%%%%%%%%%%%%%%%%%%%%%%%%%%%%%

The graphical representation of the coefficients $u^{(k)}_{\vnu}$ is
almost exactly like in Ref.~\cite{Gentile}. We advise the reader to see
Section~4 of Ref.~\cite{Gentile} for details. The only two essential
differences are the following: (1) Here we represent as black bullets
the factors $R_{\vnu}$, while in Ref.~\cite{Gentile}
they were associated to $(Q^{-1})_{\vnu}$. (2) The order of a
tree here is given only by the sum of vertices plus the sum of
the order labels of the white bullets, while in Ref.~\cite{Gentile}
the number of black bullets was also counted in the order.
There is also here a slight modification of notation:
while in Ref.~\cite{Gentile} $u^{(k)}_{\vz} := c^{(k)}$, here 
$u^{(k)}_{\vz} := \alpha^{(k)}$, for all $k \geq 0$. We give below the
pertinent definitions of Ref.~\cite{Gentile} adapted to our present 
case.

\begin{defi} \label{def:tree}
A tree $\theta$ is a connected set of points and lines with no cycle
such that all the lines are oriented toward a unique point called the
{\rm root}. We call {\rm nodes} all the points in a tree except the root.
The root only admits one entering line:
such a line is called the {\rm root line}. 
The orientation of the lines in a tree induces a partial ordering 
relation between the nodes. We denote by $\preceq$ this relation:
given two nodes $v$ and $w$, we shall write $w \preceq v$ 
every time $v$ is a long the path (of lines) which connects $w$ to the
root. Given a tree $\theta$, we can identify in $\theta$
the following subsets.
\begin{itemize}
\item $E(\theta)$: the set of {\rm endpoints} (final nodes) in 
$\theta$.
A node $v \in \theta$ will be an endpoint if no line enters $v$.
We denote by $|E(\theta)|$ the number of endpoints in $\theta$.
\item $E_W(\theta) \subseteq E(\theta)$:
the set of {\rm white bullets} in $\theta$. With each $v \in 
E_W(\theta)$
we associate a {\rm mode} label $\vnu_v = \vz$, an {\rm order} label
$k_v \in \Z_+$ and a {\rm node factor} $F_v = \alpha^{(k_v)}$.
We denote by $|E_W(\theta)|$ the number of white bullets in $\theta$.
\item $E_B(\theta) = E(\theta) \setminus E_W(\theta)$:
the set of {\rm black bullets} in $\theta$. With each $v
\in E_B(\theta)$ we associate a {\rm mode label} $\vnu_v \neq \vz$ and
a {\rm node factor} $F_v = R_{\vnu_v}$.
We denote by $|E_B(\theta)|$ the number of black bullets in $\theta$.
\item $V(\theta)$: the set of {\rm vertices} in $\theta$. If $v \in
V(\theta)$, then $v$ has at least one entering line.
We associate with each vertex $v\in V(\theta)$ a {\rm mode} label 
$\vnu_v \in \Z^d$ and a {\rm node factor} $F_v = Q_{\vnu_v}$. 
We denote by $|V(\theta)|$ the number of vertices in $\theta$.
\item $B(\theta) = E_B(\theta) \cup V(\theta)$:
the set of black bullets and vertices in $\theta$.
We denote by $|B(\theta)|$ the number of black bullets plus vertices
in $\theta$, i.e. $|B(\theta)| = |E_B(\theta)| + |V(\theta)|$. 
\item $L(\theta)$: the set of {\rm lines} in $\theta$. Each line $\ell
\in L(\theta)$ leaves a point $v$ and enters another one which we
shall denote by $v'$. Since $\ell$ is uniquely identified with $v$
(the point which $\ell$ leaves), we may write $\ell = \ell_v$. 
For each line $\ell$ we associate a {\rm momentum} label $\vnu_\ell
\in \Z^d$ and a {\rm propagator} $g_\ell = 1/(i\vomega \cdot 
\vnu_\ell)$
if $\vnu_\ell \neq \vz$ and $g_\ell = 1$ if $\vnu_\ell = \vz$;
we say that the momentum $\vnu_{\ell}$ flows through the line $\ell$.
The modes and the momenta are related by the following: if 
$\ell = \ell_v$ and $\ell'$, $\ell''$ are the lines entering $v$, then
\begin{equation} \label{eq:momentum}
\vnu_\ell = \vnu_v + \vnu_{\ell'} + \vnu_{\ell''} =
\sum_{\substack{ w \in B(\theta) \\ w \preceq v}} \vnu_w \, . 
\end{equation}
We denote by $|L(\theta)|$ the number of lines in $\theta$.
\end{itemize}  
We call {\rm equivalent} two trees which can be transformed into
each other by continuously deforming the lines in such a way that
they do not cross each other.
\end{defi}

\begin{defi}
Let $\calT_{k, \vnu}$ be the set of inequivalent trees $\theta$
satisfying:
\begin{enumerate}
\item for each vertex $v \in V(\theta)$, there exist exactly two
entering lines in $v$;
\item for each line $\ell$ which is not the root line one has
$\vnu_{\ell}=\vz$ if and only if $\ell$ leaves a white bullet;
\item the number of vertices and the sum of all the order labels of
the white bullets are such that defining
$k_1 := |V(\theta)|$ and $k_2 := \sum_{v
\in E_W(\theta)} k_v$, we have $k_1 + k_2 = k$;
\item the momentum flowing through the root line is $\vnu$.
\end{enumerate}
We refer to $\calT_{k, \vnu}$ as the {\rm set of trees of order
$k$ and total momentum $\vnu$.}
\end{defi}

Based on the above definitions, we write for all $k \geq 0$ and for
all $\vnu \in \Z^d, \vnu \neq \vz$:
\begin{equation} \label{eq:ukn} 
u^{(k)}_{\vnu}\; =\; \sum_{\theta \in \calT_{k,\vnu}} \Val(\theta) \, ,
\end{equation}
where $\Val: \calT_{k,\vnu} \to \C$ is called the {\it value} of the
tree $\theta$ and it is defined by
\begin{equation} \label{eq:val}
\Val(\theta) \;:=\; \left(\prod_{\ell \in L(\theta)}
g_\ell\right)\left(\prod_{v \in E(\theta) \cup V(\theta)} F_v\right)\, ,
\end{equation}
where
\begin{equation*}
g_\ell \;:=\; \left\{
\begin{array}{ll}
\displaystyle
\frac{1}{i \vomega \cdot \vnu_\ell} \, , & \vnu_\ell \neq \vz , \\ 
& \\
1 \, , & \vnu_\ell = \vz ,
\end{array}
\right.
\qquad\quad
F_v \;:= \;\left\{
\begin{array}{ll}
Q_{\vnu_v} \, , & v \in V(\theta) , \\
& \\
R_{\vnu_v} \, , & v \in E_B(\theta) , \\
& \\
\alpha^{(k_v)} \, , & v \in E_W(\theta) . \\
\end{array}
\right.
\end{equation*}
All the trees which appear in the expansion of the
coefficient $u^{(k)}_{\vnu}$ belong to $\calT_{k, \vnu}$.
Reciprocally, every tree in $\calT_{k, \vnu}$
appears in the graphical expansion of $u^{(k)}_{\vnu}$.

It is clear that the constants $u^{(k)}_{\vz} = \alpha^{(k)}$ should
be recursively fixed from conditions (\ref{eq:masteralphak}). We leave
this for next section.

%%%%%%%%%%%%%%%%%%%%%%%%%%%%%%%%%%%%%%%%%%%%%%%%%%%%%%%%%%%%%%%%%%%%%%%%%
%%%%%%%%%%%%%%%%%%%%%%%%%%%%%%%%%%%%%%%%%%%%%%%%%%%%%%%%%%%%%%%%%%%%%%%%%
\zerarcounters
\section{Analysis of the zero modes. 
 Fixing $\boldsymbol{{\alpha}}^{\boldsymbol{(k)}}$,
  $\boldsymbol{k \geq 0}$}
\label{sec:fixing}
%%%%%%%%%%%%%%%%%%%%%%%%%%%%%%%%%%%%%%%%%%%%%%%%%%%%%%%%%%%%%%%%%%%%%%%%%
%%%%%%%%%%%%%%%%%%%%%%%%%%%%%%%%%%%%%%%%%%%%%%%%%%%%%%%%%%%%%%%%%%%%%%%%%

We now analyze equations (\ref{eq:recursive}) and
(\ref{eq:masteralphak}) in order to fix $\alpha^{(k)}$, $k \geq 0$. 
One should keep in mind that these equations are of a recursive
nature. Therefore, one first starts by fixing $u^{(0)}_{\vnu}$, $\vnu \neq
\vz$, from (\ref{eq:recursive}), then one fixes $\alpha^{(0)}$ from
(\ref{eq:masteralphak}), then one goes back to (\ref{eq:recursive}) to fix 
$u^{(1)}_{\vnu}$, $\vnu \neq \vz$, and so on. Our intention here is to
obtain a general recursive expression for the zero modes coefficients 
$\alpha^{(k)}$. We shall prove that, apart from a spurious situation, the
only possible choice of constants $\alpha^{(k)}$ compatible with
(\ref{eq:masteralphak}) is $\alpha^{(k)} = 0$, for all $k \geq 0$. 
 
\begin{rmk} \label{rmk:topological}
Let $\theta \in \calT_{k,\vnu}$, $k \geq 0$, $\vnu \in \Z^d$.
Since $k = |V(\theta)| + \sum_{v \in E_W(\theta)} k_v$, one clearly
has $0 \leq |V(\theta)| \leq k$. If, e.g., $E_W(\theta)$
contains only one white bullet with order label $k$,
then $|V(\theta)| = 0$; on the other hand if
$E_W(\theta)$ contains only white bullets with order label all
equal to zero or if it is an empty set, then $|V(\theta)| = k$.
Another simple observation is that, by topological reasons,
the total number of endpoints of $\theta$ is exactly $|V(\theta)| + 1$
(this can be easily proved by induction).
So, $|E_W(\theta)| + |E_B(\theta)| = |V(\theta)| + 1$ and
one has $0 \leq |E_B(\theta)| \leq |V(\theta)| + 1$.
\end{rmk}

%%%%%%%%%%%%%%%%%%%%%%%%%%%%%%%%%%%%%%%%%%%%%%%%%%%%%%%%%%%%%%%%%%%%%%%%%
\begin{lema} \label{lemma:n2}
In $u^{(k)}_{\vnu}$, $k \geq 0$, $\vnu = (\undm, n_1, n_2) \in \Z^d$,
$n_2$ belongs to the following set of even integers: $\{-2k, -2(k-1),
\ldots, -2, 0, 2\}$.$\ES$
\end{lema}
%%%%%%%%%%%%%%%%%%%%%%%%%%%%%%%%%%%%%%%%%%%%%%%%%%%%%%%%%%%%%%%%%%%%%%%%%

\prova For $k = 0$, $n_2 = 2$ since $u^{(0)}_{\vnu} \propto
\delta_{n_2, 2}$. Now let $k \geq 1$ and $\theta \in \calT_{k,\vnu}$
be a tree contributing to $u^{(k)}_{\vnu}$. With each vertex $v \in
V(\theta)$ one associates the factor $\delta_{n_2^{(v)}, -2}$ in
$\Val(\theta)$ and with each black bullet $b \in E_B(\theta)$ the
factor $\delta_{n_2^{(b)}, 2}$. Thus, due to the conservation of
momentum (\ref{eq:momentum}), one must have the constraint
$n_2 = \sum_{v \in V(\theta)}n_2^{(v)} + \sum_{b \in E_B(\theta)}
n_2^{(b)} = 2|E_B(\theta)|-2|V(\theta)|$ in the root line. From
Remark.~\ref{rmk:topological}, one concludes that
$n_2 = -2k, -2(k-1), \ldots, -2, 0, 2$.$\EP$

\begin{defi} \label{def:tilT}
Let $\theta \in \calT_{k, \vnu}$, $k \geq 0$, $\vnu \in \Z^d$,
such that $E_W(\theta)$ is non-empty.
Let $\calA_\theta \subseteq E_W(\theta)$ be non-empty. We define
$\theta\!\setminus\!\calA_\theta$ as
the {\rm $\calA_\theta$-amputated tree} generated by amputating
the subset $\calA_\theta$ of white bullets from $\theta$.
This means that
\begin{equation*}
\Val\left(\theta\!\setminus\!\calA_\theta\right) =
\frac{\displaystyle \Val(\theta)}{\displaystyle \prod_{v \in
\calA_\theta} \alpha^{(k_v)}} \quad \text{and} \quad
k_{\theta\setminus\calA_\theta} = k - \sum_{v \in
\calA_\theta} k_v \, ,
\end{equation*}
where $k_{\theta\setminus\calA_\theta}$ denotes the order of the
$\calA_\theta$-amputated tree. We call {\rm amputated line}
any line coming out from a white bullet in $\calA_{\theta}$,
after amputation of $\calA_{\theta}$. Now let
\begin{equation*}
\calT^{(p)}_{k,\vnu} := \left\{\theta \in \calT_{k, \vnu} \; : \;
E_W(\theta) = \left\{ v \right\} \text{ with } k_{v}= p 
\, \right\} \, .
\end{equation*}
This means that a tree in $\calT^{(p)}_{k,\vnu}$ has only one white
bullet with order label $p$ (and hence $k-p$ vertices). We now
amputate the white bullet in $\calT^{(p)}_{k,\vnu}$; this gives
the definition of the set
\begin{equation*}
\widetilde{\calT}^{(p)}_{k,\vnu} := \left\{
\theta\!\setminus\!E_W(\theta) \; : \; \theta \in \calT^{(p)}_{k,
\vnu}\right\} \, .
\end{equation*}
Of course the order of a tree in $\widetilde{\calT}^{(p)}_{k,\vnu}$ is
equal to its number of vertices, which is just $k-p$. We also
introduce here the shorthand: $\widetilde{\calT}^{(0)}_{k,\vnu} =:
\widetilde{\calT}_{k,\vnu}$, for all $k \geq 1$.
\end{defi}

\begin{rmk} \label{rmk:T0Tp}
From the previous definition and from the fact that $g_{\ell} =1$
when $\ell$ leaves a white bullet, one notes that
$\widetilde{\calT}^{(p)}_{k,\vnu} = \widetilde{\calT}_{k-p,\vnu}$.
Indeed, let $\tilde{\theta} \in \widetilde{\calT}^{(p)}_{k,\vnu}$
be arbitrary. It follows that there exists $\theta \in \calT_{k,\vnu}$
with $E_W(\theta) = \left\{ v \right\}$ with $k_{v}=p$ such that
$\tilde{\theta} = \theta \setminus E_W(\theta)$.
This means that $|V(\theta)| = k - p$.
Now take $\theta' \in \calT_{k-p,\vnu}$ with $E_W(\theta')=
\left\{ v \right\}$ with $k_{v}=0$ such that
$\tilde{\theta} = \theta'\setminus E_W(\theta')$.
Since $\theta' \in \calT^{(0)}_{k-p, \vnu}$, one finds
that $\tilde{\theta} \in \widetilde{\calT}_{k-p,\vnu}$. Hence
$\widetilde{\calT}^{(p)}_{k,\vnu} \subseteq
\widetilde{\calT}_{k-p,\vnu}$. On the other hand, take an arbitrary
$\tilde{\theta} \in \widetilde{\calT}_{k-p,\vnu}$. Then
$\tilde{\theta} = \theta\setminus E_W(\theta)$ where $\theta \in
\calT_{k-p, \vnu}$ and $E_W(\theta) = \left\{ v \right\}$ with 
$k_{v}=0$.
Hence, $|V(\theta)| = k - p$, which means that one can take
$\theta' \in \calT_{k, \vnu}$ s.t. $E_W(\theta') = 
\left\{ v \right\}$ with $k_{v}=p$ and $\tilde{\theta} = 
\theta'\setminus
E_W(\theta')$. Since $\theta' \in \calT_{k,\vnu}^{(p)}$, one concludes
that $\tilde{\theta} \in \widetilde{\calT}^{(p)}_{k, \vnu}$. Therefore,
$\widetilde{\calT}_{k-p, \vnu} \subseteq
\widetilde{\calT}^{(p)}_{k, \vnu}$.
\end{rmk}

%%%%%%%%%%%%%%%%%%%%%%%%%%%%%%%%%%%%%%%%%%%%%%%%%%%%%%%%%%%%%%%%%%%%%%%%%
\begin{lema} \label{lemma:Gj}
Let $k \geq 1$, then $\left\laa [Qu^2]^{(k-1)}\right\raa =
\sum_{p=0}^{k-1} \alpha^{(p)}G_{k-p}$, where, for all $j \geq 1$,
$G_j := \sum_{\theta \in \widetilde{\calT}_{j,\vz} }
\Val(\theta)$.$\ES$
\end{lema}
%%%%%%%%%%%%%%%%%%%%%%%%%%%%%%%%%%%%%%%%%%%%%%%%%%%%%%%%%%%%%%%%%%%%%%%%%

\prova From the definition,
\begin{equation*}
\left\laa [Qu^2]^{(k-1)}\right\raa =
\displaystyle \sum_{k_1 + k_2 = k - 1} \sum_{\vnu_0 + \vnu_1 +
\vnu_2 = \vz} Q_{\vnu_0} u^{(k_1)}_{\vnu_1} u^{(k_2)}_{\vnu_2} \, ,
\end{equation*}
so that by using the definition of tree value (\ref{eq:val})
and the notations of Definition~\ref{def:tree} one
immediately realizes that one can write
\begin{equation*}
\left\laa [Qu^2]^{(k-1)}\right\raa =
\sum_{\theta \in \calT_{k, \vz}} \Val(\theta) \, .
\end{equation*}
Now let $\vnu_1 = (\undm_1, n^{(1)}_1, n^{(1)}_2) \in \Z^d$ and
$\vnu_2 = (\undm_2, n^{(2)}_1, n^{(2)}_2) \in \Z^d$. From
Lemma~\ref{lemma:n2},
\begin{eqnarray*}
n^{(1)}_2 &\in& \{-2k_1, -2(k_1-1), \ldots, -2, 0, 2\} \\
n^{(2)}_2 &\in& \{-2k_2, -2(k_2-1), \ldots, -2, 0, 2\} \, .
\end{eqnarray*}
To be more precise, let $\theta_j \in \calT_{k_j, \vnu_j}$, $j = 1,
2$, be a tree contributing to $u^{(k_j)}_{\vnu_j}$, 
then $n^{(1)}_2 = 2(b_1 - v_1)$ and $n^{(2)}_2 = 2(b_2 - v_2)$, 
where $b_j$ and $v_j$ are the number of black bullets and
the number of vertices in $\theta_j$, respectively.
From $\theta_1$ and $\theta_2$ we would like to
construct a tree $\theta \in \calT_{k, \vz}$, $k = k_1+k_2+1$,
contributing to $\left\laa [Qu^2]^{(k-1)}\right\raa$. First one must
note that the root lines of $\theta_1$ and $\theta_2$ enter a vertex
in $\theta$ with mode $\vnu_0 = (\undzero, n_1^{(0)}, -2)$.
As the line which exits this vertex (root line) carries 
zero momentum, one has the constraint $-2 + n_2^{(1)} + n_2^{(2)} =
0$. Thus, $(b_1 + b_2) - (v_1 + v_2 + 1) = 0$. This last relation
implies that $|E_B(\theta)| = |V(\theta)|$, so (see
Remark~\ref{rmk:topological}) the tree $\theta$ contributing to
$\left\laa [Qu^2]^{(k-1)}\right\raa$ must have exactly one white
bullet (with some order label $p$). Of course $|V(\theta)| + p = k$ 
and $1 \leq |V(\theta)| \leq k$, hence $0 \leq p \leq k-1$. 
Therefore, one can write
\begin{eqnarray*}
\left\laa [Qu^2]^{(k-1)}\right\raa &=& \sum_{p=0}^{k-1} \sum_{\theta 
\in
\calT^{(p)}_{k, \vz}} \Val(\theta) \; = \; \sum_{p=0}^{k-1}
\sum_{\theta \in \calT^{(p)}_{k, \vz}} \alpha^{(p)}
\Val(\theta\setminus E_W(\theta)) \nonumber \\
&=& \sum_{p=0}^{k-1} \alpha^{(p)} \sum_{\theta \in
\widetilde{\calT}^{(p)}_{k, \vz}} \Val(\theta) \; = \;
\sum_{p=0}^{k-1} \alpha^{(p)} \sum_{\theta \in
\widetilde{\calT}_{k-p, \vz}} \Val(\theta) \nonumber \\
&=:& \sum_{p=0}^{k-1} \alpha^{(p)}G_{k-p} \, ,
\end{eqnarray*}
where Remark~\ref{rmk:T0Tp} was used. Note that, by
construction, $G_j$, $j \geq 1$, is expressed by a sum of trees with
no white bullets such that they have exactly $j$ black bullets and
$j$ vertices.$\EP$

\begin{defi}
Let $k \geq 1$ and $\widetilde{\calT}_{k, \vnu}$ as in
Definition~\ref{def:tilT}. We split $\widetilde{\calT}_{k, \vnu}$ into
two disjoint sets as follows: $
\widetilde{\calT}_{k, \vnu} =: \widetilde{\calT}_{k, \vnu}^{c} \bigcup
\widetilde{\calT}_{k, \vnu}^{nc}$, where
\begin{itemize}
\item $\widetilde{\calT}_{k, \vnu}^{c}$: set of trees in
$\widetilde{\calT}_{k, \vnu}$ such that the amputated line is
\underline{connected} to the root line;
\item $\widetilde{\calT}_{k, \vnu}^{nc}$: set of trees in
$\widetilde{\calT}_{k, \vnu}$ such that the amputated line is
\underline{not connected} to the root line.
\end{itemize}
We call a tree in $\widetilde{\calT}_{k, \vnu}^{c}$ as a {\rm 
$c$-class}
tree and a tree in $\widetilde{\calT}_{k, \vnu}^{nc}$ as a
{\rm $nc$-class} tree. Note any $nc$-class tree has order $k \geq 2$.
\end{defi}

\begin{figure}[!ht]
\centering
\begin{minipage}[c]{0.45\textwidth}
\begin{center}
\includegraphics[scale=0.95]{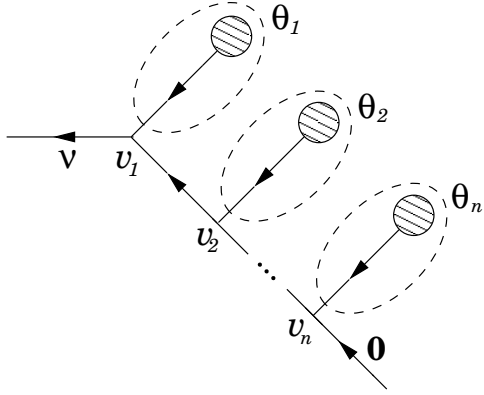}
\end{center}
\end{minipage}
%\hfill
\hspace{0.3cm}
\begin{minipage}[c]{0.45\textwidth}
\centering \caption{Canonical form of a tree in
$\widetilde{\calT}_{k, \vnu}^{nc}$. The dashed bullet represents a
general subtree containing only black bullets. Each canonical
subtree $\theta_j$, $1 \leq j \leq n$, is of order $k_{\theta_j}$
and  contains exactly $k_{\theta_j}$ vertices and $k_{\theta_j} +
1$ black bullets. Taking into account the $n$ canonical vertices
$\{v_1, \ldots, v_2\}$, one has $k = n +
\sum_{j=1}^{n}k_{\theta_j}$. Note that $2 \leq n \leq k$. The
amputation of the white bullet leaves a line with vanishing
momentum connected to the vertex $v_n$; we call amputated line
such a line.} \label{fig:canonical_form}
\end{minipage}
\end{figure}

Any tree in $\widetilde{\calT}_{k, \vnu}^{nc}$ can be transformed to
be drawn in its ``canonical form'' as depicted in
Figure~\ref{fig:canonical_form}. Indeed, let $\theta \in
\widetilde{\calT}_{k, \vnu}^{nc}$, $k \geq 2$, $\vnu \in \Z^d$, be
arbitrary. Let $v_1'$ be the vertex connected to the amputated line
of $\theta$. Define $v_2'$ as the vertex such that one of its
entering lines is exactly the line exiting $v_1'$. Define $v_j'$
inductively as the vertex such that one of its
entering lines is exactly the line exiting $v_{j-1}'$. If, for some
$j \geq 2$, $v_{j}'$ is the vertex connected to the root line, then
we set $n := j$. Now relabel the $n$ vertices defined above as
follows: $v_j = v_{n-j+1}'$, $1 \leq j \leq n$. The vertices $v_j$
will be called {\it canonical vertices}.
Set $\theta_j$, $1 \leq j \leq n-1$, as the subtree whose
root line is the one entering $v_j$ and not exiting $v_{j+1}$;
$\theta_n$ is defined as the subtree
whose root line enters $v_n$, not being the amputated line.
The subtrees $\theta_j$ will be called {\it canonical subtrees}.
Now draw the tree in such a way that the root line of each
$\theta_{j}$, $1\le j \le n$, is the upper line
entering the vertex $v_{j}$: in this way
$\theta \in \widetilde{\calT}_{k, \vnu}^{nc}$ is
as represented in Figure~\ref{fig:canonical_form}.
From now on, any tree in $\widetilde{\calT}_{k, \vnu}^{nc}$ is
thought of as being drawn in its ``canonical form''.

\begin{rmk} \label{rmk:nonzero}
Each canonical subtree $\theta_j \subset \theta$ defined above gives a
contribution to $u^{(k_{\theta_j})}_{\vnu_j}$ if $k_{\theta_j}$ is the
order of $\theta_j$ and if $\vnu_j$ is the momentum flowing through
its root line. Of course $\vnu_j \neq \vz$ since this would give a
contribution to $\alpha^{(k_j)}$ and white bullets are discarded
along the construction. More
generally, as each line in $L(\theta)$, which is neither the root line
nor the line exiting from the amputated white bullet,
can be seen as the root line of a subtree, then it must have
a momentum different from zero, as there are no other white bullets.
\end{rmk}

\begin{rmk} \label{rmk:numbercanonicalform}
Note that there are $2^{n}$ inequivalent trees in
$\widetilde{\calT}_{k, \vnu}^{nc}$ admitting the same canonical form
with $n$ canonical subtrees.
\end{rmk}

One now writes the value of a canonical subtree $\theta_j$ as
\begin{equation*}
\Val(\theta_j) = \frac{b^{(j)}_{\vnu_j}}{i\vomega\cdot\vnu_j} \equiv
B^{(j)}_{\vnu_j} \, , \quad \vnu_j \neq \vz \, , \quad 1 \leq j \leq 
n\, .
\end{equation*}
Therefore, $\theta_j$ gives a contribution to the function
\begin{equation} \label{eq:intBj}
B_j(t) = \int_0^t \der t' \, b_j(t') + C_j \, ,
\end{equation}
where the integration constant $C_j$ is chosen in such a way that
summed to the constant term arising from the definite integral gives 
the
zero Fourier mode of $B_j$:
\begin{equation} \label{eq:Cj}
C_j - \sum_{\vnu \in \Z^d_\ast}
\frac{b^{(j)}_{\vnu}}{i\vomega\cdot\vnu} = B^{(j)}_{\vz} \, .
\end{equation}
Since $B^{(j)}_{\vz}$ must be vanishing,
\begin{equation} \label{eq:int}
B_j(t) = \sum_{\vnu \in \Z^d_\ast} B^{(j)}_{\vnu}
e^{i \vomega \cdot \vnu t} =: \int b_j \, ,
\end{equation}
where the above integral has to be interpreted as a shorthand
notation for (\ref{eq:intBj}) with $C_j$ fixed from (\ref{eq:Cj}) by
imposing $B^{(j)}_{\vz} = 0$. One should think of it as just a zero
average primitive of $b_j$.

%%%%%%%%%%%%%%%%%%%%%%%%%%%%%%%%%%%%%%%%%%%%%%%%%%%%%%%%%%%%%%%%%%%%%%%%%
\begin{lema} \label{lemma:identity}
Let $\theta \in \widetilde{\calT}^{nc}_{k, \vz}$ be a $nc$-class tree
as the one in Figure~\ref{fig:canonical_form} with order $k \geq 2$
and $2 \leq n \leq k$ canonical subtrees $\theta_1, \ldots, \theta_n$.
Let $\calN_{\vz} := \left\{ \vnu_v \in \Z^d_\ast \; : \; v \in
B(\theta) \right\}$ and $a_j(t) := Q(t) B_j(t)$,
$1 \leq j \leq n$. Then
\begin{equation} \label{eq:avint}
\sum_{\calN_{\vz}} \Val(\theta) = \left\laa a_1 \int a_2 \int \cdots 
\int
a_{n-1} \int a_n \right\raa \, ,
\end{equation}
where all the integrals are in the sense of (\ref{eq:int}).$\ES$
\end{lema}
%%%%%%%%%%%%%%%%%%%%%%%%%%%%%%%%%%%%%%%%%%%%%%%%%%%%%%%%%%%%%%%%%%%%%%%%%

\prova Let $1 \leq j \leq n$ and denote by $\vnu_{0, j}$ the
Fourier mode of the canonical vertex $v_j$ and by $\vnu_j$ the
momentum flowing through the root line of the canonical subtree
$\theta_j$. Note that $\vnu_{0, j} \neq \vz$, since this would
give $Q_{\vz} = 0$. Also $\vnu_j \neq \vz$ and more generally,
$\vnu_\ell \neq \vz$, for all $\ell \in L(\theta)$ different
from the root line and the line leaving the amputated white bullet
(see Remark~\ref{rmk:nonzero}). Now the momentum flowing through the
root line is zero, which means that
\begin{equation*}
\sum_{j=1}^n \vnu_j + \sum_{j=1}^n \vnu_{0, j} = \vz \, .
\end{equation*}
Therefore, by an explicit computation,
\begin{eqnarray} \label{eq:sumN0}
\sum_{\calN_{\vz}} \Val(\theta) &=& \prod_{r=1}^n \sum_{\vnu_r}
\sum_{\vnu_{0, r}} \frac{\prod_{j=1}^{n} Q_{\vnu_{0,j}} \,
b^{(j)}_{\vnu_j}}{\prod_{j=1}^{n} (i \vomega \cdot \vnu_j)
\sum_{p=1}^{j} i \vomega \cdot
\left(\vnu_{n-p+1, 0} + \vnu_{n-p+1}\right)} \nonumber \\
&=& \prod_{r=1}^n \sum_{\vnu_r} \sum_{\vnu_{0, r}}
\frac{\prod_{j=1}^{n} Q_{\vnu_{0,j}} \,
B^{(j)}_{\vnu_j}}{\prod_{j=1}^{n} \sum_{p=1}^{j} i \vomega \cdot
\left(\vnu_{n-p+1, 0} + \vnu_{n-p+1}\right)} \nonumber \\
&=& \left\laa (Q B_1) \int (Q B_2) \int \cdots \int (Q B_{n-1}) \int
(Q B_n) \right\raa \, ,
\end{eqnarray}
which proves the statement.$\EP$

\vspace{0.4cm}

Let $\theta \in \widetilde{\calT}^{nc}_{k, \vz}$ be a $nc$-class tree
as the one in Figure~\ref{fig:canonical_form} with order $k \geq 2$ 
and $n \leq k$ canonical subtrees $\theta_1, \ldots, \theta_n$. 
Of course if some two canonical trees $\theta_i$, $\theta_j$ are
equivalent, then one gets the same contribution in (\ref{eq:avint}) by
permuting $a_i$ with $a_j$. This motivate us to give the following
definitions: let $\Theta = \{\theta_1, \ldots \theta_n\}$ be the
collection of all canonical subtrees of 
$\theta \in \widetilde{\calT}^{nc}_{k, \vz}$. We split $\Theta$ into 
$1 \leq m \leq n$ disjoint subsets $E_j$, $1 \leq j \leq m$,
such that $E_1$ is composed by all trees in $\Theta$ which are equivalent
to $\theta_1$, $E_2$ is composed by all trees in $\Theta \setminus
E_1$ which are equivalent to the first tree of $\Theta \setminus
E_1$ and so on. In this way, $\Theta = \bigcup_{j=1}^{m} E_j$, where 
each
$E_j$ collects together all trees which are equivalent to each other.
Of course each subset $E_j$ contains $r_j = |E_j|$ (equivalent) trees
such that $\sum_{j=1}^{m} r_j = n$. The contribution to (\ref{eq:avint}) of
all trees within the same $E_j$ is denoted by $a_{E_j}$, where it
represents the function $a_p$ associated to the tree $\theta_p$ which
is equivalent to all trees in $E_j$. Now, let $S_n$ denote the usual 
permutation group of $n$ elements. We define $S_n^{\Theta} := 
S_n \setminus \{\pi \in S_n \; : \; \pi(i) = j \text{ if } i 
\neq j \text{ and } \theta_i, \theta_j \in E_p \text{ for some } 
1 \leq i, j, p \leq n\}$. The set $S_n^\Theta$ will be called the set
of all valid permutations within $\Theta$.

%%%%%%%%%%%%%%%%%%%%%%%%%%%%%%%%%%%%%%%%%%%%%%%%%%%%%%%%%%%%%%%%%%%%%%%%%
\begin{lema} \label{lemma:derivative}
Let $\theta \in \widetilde{\calT}^{nc}_{k, \vz}$ be a $nc$-class tree
as the one in Figure~\ref{fig:canonical_form} with order $k \geq 2$ 
and $n \leq k$ canonical subtrees $\theta_1, \ldots, \theta_n$. Then
\begin{equation*}
\sum_{\pi \in S_n^{\Theta}} \left\laa a_{\pi(1)} \int a_{\pi(2)} \int
a_{\pi(3)} \cdots \int a_{\pi(n)} \right\raa = \frac{(-1)^{n+1}}{r_1!
\cdots r_m!} \left\laa \frac{\der}{\der t}
(A_{E_1}^{r_1} \cdots A_{E_m}^{r_m})\right\raa  = 0 \, ,
\end{equation*}
where $A_{E_j} = \int a_{E_j}$, for all $1 \leq j \leq m$.$\ES$
\end{lema}
%%%%%%%%%%%%%%%%%%%%%%%%%%%%%%%%%%%%%%%%%%%%%%%%%%%%%%%%%%%%%%%%%%%%%%%%%

\prova First let us assume that all the subtrees $\theta_1, \ldots
\theta_n$ are different. Therefore we have $a_1 \neq a_2 \neq \cdots
\neq a_n$ in (\ref{eq:avint}). With this assumption $\Theta =
\{\theta_1, \ldots, \theta_n\} = \bigcup_{j=1}^{n} E_j$ with $E_j =
\{\theta_j\}$ and $r_j = 1$ for all $1 \leq j \leq n$. By an
integration by parts, one has
\begin{equation*}
\left\langle a_{1} \int a_{2} \int a_{3} \ldots \int a_{n}
\right\rangle = - \left\langle \left( \int a_{3} \ldots \int a_{n}
\right) \left( a_{2} \int a_{1} \right) \right\rangle \, ,
\end{equation*}
so that by summing also the term $1 \leftrightarrow 2$ and performing
another integration by parts, one obtains
\begin{multline*}
\left\langle a_{1} \int a_{2} \int a_{3} \ldots \int a_{n}
\right\rangle + \left\langle a_{2} \int a_{1} \int a_{3} \ldots \int 
a_{n}
\right\rangle \\ = - \left\langle \left( \int a_{3} \ldots \int a_{n}
\right) \frac{\der}{\der t} \left( A_{1} A_{2}  \right) \right\rangle 
= \left\langle \left( \int a_{4} \ldots \int a_{n}
\right) \left( A_{1} A_{2} \, a_3  \right)
\right\rangle \, . 
\end{multline*}
Note that to construct the derivative of $(A_1A_2)$ above we  have
used the $2! = 2$ permutations of $a_1$, $a_2$:  
$(1\,2\,3\,\cdots\,n)$ and $(2\,1\,3\,\cdots\,n)$. 
So, by using also $(1\,3\,2\,\cdots\,n)$, $(3\,1\,2\,\cdots\,n)$ and
$(3\,2\,1\,\cdots\,n)$, $(2\,3\,1\,\cdots\,n)$, one gets
\begin{equation*}
\left\langle \left( \int a_{4} \ldots \int a_{n}
\right) \left( A_{1} A_{3} \, a_2  \right)
\right\rangle \quad \text{and} \quad
\left\langle \left( \int a_{4} \ldots \int a_{n}
\right) \left( A_{3} A_{2} \, a_1  \right)
\right\rangle \, . 
\end{equation*}
Therefore, the sum of the $3! = 6$ terms obtained by the permutation
of $a_1, a_2, a_3$, gives
\begin{equation*}
\left\langle \left( \int a_{4} \ldots \int a_{n} \right)
\frac{\der}{\der t} \left( A_{1} A_{2} A_{3} \right) \right\rangle =
- \left\langle \left( \int a_{5} \ldots \int a_{n}
\right) \left( A_{1} A_{2} A_{3} a_4 \right) \right\rangle \, .  
\end{equation*}
We now go on and sum the $4! = 24$ terms obtained by the permutation
of $a_1, a_2, a_3, a_4$ to obtain a derivative of $(A_1 A_2 A_3
A_4)$. We iterate this procedure until exhausting the $n!$
permutations of $a_1, \ldots, a_n$, giving
\begin{equation*}
\sum_{\pi \in S_n}
\left\langle a_{\pi(1)} \int a_{\pi(2)} \int a_{\pi(3)} \ldots
\int a_{\pi(n)} \right\rangle =
(-1)^{n+1} \left\langle
\frac{\der}{\der t} \left( A_{1} A_{2} \ldots A_{n} \right)
\right\rangle = 0 \, , 
\end{equation*}
which is the statement of the lemma in the case where all $a_j$ are
different.

Now assume the more general situation where $\Theta = \{\theta_1,
\ldots, \theta_n\} = \bigcup_{j=1}^m E_j$, for some $m < n$. Then we
can permute $i \leftrightarrow j$ iff $a_i \neq a_j$ (we call this a
valid permutation). The set of all valid permutations within $\Theta$
is what we have denoted by $S^\Theta_n$ above. The total number of
valid permutations is $\frac{n!}{r_1! \cdots r_m!}$. Therefore, by
using the result of last formula, one arrives at the general
statement.$\EP$

\begin{rmk} \label{rmk:derivative}
Note that the cancellation described by Lemma~\ref{lemma:derivative}
occurs at fixed values of the mode labels. In other words, if we
consider a fixed set of mode labels in $\calN_{0}$
contributing to the sum in (\ref{eq:sumN0}), and hence we
replace each $a_{j}=QB_{j}$ in the last line with the
corresponding harmonic $a_{j,\vnu_{j}}\,e^{i\vomega\cdot\vnu_{j}t}$,
we immediately realize that the argument given in the proof
applies unchanged.
\end{rmk}

%%%%%%%%%%%%%%%%%%%%%%%%%%%%%%%%%%%%%%%%%%%%%%%%%%%%%%%%%%%%%%%%%%%%%%%%%
\begin{lema} \label{lemma:Gjc}
For all $k \geq 2$ one has the identity
$\sum_{\theta \in \widetilde{\calT}^{nc}_{k, \vz}} \Val(\theta) =
0$. Therefore, for all $j \geq 1$, $G_j = \sum_{\theta \in
\widetilde{\calT}^{c}_{j, \vz}}\Val(\theta)$.$\ES$
\end{lema}
%%%%%%%%%%%%%%%%%%%%%%%%%%%%%%%%%%%%%%%%%%%%%%%%%%%%%%%%%%%%%%%%%%%%%%%%%

\prova Let $k \geq 2$. The result follows by a combination of 
Lemma~\ref{lemma:identity} and Lemma~\ref{lemma:derivative}. Indeed,
the sum of all possible trees in $\widetilde{\calT}^{nc}_{k, \vz}$
(including the sum over the Fourier modes) means that we have to sum
all valid permutations of $a_1, \ldots, a_n$ in (\ref{eq:avint}) for
all trees with $2 \leq n \leq k$ canonical subtrees. Since this sum
gives the average of a total derivative, one concludes that 
$\sum_{\theta
\in \widetilde{\calT}^{nc}_{k, \vz}} \Val(\theta) = 0$. Now, 
since $\widetilde{\calT}_{1,\vz}$ contain only $c$-class trees, 
one concludes that
\begin{equation*}
G_j = \sum_{\theta \in \widetilde{\calT}_{j, \vz}} \Val(\theta) = 
\sum_{\theta \in \widetilde{\calT}^{c}_{j, \vz}} \Val(\theta)
+ \sum_{\theta \in \widetilde{\calT}^{nc}_{j, \vz}} \Val(\theta) = 
\sum_{\theta \in \widetilde{\calT}^{c}_{j, \vz}} \Val(\theta) \, ,
\end{equation*}
for all $j \geq 1$.$\EP$

%%%%%%%%%%%%%%%%%%%%%%%%%%%%%%%%%%%%%%%%%%%%%%%%%%%%%%%%%%%%%%%%%%%%%%%%%
\begin{prop} \label{prop:alphak}
Let $G_j$, $j \geq 1$, be as the previous lemma. Suppose that $G_{j_0}
\neq 0$ for some $j_0 \geq 1$. Then, (\ref{eq:masteralphak}) holds iff
$\alpha^{(k)} = 0$ for all $k \geq 0$.$\ES$
\end{prop}
%%%%%%%%%%%%%%%%%%%%%%%%%%%%%%%%%%%%%%%%%%%%%%%%%%%%%%%%%%%%%%%%%%%%%%%%%

\prova By Lemma~\ref{lemma:Gj} condition (\ref{eq:masteralphak}) reads
\begin{equation} \label{eq:zero}
0 = \left\laa [Qu^2]^{(k-1)}\right\raa = \sum_{p=0}^{k-1}
\alpha^{(p)}G_{k-p} \, , \quad \forall \, k \geq 1 \, .
\end{equation}
We shall prove by induction that $\alpha^{(p)} = 0$, $p \geq 0$, is
the unique solution of (\ref{eq:zero}) if $G_{j_0} \neq 0$ for some 
$j_0 \geq 1$. Indeed, let $j_0 \geq 1$ be such that $G_1 = \cdots =
G_{j_0-1} = 0$ and $G_{j_0} \neq 0$. Then, equation (\ref{eq:zero}) is
automatically satisfied for all $1 \leq k \leq j_0-1$. For $k = j_0$,
one has 
\begin{equation*}
0 = \alpha^{(0)} G_{j_0} + \sum_{p=0}^{j_0-1} \alpha^{(p)} G_{j_0-p}
= \alpha^{(0)} G_{j_0} \, .
\end{equation*}
Therefore, $\alpha^{(0)} = 0$. Now suppose that $\alpha^{(0)} = \cdots
= \alpha^{(k_0)} = 0$ for some $k_0 \geq 1$ and let us prove that
$\alpha^{(k_0+1)} = 0$. Using (\ref{eq:zero}) for $k = j_0 + k_0 + 1$,
we have
\begin{equation*}
0 = \sum_{p=0}^{j_0 + k_0} \alpha^{(p)} G_{j_0 + k_0 + 1 - p}
= \sum_{p=0}^{k_0} \alpha^{(p)} G_{j_0 + k_0 + 1 - p}
+ \sum_{p=k_0+1}^{j_0 + k_0} \alpha^{(p)} G_{j_0 + k_0 + 1 - p}
= \alpha^{(k_0+1)} G_{j_0} \, ,
\end{equation*}
which implies that $\alpha^{(k_0+1)} = 0$.$\EP$

\begin{rmk} \label{rmk:p10=0}
Note that one can always suppose that the function $p_{1}$
in (\ref{eq:hill}) has zero average (i.e. $P^{(1)}_{\undzero}=0$),
by an appropriate choice of the average of $p_{0}$. In such a case,
since one has
\begin{equation} \label{eq:G1}
G_{1} = \left\laa Q\int R \right\raa = 2 \sum_{n_{1}\in \Z}
P^{(1)}_{\undzero} \frac{\calF_{n_1}^{(-2)}
\calF_{-n_1}^{(2)}}{i(2\Omega_0-n_1\omega_0)} 
\end{equation}
one finds $G_{1}=0$. This shows that it is important
to consider the possibility that the first non-vanishing $G_{j}$
has $j > 1$.
\end{rmk}

%%%%%%%%%%%%%%%%%%%%%%%%%%%%%%%%%%%%%%%%%%%%%%%%%%%%%%%%%%%%%%%%%%%%%%%%%
\begin{prop} \label{prop:MeQu}
Let $G_j$, $j \geq 1$, be as the previous lemma. Then, 
\begin{itemize}
\item[{\bf (a)}] $\laa Qu \raa
= \frac{1}{2} \sum_{k=0}^{\infty} \eps^k G_{k+1}$.
\item[{\bf (b)}] $\Omega_\eps = \Omega_0 \Leftrightarrow G_k = 0 \, , 
\;
\forall \, k \geq 1$. 
\item[{\bf (c)}] $\Omega_\eps \in \R \Leftrightarrow 
\overline{G_k} = -G_k \, , \; \forall \, k \geq 1$.
\end{itemize}
In {\bf (a)} the equality is in the sense  of formal power series
(that is it holds order by order).$\ES$
\end{prop}
%%%%%%%%%%%%%%%%%%%%%%%%%%%%%%%%%%%%%%%%%%%%%%%%%%%%%%%%%%%%%%%%%%%%%%%%%

\prova Let us first write $\laa Q u \raa$ in Fourier space:
\begin{equation*}
\laa Q u \raa = \sum_{k=0}^{\infty} \eps^k \laa [Q u]^{(k)} \raa \, ,
\quad \text{ where } \quad 
\laa [Q u]^{(k)} \raa = \sum_{\vnu_0 + \vnu_1 = \vz} 
Q_{\vnu_0} u^{(k)}_{\vnu_1} \, , \quad \forall \, k \geq 1.
\end{equation*}
Now let $\vnu_1 = (\undm_1, n^{(1)}_1, n^{(1)}_2) \in \Z^d$. Of course
$\vnu_1 \neq \vz$ since $\vnu_0 = -\vnu_1$ and $Q_{\vz} = 0$.
From Lemma~\ref{lemma:n2},
\begin{equation*}
n^{(1)}_2 \; \in \; \{-2k, -2(k-1), \ldots, -2, 0, 2\} \, .
\end{equation*}
To be more precise, let $\theta_1 \in \calT_{k, \vnu_1}$ be a 
tree contributing to $u^{(k)}_{\vnu_1}$; then $n^{(1)}_2 = 2(b_1 -
v_1)$, where $b_1$ and $v_1$ are the number of black bullets and
the number of vertices in $\theta_1$, respectively.
From $\theta_1$ we would like to construct a
tree $\theta \in \calT_{k+1,\vz}$ contributing to $\left\laa
[Qu]^{(k)}\right\raa$. We do this as follows. Take the root line of
$\theta_1$ entering a vertex $v$ with mode $\vnu_0 =
(\undzero, n^{(0)}_1,-2)$.
Add a line, with zero momentum, entering such a vertex.
We do not associate any propagator with this line,
which means that it works as an amputated line (we call this
the amputated line of $\theta$); note that we can consider
such a tree as a tree amputated of a white bullet.
Finally, we let the root line of
$\theta$ be the line exiting the vertex $v$ carrying zero momentum. 
Thus, $\vnu_0 + \vnu_1 = \vz$. This last relation implies that $-2 + 
n^{(1)}_2 = 0$, which means that $b_1 - (v_1 + 1) = 0$. Therefore, 
$|E_B(\theta)| = |V(\theta)|$, leading to the conclusion that $\theta$
must have only one endpoint which is not a black bullet 
(see Remark~{\ref{rmk:topological}}). This leaves room only for 
the amputated line of $\theta$, so that $\theta_1$ must have only black
bullets. Finally, one concludes that if $\theta$ contributes to 
$\left\laa
[Qu]^{(k)}\right\raa$, then $\theta \in 
\widetilde{\calT}_{k+1,\vz}^{c}$.
On the other hand, only half of the trees in 
$\widetilde{\calT}_{k+1,\vz}^{c}$ contributes to $\left\laa
[Qu]^{(k)}\right\raa$ since in $\widetilde{\calT}_{k+1,\vz}^{c}$ we
take into account two possibilities for amputating the leg connected to
the root line. Therefore, by Lemma~\ref{lemma:Gjc}, one can write 
\begin{equation} \label{eq:MeQu}
\laa [Q u]^{(k)} \raa = \frac{1}{2} \sum_{\theta \in 
\widetilde{\calT}_{k+1,\vz}^{c}} \Val(\theta) = \frac{1}{2} G_{k+1}
\quad \text{ and } \quad \laa Q u \raa = \frac{1}{2}
\sum_{k=0}^{\infty} \eps^k G_{k+1} \, ,
\end{equation}
where the last formula holds as equality between formal series.
This proves {\bf (a)}. Items {\bf (b)} and {\bf (c)} follows
immediately from  {\bf (a)} remembering that $\Omega_\eps = \Omega_0
+ \laa g \raa =  \Omega_0 + i\eps \laa Q u \raa$.$\EP$

%%%%%%%%%%%%%%%%%%%%%%%%%%%%%%%%%%%%%%%%%%%%%%%%%%%%%%%%%%%%%%%%%%%%%%%%%%%
%%%%%%%%%%%%%%%%%%%%%%%%%%%%%%%%%%%%%%%%%%%%%%%%%%%%%%%%%%%%%%%%%%%%%%%%%%%
\zerarcounters 
\section{Analysis of the non-renormalized expansion}
\label{sec:formal}
%%%%%%%%%%%%%%%%%%%%%%%%%%%%%%%%%%%%%%%%%%%%%%%%%%%%%%%%%%%%%%%%%%%%%%%%%%%
%%%%%%%%%%%%%%%%%%%%%%%%%%%%%%%%%%%%%%%%%%%%%%%%%%%%%%%%%%%%%%%%%%%%%%%%%%%

One of the main results of last section (see
Proposition~\ref{prop:alphak}) tell us that all $\alpha^{(k)} = 0$ if
some $G_{j_0} \neq 0$, a condition which we henceforth
assume; we shall come back to this in the last section. 
Therefore, one should not worry about white bullets and
trivial propagators. For all $k \geq 0$ and all $\vnu \in \Z^d_\ast$,
define $\matT_{k,\vnu} := \{\, \theta \in \calT_{k, \vnu} \; : \;
E_W(\theta) = \emptyset \, \}$. Then,
\begin{equation} \label{eq:exp} 
u^{(k)}_{\vnu} = \sum_{\theta \in \calmT_{k,\vnu}} \Val(\theta) \, ,
\qquad
\Val(\theta) = \left(\prod_{\ell \in L(\theta)}
g_\ell\right)\left(\prod_{v \in B(\theta)} F_v\right)\, ,
\end{equation}
where
\begin{equation} \label{eq:expdefs}
g_\ell = \frac{1}{i \vomega \cdot \vnu_\ell} \, , \;\;\; \vnu_\ell \in
\Z^d_\ast \quad \text{ and } \quad
F_v =  \left\{
\begin{array}{ll}
Q_{\vnu_v} \, , & v \in V(\theta) \\
& \\
R_{\vnu_v} \, , & v \in E_B(\theta)
\end{array}
\right. \, .
\end{equation}
Moreover, $u^{(k)}_{\vz} = \alpha^{(k)} = 0$ for all $k \geq 0$. 

%%%%%%%%%%%%%%%%%%%%%%%%%%%%%%%%%%%%%%%%%%%%%%%%%%%%%%%%%%%%%%%%%%%%%%%%%
\begin{lema}
Let $k \geq 0$ and $\vnu \in \Z^d_\ast$, then $|u^{(k)}_{\vnu}| \leq A
B^k (k!)^\beta e^{-\kappa'|\vnu|}$, for positive constants $A, B,
\beta$ and $\kappa' < \kappa$.$\ES$
\end{lema}
%%%%%%%%%%%%%%%%%%%%%%%%%%%%%%%%%%%%%%%%%%%%%%%%%%%%%%%%%%%%%%%%%%%%%%%%%

\prova Let $\theta \in \calmT_{k, \vnu}$, $k \geq 0$, $\vnu \in
\Z^d_\ast$. From (\ref{eq:exp}), 
(\ref{eq:expdefs}), (\ref{eq:decQR}) (see Remark~\ref{rmk:decQR}) 
and from the Diophantine condition (\ref{eq:Diophantine}) 
(see Remark~\ref{rmk:Diophantine}), we write
\begin{equation*}
|\Val(\theta)| \leq \calQ^{|B(\theta)|} C_0^{-|L(\theta)|} 
\left(\prod_{\ell \in L(\theta)} |\vnu_\ell|^{\tau}\right) 
\left(\prod_{v \in B(\theta)} e^{-\kappa|\vnu_v|}\right) \, .
\end{equation*}
Now, since $|B(\theta)|=|L(\theta)|$ and
\begin{equation*}
\sum_{\ell \in L(\theta)} |\vnu_\ell| \:=\: \sum_{\ell \in L(\theta)}
\Big|\sum_{\substack{w \in B(\theta) \\ w \preceq v \,: \, \ell = 
\ell_v}}
\vnu_w\Big| \:\leq\: \sum_{\ell \in L(\theta)} \sum_{w \in B(\theta)}
|\vnu_w| \:=\: |L(\theta)| \sum_{v \in B(\theta)} |\vnu_v| \, ,
\end{equation*}
it follows that
\begin{equation} \label{eq:fact1}
\prod_{v \in B(\theta)} e^{-\kappa|\vnu_v|} \leq \left(\prod_{v \in
B(\theta)} e^{-\kappa|\vnu_v|/2}\right)\left(\prod_{\ell \in L(\theta)} 
e^{-\kappa|\vnu_\ell|/2|L(\theta)|}\right) \, .
\end{equation}
On the other hand, for each line $\ell \in L(\theta)$, one 
has\footnote{Due to the inequality $|x|^\sigma \leq
e^{\delta|x|}(\sigma/e\delta)^\sigma$, $ \forall x \in \C$, and to
the Stirling relation.}
\begin{equation*}
|\vnu_\ell|^\tau e^{-\kappa|\vnu_\ell|/2|L(\theta)|} \leq
\tau!\left(\frac{2|L(\theta)|}{\kappa}\right)^\tau \, .
\end{equation*}
Hence
\begin{equation} \label{eq:fact2}
\left(\prod_{\ell \in L(\theta)}|\vnu_\ell|^\tau\right)
\left(\prod_{\ell \in L(\theta)}
e^{-\kappa|\vnu_\ell|/2|L(\theta)|}\right) \leq
(\tau!)^{|L(\theta)|}\left(\frac{2|L(\theta)|}
{\kappa}\right)^{\tau|L(\theta)|} \,,
\end{equation}
which, together with the fact that for all $\theta \in
\calmT_{k,\vnu}$ one has $|L(\theta)| = |B(\theta)| = 2k+1$, yields
\begin{eqnarray} \label{eq:fact3}
|\Val(\theta)| &\leq& \left(\frac{\calQ 2^\tau \tau!}{C_0
\kappa^\tau}\right)^{2k+1} \left((2k+1)^{2k+1}\right)^\tau 
\left(\prod_{v \in B(\theta)}
e^{-\kappa|\vnu_v|/2}\right) \nonumber \\
&\leq& \Gamma_1 \Gamma_2^k (k!)^\beta \, e^{-\kappa|\vnu|/4} \, 
\left(\prod_{v \in B(\theta)}
e^{-\kappa|\vnu_v|/4}\right) \, ,
\end{eqnarray}
where $\Gamma_1, \Gamma_2, \beta$ are suitable positive constants.
In the last step above we used the well known Stirling relation to
express $k^k$ in terms of $k!$ and the fact that 
\begin{equation} \label{eq:modvnu} 
\sum_{v \in B(\theta)} |\vnu_v| \geq \Big|\sum_{v \in B(\theta)}
\vnu_v \Big| = |\vnu| \, .
\end{equation}
We now take into account the fact that the number of trees of fixed
order is bounded by $\Gamma_3^k$, for some positive constant
$\Gamma_3$~\cite{Gentile}. Thus, we finally obtain
\begin{equation*}
|u_{\vnu}^{(k)}| \; \leq \;  \sum_{\theta \in \matT_{k, \vnu}}
|\Val(\theta)| \leq \Gamma_1' \Gamma_2^k \Gamma_3^k
(k!)^\beta e^{-\kappa|\vnu|/4} \, ,
\end{equation*}
with $\Gamma_{1}'$ a suitable positive constant.
This completes the proof of the lemma.$\EP$

%%%%%%%%%%%%%%%%%%%%%%%%%%%%%%%%%%%%%%%%%%%%%%%%%%%%%%%%%%%%%%%%%%%%%%%%%%%
%%%%%%%%%%%%%%%%%%%%%%%%%%%%%%%%%%%%%%%%%%%%%%%%%%%%%%%%%%%%%%%%%%%%%%%%%%%
\zerarcounters
\section{Renormalization}
\label{sec:renormalization}
%%%%%%%%%%%%%%%%%%%%%%%%%%%%%%%%%%%%%%%%%%%%%%%%%%%%%%%%%%%%%%%%%%%%%%%%%%%
%%%%%%%%%%%%%%%%%%%%%%%%%%%%%%%%%%%%%%%%%%%%%%%%%%%%%%%%%%%%%%%%%%%%%%%%%%%

The main problem with the previous proof is that it does not treat
conveniently the small denominators $1/i\vomega \cdot \vnu_\ell$
which appear in the expansion through the propagators $g_\ell$.
As a result, we end up with a crude estimate for the coefficients
$u^{(k)}_{\vnu}$, which complicates the task of studying the absolute
convergence of the series for $u$.

To overcome the problem of small denominators, we shall adopt a method
well known from the analysis of the Lindstedt series for KAM type
problems (see Ref.~\cite{Gentile_Mastropietro} and references
therein).  All the complication lies in the fact that $\vomega \cdot
\vnu$ can be arbitrarily small for certain $\vnu$ with sufficiently
large $|\vnu|$. The idea, then, is to separate the ``small'' parts of
$\vomega \cdot \vnu$ and to resume the corresponding terms in a
suitable form, obtaining then a result which can be better estimated.
The process of ``separation'' of the ``small'' parts of $\vomega \cdot
\vnu$ is implemented via a technique known as the {\it multiscale
  decomposition of the propagators}. We stress that
this technique is genuine from methods of the Renormalization Group
introduced to deal with related problems in field theories.

%%%%%%%%%%%%%%%%%%%%%%%%%%%%%%%%%%%%%%%%%%%%%%%%%%%%%%%%%%%%%%%%%%%%%%%%%%%
\subsection{Multiscale decomposition of the propagators}
%%%%%%%%%%%%%%%%%%%%%%%%%%%%%%%%%%%%%%%%%%%%%%%%%%%%%%%%%%%%%%%%%%%%%%%%%%%

\begin{figure}[h!]
\centering{\epsfig{figure = 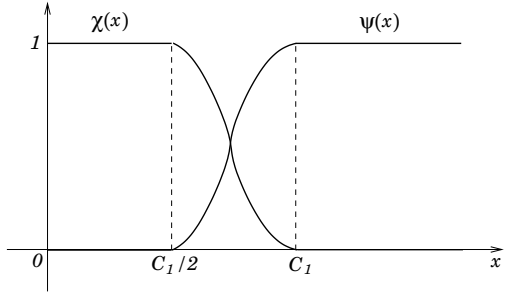, scale=0.55}
\vskip-0.15cm 
\caption{{\small Possible graphs of the functions $\chi(x)$ and
$\psi(x)$. Note that $\psi(x)+\chi(x) = 1$.}}\label{fig:chipsi}}
\end{figure}

We begin by introducing a bounded non-decreasing $C^{\infty}(\R)$
function $\psi(x)$, defined in $\R_+$, such that
\begin{equation*}
\psi(x) = \left\{
\begin{array}{ll}
1 \, , & \text{for } x \geq C_{1} \, , \\
0 \, , & \text{for } x \leq C_{1} /2 \, ,
\end{array} 
\right.
\end{equation*}
where $C_{1}\leq C_{0}$ is to be fixed,
with $C_{0}$ the Diophantine constant which appears
in (\ref{eq:Diophantine}), and setting $\chi(x) := 1-\psi(x)$.
An example of $\chi(x)$ and $\psi(x)$ with the
above properties is found in Figure~\ref{fig:chipsi}. We also define,
for all $n \in \Z_+$, $\chi_n(x) := \chi(2^nx)$ and $\psi_n(x) :=
\psi(2^nx)$. It is clear that $\chi_0(x) = \chi(x)$, 
$\psi_0(x) = \psi(x)$ and $\psi_n(x) + \chi_n(x) = 1$, $\forall \, n
\geq 0$.

Functions $\chi_n(x)$ and $\psi_n(x)$ allow us to write the propagator
$g_\ell$, for all $\ell \in L(\theta)$ and $\vnu_\ell \neq \vz$, 
as\footnote{Due to the identity $\psi_0(x) + 
\sum_{n=1}^{\infty}\psi_{n}(x)\chi_{n-1}(x) = 1$, for all $x \in \R_+$
(see Remark~\ref{obs:boundprop} below).}
\begin{equation*}
g_\ell = \frac{1}{i\vomega\cdot\vnu_\ell} = \frac{\psi_0(|\vomega
\cdot \vnu_\ell|)}{i\vomega\cdot\vnu_\ell} + \sum_{n=1}^{\infty} 
\frac{\psi_n(|\vomega \cdot \vnu_\ell|) \chi_{n-1}(|\vomega \cdot
\vnu_\ell|)}{i\vomega\cdot\vnu_\ell} \, .
\end{equation*}
We can still write
\begin{equation} \label{eq:escalas}
\begin{array}{l}
\displaystyle
g_\ell = \sum_{n=0}^{\infty} g_\ell^{(n)} \, , \\ \\
\displaystyle
g_\ell^{(0)} := \frac{\psi_0(|\vomega \cdot \vnu_\ell|)}{i \vomega
\cdot \vnu_\ell} \, , \\ \\
\displaystyle
g_\ell^{(n)} := \frac{\psi_n(|\vomega \cdot \vnu_\ell|)
\chi_{n-1}(|\vomega \cdot \vnu_\ell|)}{i \vomega \cdot \vnu_\ell} \, ,
\quad \forall \, n \geq 1 \, .
\end{array}
\end{equation}
We set $g^{(n)}_\ell = g^{(n)}(\vomega\cdot\vnu_\ell)$.

\begin{rmk} \label{obs:boundprop}
Note that for fixed $x = \vomega\cdot\vnu$, we have $g^{(n)}(x) \neq 0$
only for two values of $n$. This means that the series
(\ref{eq:escalas}) is, in fact, finite. Note also that $g^{(n)}(x)
\neq 0$ only if $2^{-n-1} C_{1} < |x| < 2^{-n+1} C_{1}$ for $n \geq
1$ and only if $|x| > 2^{-1}C_{1}$ for $n = 0$. Hence, 
\begin{equation} \label{eq:boundprop}
g_\ell^{(n)} \neq 0 \; \Rightarrow \; |g^{(n)}_\ell| 
\leq C_{1}^{-1} 2^{n+1} \, .
\end{equation}
\end{rmk}

To each line $\ell \in L(\theta)$ with $\vnu_\ell \neq \vz$ we
associate a new label $n_\ell = 0, 1, 2, \ldots$ called the {\it scale
label} of line $\ell$. It is important to stress, based on 
Remark~\ref{obs:boundprop}, that the scale label $n_\ell$ of a line
$\ell$ tells, essentially, what is the size of the associated
propagator $g_\ell^{(n_\ell)}$. This is an useful device for
``isolating'' the contribution of trees containing propagators
with too large scales. We shall do this carefully in what follows.

\begin{defi}
We define $\Theta_{k, \vnu}$ as the set of trees which differ from
those in $\calmT_{k, \vnu}$ by the introduction of the scale labels
in the propagators.
%That is, the trees in $\Theta_{k, \vnu}$
%are exactly the same  as in $\calmT_{k, \vnu}$,
%except from the introduction of the scale labels in the propagators.
\end{defi}

With the above definitions, expression (\ref{eq:exp}) now reads as
\begin{equation} \label{eq:exp2} 
u^{(k)}_{\vnu} = \sum_{\theta \in \Theta_{k,\vnu}} \Val(\theta) \, ,
\qquad
\Val(\theta) = \left(\prod_{\ell \in L(\theta)}
g_\ell^{(n_\ell)}\right)\left(\prod_{v \in B(\theta)} F_v\right)\, ,
\end{equation}
where the sum over all the trees in $\Theta_{k, \vnu}$ implies a
further sum over all the possible scale labels for each one of the
propagators. Thus, for all $\theta \in \Theta_{k, \vnu}$, 
if $N_n(\theta)$ denotes the number of lines in $\theta$ on scale $n$,
by using (\ref{eq:exp2}), (\ref{eq:boundprop}), (\ref{eq:decQR}) and
the fact that $|L(\theta)| = |B(\theta)| = 2k+1$, we obtain
\begin{eqnarray} \label{eq:Nn}
|\Val(\theta)| &\leq& (2C_{1}^{-1}\calQ)^{2k+1} \left(\prod_{v \in
B(\theta)} e^{-\kappa|\vnu_v|}\right) 
\left(\prod_{n=0}^{\infty}2^{nN_n(\theta)}\right) \nonumber \\
&\leq& (2C_{1}^{-1}\calQ 2^{n_{1}})^{2k+1}
e^{-\kappa\sum_{v \in B(\theta)}
|\vnu_v|} \left(\prod_{n=n_{1}}^{\infty}2^{nN_n(\theta)}\right) \, ,
\end{eqnarray}
where in the last step we have introduced a (so far) arbitrary
positive integer $n_{1}$ and used the obvious fact that
$N_{n}(\theta) \leq |L(\theta)| = 2k+1$, $\forall \, n \geq 0$.

Our problem now is to estimate $N_n(\theta)$. To solve this, we need
to introduce some useful definitions.

\begin{figure}[!ht]
\centering
\begin{minipage}[c]{0.55\textwidth}
\begin{center}
\includegraphics[scale=0.55]{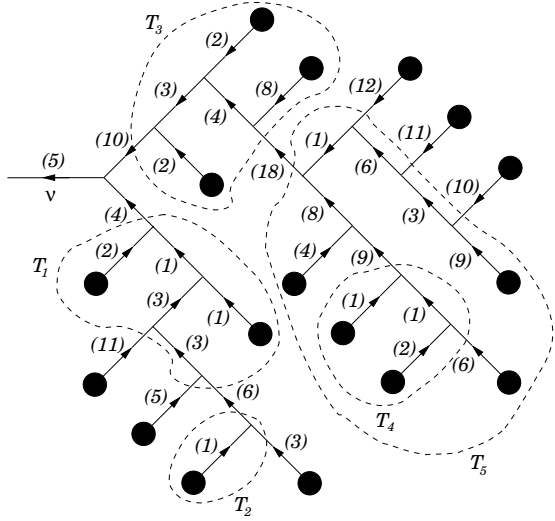}
\end{center}
\end{minipage}
%\hfill
\hspace{0.2cm}
\begin{minipage}[c]{0.35\textwidth}
\centering \caption{
Examples of clusters in a tree of order $k = 16$. The
number between parentheses above each line denotes the scale of the
propagator. Thus, we have $n_{T_1} = 3$, $n_{T_2} = 1$, $n_{T_3} = 8$,
$n_{T_4} = 2$ and $n_{T_5} = 9$. Note that $T_4 \subset T_5$ and
therefore $n_{T_4} < n_{T_5}$. Of course there are other
clusters in the example considered which are not
shown.}\label{fig:cluster}
\end{minipage}
\end{figure}

\begin{defi}[Cluster]
A {\rm cluster} $T$ on scale $n$ is a maximal connected subset of a tree 
$\theta$ such that all its lines have scale $n' \leq n$ and there is
at least one line on scale $n$. The lines entering a cluster $T$ and
the one (if any) exiting it are called the {\rm external lines} of $T$.
Given a cluster $T$ on scale $n$, we denote by $n_T = n$ the scale of
$T$. Moreover, $V(T)$, $E_B(T)$, $B(T)$, and $L(T)$ denote,
respectively, the set of vertices, black bullets, vertices plus black
bullets, and lines contained in $T$; the external lines of $T$ do not
belong to $L(T)$. We finally define the {\rm momentum of
the cluster $T$} as $\vnu_T := \sum_{v\in B(T)} \vnu_v$.
We shall call $k_{T} := |V(T)|$ the {\rm order} of $T$.
Some examples of clusters are presented in Figure~\ref{fig:cluster}.
\end{defi}

\begin{figure}[!ht]
\centering
\begin{minipage}[c]{0.45\textwidth}
\begin{center}
\includegraphics[scale=0.55]{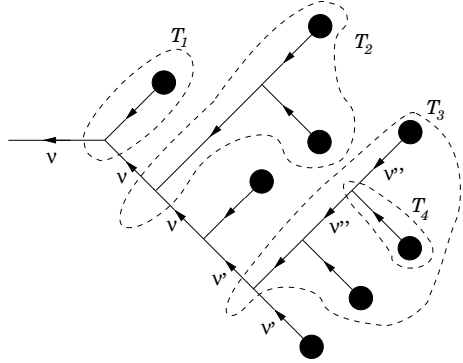}
\end{center}
\end{minipage}
%\hfill
\hspace{0.2cm}
\begin{minipage}[c]{0.45\textwidth}
\centering \caption{
Examples of self-energy graphs in a tree of order 
$k = 7$. Note that, in accordance with the definition, there is only
one entering line and one exiting line (carrying the same momentum) 
in the self-energy graphs $T_1$, $T_2$, $T_3$ and $T_4$. It is clear
that the scales on the lines of $T_1$, $T_2$, $T_3$, $T_4$ are
strictly less then the scales on their external lines (after all,
self-energy graphs are clusters).}\label{fig:self-energy}
\end{minipage}
\end{figure}

\begin{defi}[Self-Energy Graph]
We call {\rm self-energy graph} any cluster $T$ of a tree $\theta$
which satisfies
\begin{enumerate}
\item $T$ has only one entering line $\ell_T^{\rm in}$ and only one
exiting line $\ell_T^{\rm out}$;
\item The momentum of $T$ is zero, i.e. $\vnu_T = \sum_{v \in B(T)}
\vnu_v = \vz$. This means that $\vnu_{\ell_T^{\rm in}} = 
\vnu_{\ell_T^{\rm out}}$. 
\end{enumerate}
We call {\rm self-energy line} any line $\ell_T^{\rm out}$ which
exits from a self-energy graph $T$. We call {\rm normal line} any
line which is not a self-energy line. Note that if $T$ is a
self-energy graph, then $\ell_T^{\rm in}, \ell_T^{\rm out} \not\in
L(T)$, so that $|L(T)| = 2k_{T}-1$ and $|B(T)| = 2k_{T}$. Some examples 
of
self-energy graphs are depicted in Figure~\ref{fig:self-energy}. 
\end{defi}

\begin{rmk} \label{rmk:inout}
It is important to stress that due to the condition $\vnu_{\ell_T^{\rm
in}} = \vnu_{\ell_T^{\rm out}}$, the scales on the entering and
exiting lines of a self-energy graph $T$ must differ at most by one
unit, i.e. $|n_{\ell_T^{\rm in}} - n_{\ell_T^{\rm out}}| \leq
1$ (see Remark~\ref{obs:boundprop}). Moreover, due to the fact that  
$T$ defines a cluster, we must have $n_T + 1 \leq
\min\{n_{\ell_T^{\rm in}}, n_{\ell_T^{\rm out}}\}$, which is
equivalent of saying that all the lines within $T$ have scale
strictly less then the scale on the external lines $\ell_T^{\rm in}$
and $\ell_T^{\rm out}$.
\end{rmk}

%Remark~\ref{rmk:inout} is very important because it follows from it 
Due to the presence of self-energy graphs one can have
accumulation of small divisors. The heuristic explanation for 
this is as follows: imagine we have a line $\ell$ on a large scale 
$n_\ell \gg 1$ entering a self-energy graph $T$. This line $\ell'$
exiting from $T$ could enter another self-energy graph $T'$. Note that
such a line $\ell'$ is also on scale $n_{\ell'} \gg 1$.
This process could repeat itself several times, resulting at the
end in a bunch of lines $\ell_{1},\ldots,\ell_{N}$ on scales
$n_{\ell_{i}} \gg 1$, i.e. we end up with an
accumulation of small divisors\footnote{We remind that the ``size'' of
a propagator grows exponentially with its scale (see
Remark~\ref{obs:boundprop}).}.

Actually, from a more precise point of view, the whole problem with the
self-energy graphs is that we are not able to give a satisfactory bound
on the number of self-energy lines in a given tree $\theta$. On the
other hand, it is easy to show that (see,
e.g., Ref.~\cite{Gallavotti_Gentile}) if we denote by $N_n^{\rm
norm}(\theta)$ the number of normal lines in a tree $\theta$, then
there exists a positive constant $c$ such that 
\begin{equation} \label{eq:Nnnorm}
N_n^{\rm norm}(\theta) \leq c \, 2^{-n/\tau} \sum_{v \in B(\theta)}
|\vnu_v| \, ,
\end{equation}
where $\tau$ is one of the Diophantine constants appearing in
(\ref{eq:Diophantine}). Thus, suppose we could neglect all
the self-energy lines within any tree $\theta$, i.e. suppose
that we could substitute $N_n(\theta)$ in (\ref{eq:Nn}) by
$N_n^{\rm norm}(\theta)$ with the above estimate. Then, we would have
\begin{equation} \label{eq:boundvalue}
|\Val(\theta)| \leq (2 C_{1}^{-1} \calQ 2^{n_{1}})^{2k+1}
e^{-\kappa \sum_{v \in B(\theta)}|\vnu_v|} \, 
e^{\left(c \, \log{2} \sum_{n=n_{1}}^{\infty} n 2^{-n/\tau}\right) 
\sum_{v \in B(\theta)} |\vnu_v|} \, ,
\end{equation}
for all $n_{1} \geq 0$. Thus, picking $n_{1} = n_{1}(\kappa, c,
\tau)$ such that\footnote{Note that this is always possible by
choosing $n_{1}$ sufficiently large due to the fact that the sum over
$n$ above is convergent and exponentially small in $n_{1}$.}
\begin{equation} \label{eq:choicen1}
-\frac{\kappa}{2} + c \, \log{2}
\sum_{n=n_{1}}^{\infty} n 2^{-n/\tau} < 0 \, ,
\end{equation}
we would obtain
\begin{equation*}
|\Val(\theta)| \leq 
(2 C_{1}^{-1} \calQ 2^{n_{1}(\kappa, c, \tau)})^{2k+1}
e^{-\frac{\kappa}{2} \sum_{v \in B(\theta)}|\vnu_v|}
\leq \Lambda_1 \Lambda_2^k \, 
e^{-\kappa|\vnu|/4} \, \left(\prod_{v\in B(\theta)}
e^{-\kappa|\vnu_v|/4}\right) ,
\end{equation*}
where, in the last inequality, we have used (\ref{eq:modvnu}).
Therefore, summing over all the trees (whose number grows at most as
$\Lambda_3^k$, for some positive $\Lambda_3$) and all the Fourier 
labels, 
\begin{equation} \label{eq:sup} 
|u_{\vnu}^{(k)}| \; \leq \;  \sum_{\theta \in \Theta_{k, \vnu}}
|\Val(\theta)| \leq \Lambda'_1 \Lambda_2^k \Lambda_3^k 
e^{-\kappa|\vnu|/4}
\end{equation}
what would imply in the convergence of expansion
(\ref{eq:expanalitico}) provided $|\eps| < (\Lambda_2\Lambda_3)^{-1}$.

It is clear that the above result is false since we cannot simply
forget the self-energy graphs. The estimate obtained just illustrates
the fact that all the problem concerning the convergence of the series
(\ref{eq:expanalitico}) lies in the existence of self-energy graphs
(small divisors). We have to overcome this difficult with some
different approach.

%%%%%%%%%%%%%%%%%%%%%%%%%%%%%%%%%%%%%%%%%%%%%%%%%%%%%%%%%%%%%%%%%%%%%%%%
\subsection{Renormalized expansion}
%%%%%%%%%%%%%%%%%%%%%%%%%%%%%%%%%%%%%%%%%%%%%%%%%%%%%%%%%%%%%%%%%%%%%%%%%%%

The problem with the self-energy graphs can be solved by a suitable
resummation procedure of the formal series obtained from the
coefficients (\ref{eq:exp2}). The basic idea is to ``dress'' the
propagators $g_\ell^{(n_\ell)}$ in such a way that they could
harbour all the malign contribution deriving from the self-energy
graphs. The next step is to define an expansion in terms of only 
non-self-energy graphs and renormalized propagators which we hope to
give an estimate like (\ref{eq:Nnnorm}). This is something analogous 
to the procedure of mass renormalization in field theories. We shall 
therefore iteratively define new propagators $g_\ell^{[n_\ell]}$ 
(renormalized propagators).

\begin{defi}[Self-Energy Value]
Suppose that the renormalized propagators $g_\ell^{[n_\ell]}$ are 
given.
For a self-energy graph $T$ which does not contain any other
self-energy graph, we define the {\rm self-energy value}
associated with $T$ as
\begin{equation} \label{eq:calVT}
\calV_T(\vomega\cdot\vnu; \eps) := \eps^{k_T} \left(\prod_{\ell \in
L(T)} g_\ell^{[n_\ell]}\right) \left(\prod_{v \in B(T)} F_v\right) \, , 
\end{equation}
where $\vnu$ is the momentum which enters $T$ through the external
line $\ell_T^{\rm in}$, $k_T = |B(T)|$, and $F_v$ is defined as in
(\ref{eq:expdefs}). Note that $\calV_T(\vomega\cdot\vnu; \eps)$
depends on $\vomega\cdot\vnu$ through the propagators in $L(T)$.
\end{defi}  

By setting $x=\vomega\cdot\vnu$ in (\ref{eq:calVT}) and
$x_{\ell}=\vomega\cdot\vnu_{\ell}$ for each line $\ell\in L(T)$,
one can write $x_{\ell}=x_{\ell}^{0} + \sigma_{\ell} x$, where
\begin{equation} \label{eq:xell0}
x_{\ell}^{0} = \vomega\cdot\vnu_{\ell}^{0} , \qquad \vnu_{\ell}^{0} =
\sum_{\substack{ w \in B(T) \\ w \preceq v \, : \, \ell = \ell_v}} 
\vnu_w \, ,
\end{equation}
and $\sigma_{\ell}=1$ if $\ell$ is along the path of lines
connecting the external lines of the self-energy $T$,
and $\sigma_{\ell}=0$ otherwise.

\begin{rmk} \label{rmk:calVx}
The value $\calV_{T}(x;\eps)$ of a
self-energy graph $T$ can depend on $x$ only if $k_{T} \geq 2$.
\end{rmk}

\begin{defi}
We define $\Theta_{k,\vnu}^\calR$ as the set of
{\rm renormalized trees}, that is of trees which do not
contain any self-energy graph. We also define
$\calS_{k, n}^\calR$ as the set of self-energy graphs of order $k$
which do not contain any other self-energy graph and such that the
maximum of scales of the lines in $T \in \calS_{k, n}^\calR$ is
exactly $n$, and we call them the {\rm self-energy renormalized graphs
of order $k$ and on scale $n$}. We stress that the propagators
associated with the lines in $\Theta_{k,\vnu}^\calR$ and $\calS_{k, 
n}^\calR$
are the renormalized ones, $g_\ell^{[n_\ell]}$.
\end{defi}

Then we can define the {\it renormalized propagators}
$g^{[n]}_{\ell} := g^{[n]}(\vomega\cdot\vnu_{\ell};\eps)$ and the
quantities $M^{[n]}(\vomega\cdot\vnu_{\ell};\eps)$ recursively
as follows. For $n_{0} \in \Z_{+}$ we set
\begin{eqnarray} \label{eq:renprop0}
M^{[n_{0}-1]}(x;\eps) & := & 0 , \nonumber \\
g^{[n_{0}]}(x;\eps) & := & \frac{\psi_{n_{0}}(|x|)}{ix} ,  \nonumber \\
M^{[n_{0}]}(x;\eps) & := & \sum_{k=1}^{\infty}
\sum_{T \in \calS^{\calR}_{k,n_{0}} } \calV_{T}(x;\eps) , \nonumber \\
\calM^{[n_{0}]}(x;\eps) & := & \chi_{n_{0}}(|x|) \, M^{[n_{0}]}(x;\eps) 
,
\end{eqnarray}
while, for $n\ge n_{0}+1$, by writing
\begin{eqnarray*}
\Xi_{n}(x;\eps) & := & \chi_{n_{0}}(|x|) \ldots
\chi_{n-1}(|ix - \calM^{[n-2]}(x;\eps)|) \,
\chi_{n}(|ix - \calM^{[n-1]}(x;\eps)|) , \nonumber \\
\Psi_{n}(x;\eps) & := & \chi_{n_{0}}(|x|) \ldots
\chi_{n-1}(|ix - \calM^{[n-2]}(x;\eps)|) \,
\psi_{n}(|ix - \calM^{[n-1]}(x;\eps)|) ,
\end{eqnarray*}
we define
\begin{eqnarray} \label{eq:renpropn}
g^{[n]}(x;\eps) & := &
\frac{\Psi_{n}(x;\eps)}{ix - \calM^{[n-1]}(x;\eps) } , \nonumber \\
M^{[n]}(x;\eps) & := & \sum_{k=1}^{\infty}
\sum_{T \in \calS^{\calR}_{k,n} } \calV_{T}(x;\eps) , \nonumber \\
\calM^{[n]}(x;\eps) & := &
\calM^{[n-1]}(x;\eps) + \Xi_{n}(x;\eps) \, M^{[n]}(x;\eps) \nonumber \\
& = & \sum_{j=n_{0}}^{n} \Xi_{j}(x;\eps) \, M^{[j]}(x;\eps) \, , 
 \end{eqnarray}
where $\calV_{T}(x;\eps)$ is defined as in (\ref{eq:calVT}).

One should now realize, from the above definitions, that if $\theta$
is a tree in $\Theta^\calR_{k, \vnu}$ or $\calS^\calR_{k, n}$, all of
its lines are on scale $\geq n_0$. In particular, if $T \in 
\calS^\calR_{k, n_0}$, all lines in $T$ are exactly on the scale $n_0$
and, hence, for all $\ell \in L(T)$, the propagators are
$g^{[n_0]}(x_\ell;\eps)$, as in (\ref{eq:renprop0}).

\begin{rmk} \label{rmk:scales}
Note that if a line $\ell$ is on scale $n \geq n_{0}+1$ and,
by setting $x=\vomega\cdot\vnu_{\ell}$, one has $g^{[n]}(x;\eps)
\neq 0$, this requires $\chi_{n_{0}}(|x|)\neq 0$,
$\chi_{n_{0}+1}(|ix-\calM^{[n_{0}]}(x;\eps)|)\neq 0$, $\ldots$,
$\chi_{n-1}(|ix-\calM^{[n-2]}(x;\eps)|)\neq0$ and
$\psi_{n}(|ix-\calM^{[n-1]}(x;\eps)|)\neq0$, which means
\begin{eqnarray} \label{eq:support}
& & |x| \leq 2^{-n_{0}} C_{1} , \nonumber \\
& & |ix-\calM^{[n_{0}]}(x;\eps)| \leq 2^{-(n_{0}+1)} C_{1} , \nonumber 
\\
& & |ix-\calM^{[n_{0}+1]}(x;\eps)| \leq 2^{-(n_{0}+2)} C_{1} , 
\nonumber \\
& & \ldots \ldots \ldots \nonumber \\
& & |ix-\calM^{[n-2]}(x;\eps)| \leq 2^{-(n-1)} C_{1} , \nonumber \\
& & |ix-\calM^{[n-1]}(x;\eps)| \geq 2^{-(n+1)} C_{1} , \nonumber
\end{eqnarray}
so that, in particular, one has $|g^{[n]}_{\ell}|\le
C_{1}^{-1}2^{n+1}$. If $\ell$ is on scale $n_0$ and $g^{[n_0]}(x;\eps)
\neq 0$, then $\psi_{n_0}(|x|) \neq 0$, which implies that
$|g^{[n_0]}_\ell| \leq C_1^{-1} 2^{n_0+1}$.
\end{rmk}

Then we define, formally, for $\vnu\neq\vz$,
\begin{equation} \label{eq:uknren}
u^{[k]}_{\vnu} = \sum_{\theta\in\Theta^{\calR}_{k,\vnu}}
\Val(\theta) , \qquad \Val(\theta) =
\left( \prod_{\ell\in L(\theta)} g^{[n_{\ell}]}_{\ell} \right)
\left( \prod_{v\in B(\theta)} F_{v} \right) ,
\end{equation}
while, for $\vnu=\vz$, one has $u^{[k]}_{\vz}=0$, and we write
\begin{equation} \label{eq:uren}
\olu (t) = \sum_{k=0}^{\infty} \eps^{k} u^{[k]}(t) =
\sum_{k=1}^{\infty} \eps^{k} \sum_{\vnu \in \Z^{d}}
e^{i\vnu\cdot\vomega t} u^{[k]}_{\vnu} ,
\end{equation}
where the coefficients $u^{[k]}_{\vnu}$ depend on $\eps$
(as the propagators do); note that the order $k$ of a renormalized
tree $\theta$ is still defined as $k=|B(\theta)|$,
but it does not correspond to the perturbative order any more.

\begin{defi} \label{def:epsadm}
Let $\vomega$ satisfy the Diophantine conditions
(\ref{eq:Diophantine}). Fix $\eps$ such that one has
\begin{equation} \label{eq:Melnikov}
\left| i\vomega\cdot\vnu - \calM^{[n]}(\vomega\cdot\vnu;\eps) \right|
\ge C_{1} |\vnu|^{-\tau_{1}} \qquad
\forall \, \vnu\in\Z^{d}_{*} \; \text{ and } \; \forall \, n \ge n_{0} 
,
\end{equation}
with Diophantine constants $C_{1}$ and $\tau_{1}$,
where $\tau_{1}>\tau$ and $C_{1}<C_{0}$ are to be fixed later.
We call $\calE_{*}$ the set of $\eps$ for which the Diophantine
conditions (\ref{eq:Melnikov}) are satisfied, and we shall refer
to it as the {\rm set of admissible values} of $\eps$.
\end{defi}

We shall see in next section that for $\eps\in\calE_{*}$ we shall be
able to give a meaning to the (so far formal)
renormalized expansion (\ref{eq:uren}), hence we shall
prove that the set $\calE_{*}$ has positive Lebesgue measure,
provided that $\tau_{1}$ and $C_{1}^{-1}$ are chosen large enough.

Fix $\ov\eps$ such that the series obtained from (\ref{eq:uren})
by replacing $g^{[n_{\ell}]}_{\ell}$ in (\ref{eq:uknren}) with the 
bound
$2^{n_{\ell}+1}C_{1}^{-1}$ converges for $|\eps|\leq \ov\eps$,
and fix $\eps_{0} \leq \ov\eps$ small enough (how small
will be determined by the forthcoming analysis). In the following
we shall consider the interval $[0,\eps_{0}]$;
the interval $[-\eps_{0},0]$ can be studied in the same way.

It will be convenient to split the interval $[0,\eps_{0}]$ into
infinitely many disjoint intervals by setting
\begin{equation} \label{eq:em}
[0,\eps_{0}] =\{0\}\cup\bigcup_{m=0}^{\infty} \calE_{m} ,
\qquad \calE_{m} :=
\left( 2^{-(m+1)}\eps_{0},2^{-m}\eps_{0} \right],
\end{equation}
and to study separately each interval $\calE_{m}$.
We shall prove that for each $m$ the admissible values
of $\eps$ inside $\calE_{m}$ have large measure,
and that the their relative measure
${\rm meas}(\calE_{m} \cap \calE_{*})/
{\rm meas}(\calE_{m})$ tends to 1 as $m$ tends to infinity.

Therefore in the following we imagine we have fixed $m$, and we set
$\eps_{m}=2^{-m}\eps_{0}$, so that we can write $\calE_{m}=
(\eps_{m}/2,\eps_{m}]$.

%%%%%%%%%%%%%%%%%%%%%%%%%%%%%%%%%%%%%%%%%%%%%%%%%%%%%%%%%%%%%%%%%%%%%%%%%
%%%%%%%%%%%%%%%%%%%%%%%%%%%%%%%%%%%%%%%%%%%%%%%%%%%%%%%%%%%%%%%%%%%%%%%%%
\zerarcounters
\section{Properties of the self-energy values}
\label{sec:properties}
%%%%%%%%%%%%%%%%%%%%%%%%%%%%%%%%%%%%%%%%%%%%%%%%%%%%%%%%%%%%%%%%%%%%%%%%%
%%%%%%%%%%%%%%%%%%%%%%%%%%%%%%%%%%%%%%%%%%%%%%%%%%%%%%%%%%%%%%%%%%%%%%%%%

Given a self-energy $T \in \calS^{\calR}_{k,n_{0}}$ define
\begin{equation} \label{eq:ovcalVT}
\ov{\calV}_{T}(x;\eps) := \eps^{k_{T}}
\left(\prod_{\ell \in L(T)} \frac{1}{i x_{\ell}} \right)
\left(\prod_{v \in B(T)} F_v\right) \, , 
\end{equation}
which differs from $\calV_{T}(x;\eps)$ as $\psi_{n_{0}}(|x_{\ell}|)$
is replaced with $1$ for all $\ell \in L(T)$, and set
\begin{eqnarray} \label{eq:ovcalM}
\calM^{[n_{0}]}_{j}(x;\eps) = \chi_{n_{0}}(|x|)
M^{[n_{0}]}_{j}(x;\eps) , \qquad
M^{[n_{0}]}_{j}(x;\eps) = \sum_{k=1}^{j} \eps^{k}
\sum_{T \in \calS^{\calR}_{k,n_{0}} } \calV_{T}(x;\eps) , & &
\nonumber \\
\ov{\calM}^{[n_{0}]}_{j}(x;\eps) = \chi_{n_{0}}(|x|)
\ov{M}^{[n_{0}]}_{j}(x;\eps) , \qquad
\ov{M}^{[n_{0}]}_{j}(x;\eps) = \sum_{k=1}^{j} \eps^{k}
\sum_{T \in \calS^{\calR}_{k,n_{0}} } \ov{\calV}_{T}(x;\eps) . & &
\end{eqnarray}
This allows us to decompose
\begin{equation} \label{eq:decemme0}
\calM^{[n_{0}]}_{j}(0;\eps) =
\ov{\calM}^{[n_{0}]}_{j}(0;\eps) +
\left( \calM^{[n_{0}]}_{j}(0;\eps) -
\ov{\calM}^{[n_{0}]}_{j}(0;\eps) \right) .
\end{equation}
where $\ov{\calM}^{[n_{0}]}_{j}(0;\eps)$
depends neither on $x$ nor on $n_{0}$. Note that one has
$\calM^{[n_{0}]}_{j}(0;\eps) = M^{[n_{0}]}_{j}(0;\eps)$ and
$\ov{\calM}^{[n_{0}]}_{j}(0;\eps) = \ov{M}^{[n_{0}]}_{j}(0;\eps)$
as $\chi_{n_{0}}(0)=1$ for all $n_{0}\geq 0$.

%%%%%%%%%%%%%%%%%%%%%%%%%%%%%%%%%%%%%%%%%%%%%%%%%%%%%%%%%%%%%%%%%%%%%%%%%
\begin{lema} \label{lemma:MG}
Let $G_j$, $j \geq 1$, be as the previous sections. Then one has
\begin{equation} \label{eq:MG}
\ov{\calM}^{[n_{0}]}_{j_{0}}(0;\eps) =
\sum_{k=1}^{j_{0}} \eps^{k} G_{k} .
\end{equation}
for all $n_{0}$ and all $j_{0}$.$\ES$
\end{lema}
%%%%%%%%%%%%%%%%%%%%%%%%%%%%%%%%%%%%%%%%%%%%%%%%%%%%%%%%%%%%%%%%%%%%%%%%%

\prova By setting $x=0$ any self-energy graph $T$ in
$\calS^{\calR}_{k,n_{0}}$ contributing to
$\ov{\calM}^{[n_{0}]}_{j_{0}}(x;\eps)$ looks like a tree $\theta$
in $\widetilde{\calT}_{k, \vz}$, except for the presence
of the scale labels (compare (\ref{eq:ovcalVT}) with (\ref{eq:calVT}):
if $T\in\calS^{\calR}_{k,n_{0}}$ each line $\ell\in L(T)$
has scale $n_{\ell}=n_{0}$). Nevertheless the corresponding
propagators do not depend on the scales. Hence
\begin{equation*}
\sum_{T \in \calS^{\calR}_{k,n} } \ov{\calV}_{T}(0;\eps)
= \sum_{\theta \in \widetilde{\calT}_{k,\vz} } \Val(\theta) ,
\end{equation*}
so that the assertion follows from the definition of $G_{k}$
(see Lemma~\ref{lemma:Gj}).$\EP$

%%%%%%%%%%%%%%%%%%%%%%%%%%%%%%%%%%%%%%%%%%%%%%%%%%%%%%%%%%%%%%%%%%%%%%%%%
\begin{lema} \label{lemma:chi}
For any self-energy $T$ one has
\begin{equation*}
1 - \prod_{\ell\in L(T)} \left( 1 - \chi_{n_{0}}(|x_{\ell}|) \right)
\leq \sum_{\ell \in L(T)} \chi_{n_{0}}(|x_{\ell}|) ,
\end{equation*}
and the same result holds if each $x_{\ell}$ is
replaced with $x_{\ell}^{0}$.$\ES$
\end{lema}
%%%%%%%%%%%%%%%%%%%%%%%%%%%%%%%%%%%%%%%%%%%%%%%%%%%%%%%%%%%%%%%%%%%%%%%%%

\prova It follows from the identity
\begin{equation*}
\prod_{j=1}^{n} \left( 1 - a_{j} \right) =
1 - a_{1} - \sum_{j=2}^{n} a_{j} \prod_{i=1}^{j-1}
\left( 1 - a_{i} \right)
\end{equation*}
with $n\ge 2$ and $0 \le a_{j} \le 1$, which can be easily proved
by induction.$\EP$

%%%%%%%%%%%%%%%%%%%%%%%%%%%%%%%%%%%%%%%%%%%%%%%%%%%%%%%%%%%%%%%%%%%%%%%%%
\begin{lema} \label{lemma:M-ovM}
For all $n_{0}$ one has
\begin{eqnarray*}
\left| \calM^{[n_{0}]}_{j_{0}}(0;\eps) -
\ov{\calM}^{[n_{0}]}_{j_{0}}(0;\eps) \right|
& \leq & B_{1} \left| \eps \right|
e^{-B_{2} 2^{n_{0}/\tau_{1}}} , \nonumber \\
\left| \partial_{\eps} \left( \calM^{[n_{0}]}_{j_{0}}(0;\eps) -
\ov{\calM}^{[n_{0}]}_{j_{0}}(0;\eps) \right) \right|
& \leq & B_{1} e^{-B_{2} 2^{n_{0}/\tau_{1}}} , \nonumber
\end{eqnarray*}
for suitable constants $B_{1}$ and $B_{2}$,
depending on $j_{0}$ but independent of $n_{0}$.$\ES$
\end{lema}
%%%%%%%%%%%%%%%%%%%%%%%%%%%%%%%%%%%%%%%%%%%%%%%%%%%%%%%%%%%%%%%%%%%%%%%%%

\prova
One can write $M^{[n_{0}]}(0;\eps)$ as in (\ref{eq:renprop0})
with $x=0$, where $\calV_{T}(0;\eps)$ is given by (\ref{eq:calVT})
with $n_{\ell}=n_{0}$ and
\begin{equation*}
g^{[n_{\ell}]}_{\ell} =
\frac{\psi_{n_{0}}(|x_{\ell}^{0}|)}{ix_{\ell}^0}
=\frac{1-\chi_{n_{0}}(|x_{\ell}^{0}|)}{ix_{\ell}^0} .
\end{equation*}
Furthermore $\calM^{[n_{0}]}_{j_{0}}(x;\eps)$
and $\ov{\calM}^{[n_{0}]}_{j_{0}}(0;\eps)$ are polynomials
of degree $j_{0}$ in $\eps$, hence one has
\begin{equation*}
\calM^{[n_{0}]}_{j_{0}}(0;\eps) -
\ov{\calM}^{[n_{0}]}_{j_{0}}(0;\eps) =
\sum_{k=1}^{j_{0}} \sum_{T \in \calS^{\calR}_{k,n_{0}}}
\left( \calV_{T}(0;\eps) - \ov{\calV}_{T}(0;\eps) \right) ,
\end{equation*}
which is trivially differentiable with respect to $\eps$.
By applying Lemma~\ref{lemma:chi}, we obtain
\begin{eqnarray} \label{eq:M-ovMsum}
& & \left| \left( \calM^{[n_{0}]}_{j_{0}}(0;\eps) -
\ov{\calM}^{[n_{0}]}_{j_{0}}(0;\eps) \right) \right|
\\
& & \qquad \qquad \leq \sum_{k=1}^{j_{0}} \left| \eps \right|^{k}
\sum_{T \in \calS^{\calR}_{k,n_{0}}}
\sum_{\ell' \in L(T)} \chi_{n_{0}}(|x_{\ell'}^{0}|)
\left( \prod_{\ell \in L(T)} \frac{1}{|x_{\ell}^{0}|} \right)
\left( \prod_{v \in B(T)} \left| F_{v} \right| \right) , \nonumber 
\end{eqnarray}
and for $\partial_{\eps} (\calM^{[n_{0}]}_{j_{0}}(0;\eps) -
\ov{\calM}^{[n_{0}]}_{j_{0}}(0;\eps))$ the same bound can be
obtained with $k \left| \eps \right|^{k-1}$ instead of $|\eps|^{k}$.
In (\ref{eq:M-ovMsum}) the factor $\chi_{n_{0}}(|x_{\ell'}^{0}|)$
requires $|x_{\ell'}^{0}| \le C_{1} 2^{-n_{0}}$, so that
by the Diophantine condition (\ref{eq:Diophantine})
one has $|\vnu_{\ell'}^{0}| \geq 2^{n_{0}/\tau}$, hence in
\begin{equation*}
\left(\prod_{v \in B(T)} \left| F_v \right| \right) \leq \calQ^{|B(T)|}
\left( \prod_{v \in B(T)} e^{-\kappa|\vnu_{v}|/2} \right)
\left( \prod_{v \in B(T)} e^{-\kappa|\vnu_{v}|/4} \right)
\left( \prod_{v \in B(T)} e^{-\kappa|\vnu_{v}|/4} \right)
\end{equation*}
one can bound the third product by
\begin{equation*}
\left( \prod_{v \in B(T)} e^{-\kappa|\vnu_{v}|/4} \right)
\leq e^{-\kappa|\vnu_{\ell'}^{0}|/4}
\leq e^{-\kappa 2^{n_{0}/\tau}/4} <
e^{-\kappa 2^{n_{0}/\tau_{1}}/4} ,
\end{equation*}
if $\tau_{1}>\tau_{0}$, while using the second product to perform
the sum over the mode labels and the first one to find,
by reasoning as for (\ref{eq:fact1}) to (\ref{eq:fact3}),
\begin{equation*}
\left( \prod_{\ell \in L(T)} \frac{1}{|x_{\ell}^{0}|} \right)
\left(\prod_{v \in B(T)}
e^{-\kappa|\vnu_{v}|/2}\right) \leq
\left( \frac{\tau!}{C_{0}} \left(
\frac{2|L(T)|}{\kappa} \right)^{\tau} \right)^{|L(T)|}
\leq \ov{\Gamma}_{1} \ov{\Gamma}_{2}^{k_{T}}
\left( k_{T}! \right)^{\beta} ,
\end{equation*}
with $k_{T}=k$ for $T\in \calS^{\calR}_{k,n_{0}}$,
so that, by collecting together the bounds and inserting them
into (\ref{eq:M-ovMsum}), we prove the assertion.
In particular $B_{1}$ is proportional to $\ov{\Gamma}_{2}^{j_{0}}
(j_{0}!)^{\beta}$, while $B_{2}$ is independent of $j_{0}$. $\EP$

%%%%%%%%%%%%%%%%%%%%%%%%%%%%%%%%%%%%%%%%%%%%%%%%%%%%%%%%%%%%%%%%%%%%%%%%%
\begin{lema} \label{lemma:n0}
Let $G_j$, $j \geq 1$, be as the previous sections. Assume
that there is $j_{0}\in \N$ such that
$G_{j_{0}}\neq 0$ and $G_{j}=0$ for all $1\leq j < j_{0}$.
There exists two constants $c_{1}$ and $c_{2}$, depending on $j_{0}$,
such that for
\begin{equation} \label{eq:n0}
n_{0} \geq \tau_{1} \log_{2} \left( c_{1} +
c_{2} \log \frac{1}{|\eps|} \right)
\end{equation}
one has
\begin{equation} \label{eq:Mn0}
\left| \partial_{\eps} \calM^{[n_{0}]}_{j_{0}}(0;\eps) \right|
\geq \frac{j_{0}}{2} \left| \eps \right|^{j_{0}-1} |G_{j_{0}}| .
\end{equation}
provided $\eps$ is small enough.
If $j_{0}=1$ one can take $c_{2}=0$.$\ES$
\end{lema}
%%%%%%%%%%%%%%%%%%%%%%%%%%%%%%%%%%%%%%%%%%%%%%%%%%%%%%%%%%%%%%%%%%%%%%%%%

\prova One can write
\begin{equation} \label{eq:Mn0sum}
\ov{\calM}^{[n_{0}]}_{j_{0}}(0;\eps) =
\sum_{k=1}^{j_{0}} \eps^{k} G_{k}
= \eps^{j_{0}} G_{j_{0}} ,
\end{equation}
by Lemma~\ref{lemma:MG}, so that
\begin{equation} \label{eq:Mn0bound}
\partial_{\eps} \ov{\calM}^{[n_{0}]}_{j_{0}} (0;\eps) =
j_{0} \eps^{j_{0}-1} G_{j_{0}} .
\end{equation}
By Lemma~\ref{lemma:M-ovM}, we can bound
\begin{equation} \label{eq:Mn01bound}
\left| \partial_{\eps} \left( \calM^{[n_{0}]}_{j_{0}}(0;\eps) -
\ov{\calM}^{[n_{0}]}_{j_{0}}(0;\eps) \right) \right|
\leq \beta_{1} B_{1} e^{- \beta_{2}
B_{2}2^{n_{0}/\tau_{1}}} \leq \frac{1}{2}
|\eps|^{j_{0}-1}\left| G_{j_{0}} \right| ,
\end{equation}
where the first inequality is obtained as soon as
$\beta_{2} \leq 1$ and $\beta_{1} \geq 1$, while the second one 
requires
\begin{equation} \label{eq:n0intermediate}
n_{0} \geq \tau_{1} \log_{2} \left( \frac{1}{\beta_{2} B_{2}} \log
\frac{2\beta_{1}B_{1}}{|G_{j_{0}}|} +
\frac{j_{0}-1}{\beta_{2}B_{2}} \log\frac{1}{|\eps|} \right) ,
\end{equation}
so that the assertion follows if $c_{1}$ and $c_{2}$ are chosen
according to (\ref{eq:n0intermediate}).$\EP$

\begin{rmk} \label{rmk:betafree}
The constants $\beta_{1}$ and $\beta_{2}$
in (\ref{eq:Mn01bound}) could be taken $\beta_{1}=\beta_{2}=1$.
However in the following it will turn out useful
to have some freedom in fixing their values;
see in particular Remark~\ref{rmk:betafixed}.
\end{rmk}

\begin{rmk} \label{rmk:n0e1}
Note that if we choose $n_{0}= \tau_{1} \log_{2} (c_{1}+c_{2}
\log(2/\eps_{m}) )$ we obtain a value of $n_{0}$ which can be
used for all $\eps\in\calE_{m}$.
\end{rmk}

%%%%%%%%%%%%%%%%%%%%%%%%%%%%%%%%%%%%%%%%%%%%%%%%%%%%%%%%%%%%%%%%%%%%%%%%%
%%%%%%%%%%%%%%%%%%%%%%%%%%%%%%%%%%%%%%%%%%%%%%%%%%%%%%%%%%%%%%%%%%%%%%%%%
\zerarcounters
\section{Convergence of the renormalized expansion}
\label{sec:convergence}
%%%%%%%%%%%%%%%%%%%%%%%%%%%%%%%%%%%%%%%%%%%%%%%%%%%%%%%%%%%%%%%%%%%%%%%%%
%%%%%%%%%%%%%%%%%%%%%%%%%%%%%%%%%%%%%%%%%%%%%%%%%%%%%%%%%%%%%%%%%%%%%%%%%

We are left with the problem of proving that the series defining
the renormalized expansion (\ref{eq:uren}) converges, and of studying 
how large is the set $\calE_{*} \cap [0,\eps_{0}]$ of
admissible values of $\eps$; we shall verify that
it is a set with positive relatively large measure.

As we have fixed $m$, for notational simplicity, in the following
we shall find convenient to shorthand
$\calE^{[\infty]} \equiv \calE_{*} \cap \calE_{m}$. We shall assume
$\eps\in\calE^{[\infty]}$, and $n_{0}$ fixed as in Remark~\ref{rmk:n0e1}.

%%%%%%%%%%%%%%%%%%%%%%%%%%%%%%%%%%%%%%%%%%%%%%%%%%%%%%%%%%%%%%%%%%%%%%%%%
\begin{lema} \label{lemma:lemma8.1}
Assume that the set $\calE^{[\infty]}$ has non-zero measure and that
for all $\eps\in\calE^{[\infty]}$ and for all $n_{0} \le j < n-1$
the functions $\calM^{[j]}(x;\eps)$ are $C^{1}$
in $x$ and satisfy the bounds
\begin{equation} \label{eq:boundM}
\left| \calM^{[j]}(x;\eps) \right| \le D \sqrt{|\eps|} , \qquad
\left| \partial_{x} \calM^{[j]}(x;\eps) \right| \leq D \sqrt{|\eps|} ,
\end{equation}
for some constant $D$. There there exists a positive constant $c$,
independent of $n$, such that for any renormalized tree $\theta$
with $\Val(\theta)\neq 0$ the number $N_{j}(\theta)$
of lines on scale $j$ satisfies the bound
\begin{equation} \label{eq:Nj}
N_{j}(\theta) \le c \, 2^{-j/\tau_{1}} \sum_{v\in B(\theta)}
|\vnu_{v}| ,
\end{equation}
for all $n_{0} < j \le n-1$.$\ES$
\end{lema}
%%%%%%%%%%%%%%%%%%%%%%%%%%%%%%%%%%%%%%%%%%%%%%%%%%%%%%%%%%%%%%%%%%%%%%%%%

\prova Set $K(\theta):=\sum_{v\in B(\theta)}|\vnu_{v}|$ and call
$N_{j}^{\dagger}(\theta)$ the number of lines in $L(\theta)$ on scale
$j'\ge j$. We prove inductively on the order $k$ of the renormalized trees
the bound
\begin{equation} \label{eq:Nj*}
N^{\dagger}_{j}(\theta) \le \max\{ 0 , 2 K(\theta) 2^{(3-j)/\tau_{1}} - 
1 \} \,
\end{equation}
for all $n_{0}<j\leq n-1$

If $\theta$ has $k=0$ one has $B(\theta)=\{v\}$ and 
$K(\theta)=|\vnu_{v}|$.
The line $\ell$ exiting $v$ can be on scale $n_{\ell}\geq j$ only if
have $|ix-\calM^{[j-2]}(x;\eps)|\le 2^{-j+1}C_{1}$ (see Remark
\ref{rmk:scales}), with $x=\vomega\cdot\vnu_{v}$, hence,
by the Diophantine conditions (\ref{eq:Melnikov}),
one has $|\vnu_{v}|\ge 2^{(j-1)/\tau_{1}}$, which implies
$2 \,K(\theta)\, 2^{(3-j)/\tau_{1}} \ge 2 2^{2/\tau_{1}} \ge 2$.
Therefore in such a case  the bound (\ref{eq:Nj*}) is trivially 
satisfied.
   
If $\theta$ is a renormalized tree of order $k\geq 1$, we assume that
the bound holds for all renormalized trees of order $k'<k$.
Define $E_{j}=(2\,2^{(3-j)/\tau_{1}})^{-1}$: so we have
to prove that $N^{\dagger}_{j}(\theta)\le \max\{0, K(\theta) 
E_{j}^{-1}-1\}$.

Call $\ell$ the root line of $\theta$ and
$\ell_{1},\ldots,\ell_{m}$ the $m\ge 0$ lines on scales $\ge j$
which are the closest to $\ell$ (i.e. such that no other line
along the paths connecting the lines $\ell_{1},\ldots,\ell_{m}$
to the root line is on scale $\geq j$).

If the root line $\ell$ of $\theta$ is on scale $n_{\ell}< j$, then
\begin{equation*}
N_{j}^{\dagger}(\theta) = \sum_{i=1}^{m} N_{j}^{\dagger}(\theta_{i}) ,
\end{equation*}
where $\theta_{i}$ is the renormalized subtree with $\ell_{i}$
as root line, hence the bound follows by the inductive hypothesis.

If the root line $\ell$ has scale $n_{\ell} \geq j$,
then $\ell_{1},\ldots,\ell_{m}$ are the entering lines of a cluster 
$T$.

By denoting again with $\theta_{i}$ the renormalized subtree
having $\ell_{i}$ as root line, one has
\begin{equation} \label{eq:Njm}
N_{j}^{\dagger}(\theta) = 1 + \sum_{i=1}^{m} N_{j}^{\dagger}(\theta_{i}) ,
\end{equation}
so that the bound becomes trivial if either $m=0$ or $m\geq 2$.

If $m=1$ then one has a cluster $T$ with two external lines
$\ell$ and $\ell_{1}$, which are both with scales $\geq j$.
Set $x_{\ell}=\vomega\cdot\vnu_{\ell}$ and
$x_{\ell_{1}}=\vomega\cdot\vnu_{\ell_{1}}$; then
\begin{equation} \label{eq:ellell1}
\left| i x_{\ell} - \calM^{[j-2]}(x_{\ell};\eps) \right|
\le 2^{-j+1} C_{1} , \qquad
\left| ix_{\ell_{1}} - \calM^{[j-2]}(x_{\ell_{1}};\eps) \right|
\le 2^{-j+1} C_{1} ,
\end{equation}
and $\vnu_{\ell} \neq \vnu_{\ell_{1}}$, otherwise $T$ would be a
self-energy graph. Then, by (\ref{eq:ellell1}), one has
\begin{eqnarray*}
2^{-j+2} C_{1} & \ge & \left| i ( x_{\ell} - x_{\ell_{1}} ) -
\calM^{[j-2]}(x_{\ell};\eps) +
\calM^{[j-2]}(x_{\ell_{1}};\eps) \right| \nonumber \\
& = & \left| \left( x_{\ell}-x_{\ell_{1}} \right) \left( i -
\partial_{x}\calM^{[j-2]}(x_{*};\eps) \right) \right| \nonumber \\
& \geq & \left| x_{\ell}-x_{\ell_{1}} \right|
\left( 1 - D \sqrt{|\eps|} \right) \geq \frac{1}{2}
\left| x_{\ell}-x_{\ell_{1}} \right| , \nonumber
\end{eqnarray*}
where $x_{*}$ is a point between $x_{\ell}$
and $x_{\ell_{1}}$, and the assumption (\ref{eq:boundM})
has been used. Hence by the Diophantine condition 
(\ref{eq:Diophantine}),
one has $| \vnu_{\ell}-\vnu_{\ell_{1}} | > 2^{(j-3)/\tau_{1}}$, so that
\begin{equation*}       
\sum_{v\in B(T)} |\vnu_{v}| \geq |\vnu_{T}| =
\left| \vnu_{\ell}-\vnu_{\ell_{1}} \right| > 2^{(j-3)/\tau_{1}} > E_{j} 
,
\end{equation*}
hence $K(\theta)-K(\theta_{1})> E_{j}$,
which, inserted into (\ref{eq:Njm}) with $m=1$, gives,
by using the inductive hypothesis,
\begin{eqnarray*}
N_{j}^{\dagger}(\theta) & = & 1 + N_{j}^{\dagger}(\theta_{1})
\leq 1 + K(\theta_{1}) \, E_{j}^{-1} - 1 \nonumber \\
& \leq & 1 + \Big( K(\theta) - E_{j} \Big) E_{j}^{-1} - 1
\le K(\theta) \, E_{j}^{-1} - 1 , \nonumber
\end{eqnarray*}
hence the bound is proved also if the root line is on
scale $n_{\ell} \geq j$.$\EP$

\begin{rmk} \label{rmk:epssquare}
Let $j_{0}$ be as in Lemma~\ref{lemma:n0}.
If $\eps_{0}$ is small enough,
for all $\eps\in(\eps_{m}/2,\eps_{m}]$ and $n_{0}$ chosen according 
to Remark~\ref{rmk:n0e1}, if $j_{0} = 1$ we can bound 
\begin{equation} \label{eq:epslog1}
\left| \eps \right|^{k} 2^{(2k-1)n_{0}} \leq
c_{1}^{(2k-1)\tau_{1}} \left| \eps \right|^{k} ,
\end{equation}
while if $j_{0} > 1$ we can bound
\begin{equation} \label{eq:epslog2}
\left| \eps \right|^{k} 2^{(2k-1)n_{0}} \leq
\left( 2 c_{2} \right)^{(2k-1)\tau_{1}}
\left| \eps \right|^{k}
\left( \log \frac{2}{|\eps|} \right)^{(2k-1) \tau_{1}} ,
\end{equation}
where, under the same smallness assumption on $\eps$, one has
\begin{equation} \label{eq:epslogroot}
\left( \log \frac{2}{|\eps|} \right)^{\tau_{1}}
\leq S_{p} \left( \frac{1}{|\eps|} \right)^{p} ,
\end{equation}
for all $p>0$ and with $S_{p}$ a positive constant depending on $p$.
Hence, in (\ref{eq:epslog2}), by taking $p \leq 1/4$, one obtains
for $j_{0}>1$
\begin{equation} \label{eq:epsk2j0}
\left| \eps \right|^{k} 2^{n_{0}(2k-1)} \leq
\left( 2 c_{2} \right)^{(2k-1)\tau_{1}}
S_{p}^{2k-1} \left| \eps \right|^{k/2} .
\end{equation}
Therefore, whichever the value of $j_{0}$ is, we can bound
\begin{equation} \label{eq:epsk2}
|\eps|^{k} 2^{(2k-1)n_{0}} \le c_{3} |\eps|^{k/2} ,
\end{equation}
for all $k\geq 1$, with $c_{3}$ a suitable positive constant.
\end{rmk}

\begin{rmk} \label{rmk:pj0}
In particular one can choose $p\leq 1/(2(2j_{0}+1))$,
which implies
\begin{equation*}
\left| \eps \right|^{k-1} \left( \log \frac{2}{|\eps|}
\right)^{(2k-1)\tau_{1}}
\leq
%\left| \eps \right|^{k} \left( \log \frac{2}{|\eps|}
%\right)^{(2k+1)\tau_{1}} \leq
\left| \eps \right|^{j_{0}-1} \sqrt{|\eps|}
\left| \eps \right|^{(k-j_{0}-1)/2}
\end{equation*}
for all $k \geq j_{0}+1$, a property
which will be useful in the following.
\end{rmk}

%%%%%%%%%%%%%%%%%%%%%%%%%%%%%%%%%%%%%%%%%%%%%%%%%%%%%%%%%%%%%%%%%%%%%%%%%
\begin{lema} \label{lemma:boundMn0}
Fix $p$ as in Remark~\ref{rmk:epssquare}. Then one has
\begin{equation*}
\left| \calM^{[n_{0}]}(x;\eps) \right| \leq D \sqrt{|\eps|} ,
\qquad \left| \partial_{x} \calM^{[n_{0}]}(x;\eps)
\right| \leq D \, |\eps| ,
\end{equation*}
for suitable positive constants $D$ and $D'$.$\ES$
\end{lema}
%%%%%%%%%%%%%%%%%%%%%%%%%%%%%%%%%%%%%%%%%%%%%%%%%%%%%%%%%%%%%%%%%%%%%%%%%

\prova The first bound follows from (\ref{eq:epsk2}).

Let $j_{0}$ be as in Lemma~\ref{lemma:n0}.
If $j_{0}=1$ then $n_{0}$ does not depend on $\eps$,
and also the bound second is trivially satisfied.

If $j_{0} \geq 2$, in order to obtain the second bound,
one can discuss in a different ways contributions with
$k_{T}=1$ and contributions with $k_{T}\geq 2$.
If $k_{T}=1$ then $\calV_{T}(x;\eps)$ does not depend on $x$
(see Remark~\ref{rmk:calVx}), so that, by using
the notations (\ref{eq:ovcalM}), one has $\calM^{[n_{0}]}_{1}(x;\eps)=
\chi_{n_{0}}(|x|) M^{[n_{0}]}_{1}(0;\eps)$, and one can write
\begin{equation*} 
M^{[n_{0}]}_{1}(0;\eps) =
\ov{M}^{[n_{0}]}_{1}(0;\eps)+ \left( M^{[n_{0}]}_{1}(0;\eps) -
\ov{M}^{[n_{0}]}_{1}(0;\eps) \right) ,
\end{equation*}
where $\ov{M}^{[n_{0}]}_{1}(0;\eps)=0$ by Lemma~\ref{lemma:MG}
(and the definition of $j_{0}$), while the difference
\begin{equation*}
\calM^{[n_{0}]}_{1}(0;\eps) - \ov{\calM}^{[n_{0}]}_{1}(0;\eps) =
\chi_{n_{0}}(|x|) \left( M^{[n_{0}]}_{1}(0;\eps) -
\ov{M}^{[n_{0}]}_{1}(0;\eps) \right)
\end{equation*}
can be bounded through Lemma~\ref{lemma:M-ovM}
proportionally to $e^{-B_{2} 2^{n_{0}/\tau_{1}}}$.
Hence the derivative with respect to $x$ acts only
on the compact support function $\chi_{n_{0}}(|x|)$
and produces a factor $2^{n_{0}}$ which is controlled
by the exponentially small factor $e^{-B_{2} 2^{n_{0}/\tau_{1}}}$.
The conclusion is that the contributions with $k_{T}=1$
can be bounded proportionally to $\eps$.
The contributions with $k_{T}=2$ can be bounded relying again
on the bound (\ref{eq:epsk2}).$\EP$

%%%%%%%%%%%%%%%%%%%%%%%%%%%%%%%%%%%%%%%%%%%%%%%%%%%%%%%%%%%%%%%%%%%%%%%%%
\begin{lema} \label{lemma:lemma8.2}
Fix $p$ as in Remark~\ref{rmk:epssquare} and $n_{0}$ as
in Remark~\ref{rmk:n0e1}. For $\eps\in\calE^{[\infty]}$ and for $x$
such that $g^{[n]}(x;\eps)\neq 0$, there exist two constants $D$ and 
$D'$
such that the functions $\calM^{[j]}(x;\eps)$
are smooth functions of $x$ and satisfy the bounds
\begin{eqnarray} \label{eq:boundsMderM}
& & \left| \calM^{[j]}(x;\eps) \right| \leq D \sqrt{|\eps|} ,
\qquad \left| \partial_{x} \calM^{[j]}(x;\eps)
\right| \leq D \, |\eps| , \nonumber \\
& & \left| \calM^{[j]}(x;\eps) - \calM^{[j-1]}(x;\eps) \right|
\leq D \, |\eps| e^{-D' 2^{j/\tau_{1}}} ,
\end{eqnarray}
for all $n_{0} < j \le n-1$. Furthermore for all $T$
contributing to $\calM^{[j]}(x;\eps)$, with $n_{0}< j \leq n-1$,
one has
\begin{equation} \label{eq:NjT}
N_{j'}(T) \leq c \, 2^{-j'/\tau_{1}} \sum_{v \in B(T)} |\vnu_{v}| ,
\end{equation}
for all $j' \leq j$.$\ES$
\end{lema}
%%%%%%%%%%%%%%%%%%%%%%%%%%%%%%%%%%%%%%%%%%%%%%%%%%%%%%%%%%%%%%%%%%%%%%%%%

\prova The first bound in (\ref{eq:boundsMderM})
can be proved by induction on $n_{0}\leq j \leq n-1$.
For $j=n_{0}$ it has been already checked
(see Lemma~\ref{lemma:boundMn0}). Let us assume that it holds
for all $n_{0}\leq j'<j$.
One can proceed as for the proof of Lemma~2 in Ref.~\cite{Gentile}.
First of all one can prove for any self-energy graph
$T\in \calS^{\calR}_{k,j}$ the inequalities
\begin{equation} \label{eq:boundsNjT}
\sum_{v \in B(T)} |\vnu_{v}| > 2^{(j-4)/\tau_{1}} \, , \quad
N_{j'}(T) \leq 2\,2^{(3-j')/\tau_{1}}\sum_{v\in B(T)} |\vnu_{v}| \, ,
\quad n_{0}+1 \leq j' \leq j, 
\end{equation}
where $N_{j'}(T)$ denotes the number of lines on scales $j'$
contained in $T$. We omit the proof, as it is identical to that
given in Ref.~\cite{Gentile}.

The estimates (\ref{eq:boundsNjT}) allow us to bound
\begin{equation} \label{eq:boundcalV}
\left| \calV_{T}(x;\eps) \right| \leq
\left| \eps \right|^{k} A_{1} A_{2}^{k}
e^{-A_{3} 2^{j/\tau_{1}}} \prod_{v \in B(T)}
e^{-\kappa |\vnu_{v}|/2} .
\end{equation}
The only difference with respect to the analogous
bound (7.18) in Ref.~\cite{Gentile} is that
the constants $A_{1}$ and $A_{2}$ depend on $\eps$.
In fact given a self-energy graph $T\in\calS^{\calR}_{k,j}$,
if we express its value according to (\ref{eq:calVT}),
we can bound
\begin{equation*}
\prod_{\ell \in L(T)} \big| g^{[n_{\ell}]}_{\ell}
\big| \leq \left( 2 C_{1}^{-1} \right)^{|L(T)|}
2^{n_{0}N_{n_{0}}(T)} \left(
\prod_{n=n_{0}+1}^{j} 2^{nN_{n}(T)} \right) ,
\end{equation*}
with $N_{n_{0}}(T)\leq 2k-1$ and $N_{n}(T) \leq
c\,2^{-n/\tau_{1}}\sum_{v\in B(T)} |\vnu_{v}|$ for
all $n_{0}+1 \leq n \leq j$, as it follows from the second bound in
(\ref{eq:boundsNjT}). Hence the last product can be bounded
by using the bound on $N_{n}(T)$ and (\ref{eq:choicen1})
with $n_{1}(\kappa,c,\tau)=n_{0}$: just note that for $\eps_{0}$
small enough such a choice for $n_{1}(\kappa,c,\tau)$
automatically satisfies the inequality in (\ref{eq:choicen1}).
Then we can apply the bounds given
in Remark~\ref{rmk:epssquare} to write
\begin{equation} \label{eq:A1A2}
|\eps|^k A_{1}A_{2}^{k} = \ov{A}_{1}\ov{A}_{2}^{k}|\eps|^{k/2} ,
\end{equation}
with $\ov{A}_{1}$ and $\ov{A}_{2}$ two constants independent of $\eps$.
Then the first bound in (\ref{eq:boundsMderM}) is proven.

To obtain the third bound in (\ref{eq:boundsMderM}) we note that
one has for $j \geq n_{0}+1$
\begin{eqnarray} \label{eq:thirdbound}
\left| \calM^{[j]}(x;\eps) - \calM^{[j-1]}(x;\eps) \right|
&\leq& \left| M^{[j]}(x;\eps) \right| \; \leq \; \sum_{k=1}^{\infty}
\sum_{T\in\calS^{\calR}_{k,j}} |\calV_{T}(x;\eps)| \nonumber \\
&=& \sum_{T\in\calS^{\calR}_{1,j}} |\calV_{T}(x;\eps)| + 
\sum_{k=2}^{\infty}
\sum_{T\in\calS^{\calR}_{k,j}} |\calV_{T}(x;\eps)| , \qquad
\end{eqnarray}
where the last sum can be bounded proportionally to
$|\eps|\,e^{-A_{3}2^{j/\tau_{1}}}$, because of (\ref{eq:boundcalV})
and (\ref{eq:A1A2}). The first one can be bounded proportionally
to $|\eps|$ if $j_{0}=1$.
If $j_{0} \geq 2$ we can reason as follows.
We can bound $|\calV_{T}(x;\eps)|$
according to (\ref{eq:boundcalV}), with $k=1$, and write
$e^{-A_{3}2^{j/\tau_{1}}}=e^{-A_{3}2^{j/\tau_{1}}/2}
e^{-A_{3}2^{j/\tau_{1}}/4}e^{-A_{3}2^{j/\tau_{1}}/4}$.
The self-energy graph $T$ contains exactly one a line $\ell$
on scale $j$ (as $T\in \calS^{\calR}_{1,j}$), hence $n_{\ell}=j$
and $|g^{[n_{\ell}]}_{\ell}|\leq C_{1}^{-1}2^{j+1}$, so that we can use
that $2^{j} e^{-A_{3}2^{j/\tau_{1}}/4}$ is bounded by a constant.
Moreover we have $e^{-A_{3}2^{j/\tau_{1}}/4} \leq
e^{-\beta_{2} B_{2} 2^{n_{0}/\tau_{1}} } \leq
|G_{j_{0}}|(2\beta_{1}B_{1})^{-1}|\eps|^{j_{0}-1}$
if $\beta_{2}$ in (\ref{eq:Mn01bound})
is chosen such that $\beta_{2} B_{2} \leq A_{3}/4$
(see Remark~\ref{rmk:betafree}). Therefore, we can conclude
that if $j_{0}\geq 2$ the first sum in (\ref{eq:thirdbound})
can be bounded proportionally to $|\eps| |\eps|^{j_{0}-1}
e^{-A_{3}2^{j/\tau_{1}}/2}$. Hence the third bound in
(\ref{eq:boundsMderM}) follows for any value of $j_{0}$, 
with $D'=A_{3}/2$.

The second bound in (\ref{eq:boundsMderM}) again can be proved
by reasoning as in Ref.~\cite{Gentile} for the contributions
arising from self-energy graphs $T$ with $k_{T}\geq 2$.
The contributions arising from self-energy graphs $T$ with with $k_T = 1$ 
can be bounded as $|\eps| \calQ C_1^{-1} 2^{n+1} e^{-A_3 2^{n/\tau_1}}
e^{-\kappa|\vnu_v|/2}$ (as in the bound on the first sum in the r.h.s. of 
(\ref{eq:thirdbound})) because there is only one propagator on scale
$n$. Then, if the derivative acts on the compact support function
$\chi_{n_0}(|x|)$, one has that $2^{n_0} 2^{n+1} e^{-A_3 2^{n/\tau_1}/2}$
is bounded by a constant for all $n>n_0$.$\EP$

\begin{rmk} \label{rmk:betafixed}
As suggested by the proof of Lemma~\ref{lemma:lemma8.2}
we shall fix $\beta_{2}$ in (\ref{eq:Mn01bound}) such that one
has $D' \geq 2 \beta_{2} B_{2}$, where $D'$ is the constant appearing
in the last of (\ref{eq:boundsMderM}). For future convenience
we shall choose $\beta_{2}$ such that
$D'=4\beta_{2}B_{2}$; see (\ref{eq:diffMnimproved}).
We shall see below that it will be useful (even not necessary)
also to choose $\beta_{1}$ in (\ref{eq:Mn01bound})
such that $2D \leq \beta_{1} B_{1}$.
\end{rmk}

%%%%%%%%%%%%%%%%%%%%%%%%%%%%%%%%%%%%%%%%%%%%%%%%%%%%%%%%%%%%%%%%%%%%%%%%%
\begin{prop} \label{prop:prop8}
Assume that the set $\calE^{[\infty]}$ has non-zero measure. Then for
all $\eps\in\calE^{[\infty]}$ one has
\begin{equation} \label{eq:boundpropren}
\big| g^{[n_{\ell}]}_{\ell} \big| \leq C_{1}^{-1} 2^{n_{\ell}+1}
\end{equation}
for all lines $\ell$ in any tree or self-energy graph.
In particular the series (\ref{eq:uren}) is uniformly convergent
to a function analytic in $t$.$\ES$
\end{prop}
%%%%%%%%%%%%%%%%%%%%%%%%%%%%%%%%%%%%%%%%%%%%%%%%%%%%%%%%%%%%%%%%%%%%%%%%%

\prova It follows from Lemma~\ref{lemma:lemma8.1},
by taking the limit $n\to\infty$
and using that the constant $c$ does not depend on $n$,
that the bound (\ref{eq:Nj}) holds for all $j>n_{0}$.
Then one can bound the product of propagators as done
in the proof of Lemma~\ref{lemma:lemma8.2}, and using
part of the decaying factors $e^{-\kappa|\vnu_{v}|}$
to obtain an overall factor $e^{-\kappa|\vnu|/4}$
for any tree $\theta\in\Theta_{k,\vnu}$ contributing
to $u^{[k]}_{\vnu}$.$\EP$

%%%%%%%%%%%%%%%%%%%%%%%%%%%%%%%%%%%%%%%%%%%%%%%%%%%%%%%%%%%%%%%%%%%%%%%%%
%%%%%%%%%%%%%%%%%%%%%%%%%%%%%%%%%%%%%%%%%%%%%%%%%%%%%%%%%%%%%%%%%%%%%%%%%
\zerarcounters
\section{Measure of the set of admissible values}
\label{sec:measure}
%%%%%%%%%%%%%%%%%%%%%%%%%%%%%%%%%%%%%%%%%%%%%%%%%%%%%%%%%%%%%%%%%%%%%%%%%
%%%%%%%%%%%%%%%%%%%%%%%%%%%%%%%%%%%%%%%%%%%%%%%%%%%%%%%%%%%%%%%%%%%%%%%%%

To apply the above results we have still to construct the
set $\calE_\ast$ for which the Diophantine conditions (\ref{eq:Melnikov})
hold, and to show that such a set has positive measure.
Here and henceforth we assume that the constants
$n_{0}$ and $p$ are chosen according to Remark~\ref{rmk:n0e1}
and Remark~\ref{rmk:pj0}, respectively.

Define recursively the sets $\calE^{[n]}$ as follows.
Set $\calE^{[n_{0}]} = \calE_{m}$ and, for $n\geq n_{0}+1$,
\begin{equation} \label{eq:en}
\calE^{[n]} := \left\{ \eps \in \calE^{[n-1]} \; : \;
|i\vomega \cdot \vnu - \calM^{[n-1]}(\vomega \cdot\vnu;\eps) | >
C_{1} |\vnu|^{-\tau_{1}} \right\} ,
\end{equation}
for suitable Diophantine constants $C_{1}$ and $\tau_{1}$
(to be fixed later). It is clear that 
\begin{equation} \label{eq:ecap}
\calE_\ast \cap \calE_m = \calE^{[\infty]} 
= \bigcap_{n=n_{0}}^{\infty} \calE^{[n]} =
\lim_{n\to\infty} \calE^{[n]} .
\end{equation}
%

%%%%%%%%%%%%%%%%%%%%%%%%%%%%%%%%%%%%%%%%%%%%%%%%%%%%%%%%%%%%%%%%%%%%%%%%%
\begin{lema} \label{lemma:lemma9.1}
The functions $M^{[n]}(x;\eps)$ and their derivatives
$\partial_{x}M^{[n]}(x;\eps)$ are $C^{1}$ extendible
in the sense of Whitney outside $\calE^{[n-1]}$,
and for all $\eps,\eps'\in \calE^{[n-1]}$ one has
\begin{equation} \label{eq:Whitney}
\partial_{x}^{s} M^{[n]}(x;\eps') - \partial_{x}^{s} M^{[n]}(x;\eps)
= \left( \eps'-\eps \right)
\partial_{\eps} \partial_{x}^{s} M^{[n]}(x;\eps) + o \left(
\eps'-\eps \right) ,
\end{equation}
where $s=0,1$ and $\partial_{\eps} \partial_{x}^{s}M^{[n]}(x;\eps)$
denotes the formal derivative with respect to $\eps$
of $\partial_{x}^{s} M^{[n]}(x;\eps)$. Furthermore one has
\begin{equation} \label{eq:boundsderderM}
\left| \partial_{\eps} \partial_{x} \calM^{[n]}(x;\eps)
\right| \leq D \sqrt{|\eps|} , \qquad
\left| \partial_{\eps} \partial_{x} M^{[n]}(x;\eps)
\right| \leq D \sqrt{|\eps|} e^{-D'2^{n/\tau_{1}}} ,
\end{equation}
for all $n > n_{0}$. One can take $D$
as in Lemma~\ref{lemma:lemma8.2}.$\ES$
\end{lema}
%%%%%%%%%%%%%%%%%%%%%%%%%%%%%%%%%%%%%%%%%%%%%%%%%%%%%%%%%%%%%%%%%%%%%%%%%

\prova As the proof of Lemma~3 in Ref.~\cite{Gentile}.
In order to obtain the inequality (\ref{eq:boundsderderM})
one has to use the Remark~\ref{rmk:calVx}.
Of course, when expressing $\calM^{[n]}(x;\eps)$ in terms
of the self-energy values $\calV_{T}(x;\eps)$ we have to
bear in mind that the constant $2^{n_{0}}$ can be bounded in terms
of $\eps$, but it does not depend on $\eps$ (as far as
$\eps$ varies in $\calE_{m}$ and $n_{0}$ is chosen according
to Remark~\ref{rmk:n0e1}), so that the
derivatives with respect to $\eps$ of $\calV_{T}(x;\eps)$,
as expressed in (\ref{eq:calVT}), act only on $\eps^{k_{T}}$ and
on the quantities $\calM^{[j]}(x;\eps)$ appearing in the propagators.
Hence $\partial_{\eps}\calV_{T}(x;\eps)$
and $\partial_{\eps}\partial_{x}\calV_{T}(x;\eps)$
can be studied as in Ref.~\cite{Gentile}. We simply note that
when acting on some propagator $g^{[n_{\ell}]}_{\ell}$
the derivatives with respect to $\eps$ can rise the power of the 
divisor
$ix-\calM^{[n_{\ell}-1]}(x;\eps)$, and if $n_{\ell}=n_{0}$
we have to use part of the exponential decay $e^{-A_{3}2^{j/\tau_{1}}}$
(see (\ref{eq:boundcalV})) to take into account the extra
factors $2^{n_{0}}$. The conclusion is that essentially the
derivative with respect to $\eps$ of $\calV_{T}(x;\eps)$ 
admits the same bound (\ref{eq:boundcalV}) as $\calV_{T}(x;\eps)$,
possibly with different constants $A_{1}$ and $A_{2}$
(but still such that a bound like (\ref{eq:A1A2}) is fulfilled,
as far as their dependence on $\eps$ is concerned),
except that the exponent of $|\eps|$ is $k-1$ instead of $k$.$\EP$

\vspace{0.4cm}

Therefore for all $\eps\in\calE^{[n-1]}$ the quantities
$\calM^{[n]}(x;\eps)$ are well defined
and formally differentiable (in the sense of Whitney)
together with their derivatives with respect to~$x$.

%%%%%%%%%%%%%%%%%%%%%%%%%%%%%%%%%%%%%%%%%%%%%%%%%%%%%%%%%%%%%%%%%%%%%%%%%
\begin{lema} \label{lemma:lemma9.2}
There are two positive constants $\gotm_{1}$ and $\gotm_{2}$ such that
\begin{equation} \label{eq:boundderM}
\left| \partial_{\eps} \calM^{[n]}(x;\eps) \right| \geq
\gotm_{1} \left| \eps \right|^{j_{0}-1} -
\gotm_{2} \sqrt{|\eps|} \left| x \right| ,
\end{equation}
for all $n \geq n_{0}$.$\ES$
\end{lema}
%%%%%%%%%%%%%%%%%%%%%%%%%%%%%%%%%%%%%%%%%%%%%%%%%%%%%%%%%%%%%%%%%%%%%%%%%

\prova If we write
\begin{equation} \label{eq:Minterp}
\partial_{\eps} \calM^{[n]}(x;\eps) =
\partial_{\eps} \calM^{[n]}(0;\eps) +
\int_{0}^{x} \der x' \partial_{\eps} \partial_{x'}
\calM^{[n]}(x';\eps),
\end{equation}
we have
\begin{equation*}
\left| \partial_{\eps} \calM^{[n]}(0;\eps) \right| \geq
\left| \partial_{\eps} \calM^{[n_{0}]}(0;\eps) \right| -
\sum_{j=n_{0}+1}^{n} \left| \partial_{\eps} \left( \Xi_j(0;\eps)
M^{[j]}(0;\eps) \right) \right| 
\end{equation*}
and we can bound
\begin{eqnarray*}
\left| \partial_{\eps} \calM^{[n_{0}]}(0;\eps) \right| & \geq &
\left| \partial_{\eps} \calM^{[n_{0}]}_{j_{0}}(0;\eps) \right|
- \sum_{j=j_{0}+1}^{\infty}
\sum_{T\in \calS^{\calR}_{j,n_{0}}}
\left| \partial_{\eps} \calV_{T}(0;\eps) \right| \nonumber \\
& \geq & \frac{j_{0}}{2} \left| \eps \right|^{j_{0}-1}
\left| G_{j_{0}} \right| + O \left( |\eps|^{j_{0}-1}\sqrt{|\eps|} 
\right)
\geq \frac{j_{0}}{4} \left| \eps \right|^{j_{0}-1}
\left| G_{j_{0}} \right| \nonumber ,
\end{eqnarray*}
where we have reasoned as at the end of the proof of
Lemma~\ref{lemma:lemma9.1} in order to bound
$\partial_{\eps} \calV_{T}(0;\eps)$, and have
used Lemma~\ref{lemma:n0} and Remark~\ref{rmk:pj0}
in order to fix $p$ in (\ref{eq:epslogroot}). Hence
\begin{eqnarray*}
\left| \partial_{\eps} \calM^{[n]}(0;\eps) \right| & \geq &
\frac{j_{0}}{4} \left| \eps \right|^{j_{0}-1}
\left| G_{j_{0}} \right| +
O \left( \left| \eps \right|^{j_{0}-1} \sqrt{\eps} \right) \nonumber \\
& \geq & \frac{j_{0}}{8} \left| \eps \right|^{j_{0}-1} G_{j_{0}} \equiv
\gotm_{1} \left| \eps \right|^{j_{0}-1} , \nonumber
\end{eqnarray*}
by the second inequality in (\ref{eq:boundsderderM})
and by proceeding as at the end of the proof
of Lemma~\ref{lemma:lemma8.2} (see also Remark~\ref{rmk:betafixed}).

Furthermore one has
\begin{equation} \label{eq:Mderinterp}
\left| \int_{0}^{x} \der x' \partial_{\eps} \partial_{x'}
\calM^{[n]}(x';\eps) \right|
\le \left| x \right| \max_{x} \left| \partial_{\eps}
\partial_{x} \calM^{[n]}(x;\eps) \right| \equiv
\gotm_{2} \sqrt{|\eps|} \left| x \right| 
\end{equation}
because of Lemma~\ref{lemma:lemma9.1}, and the assertion is 
proved.$\EP$

%%%%%%%%%%%%%%%%%%%%%%%%%%%%%%%%%%%%%%%%%%%%%%%%%%%%%%%%%%%%%%%%%%%%%%%%%
\begin{lema} \label{lemma:lemma9.3}
There are two positive constants $b$ and $\xi$ such that,
for $\eps_{0}$ small enough and $\eps_{m}=2^{-m}\eps_{0}$, one has
\begin{equation}
{\rm meas}(\calE^{[\infty]}) = {\rm meas}(\calE_{m} \cap \calE_*)
\geq \frac{\eps_{m}}{2} \left( 1 - b \eps_{m}^{\xi} \right) ,
\end{equation}
where ${\rm meas}$ denotes the Lebesgue measure. The constants
$b$ and $\xi$ are independent of~$m$.$\ES$
\end{lema}
%%%%%%%%%%%%%%%%%%%%%%%%%%%%%%%%%%%%%%%%%%%%%%%%%%%%%%%%%%%%%%%%%%%%%%%%%

\prova Define $\calI^{[n_{0}]} = \emptyset$ and 
$\calI^{[n]} = \calE^{[n-1]}\setminus\calE^{[n]}$
for $n\ge n_{0}+1$; note that $\calI := \cup_{n=n_0}^{\infty} 
\calI^{[n]}
= \calE_{m} \setminus \calE^{[\infty]}$.
Recall also that we have set $\calE^{[n_{0}]}=\calE_{m}$.

For all $n\geq n_{0}+1$ and for all $\vnu\in\Z^{d}_{*}$ define
\begin{equation} \label{eq:In}
I^{[n]}(\vnu) = \left\{
\eps \in \calE^{[n-1]} \;:\;
\left| i\vomega\cdot\vnu -\calM^{[n-1]}(\vomega\cdot\vnu;\eps) \right|
\leq C_{1} |\vnu|^{-\tau_{1}}  \right\} .
\end{equation}
Each set $I^{[n]}(\vnu)$ has ``center'' in a point $\eps^{[n]}(\vnu)$,
defined implicitly by the equation
\begin{equation}
i\vomega\cdot\vnu - \calM^{[n]}(\vomega\cdot\vnu;\eps^{[n]}(\vnu)) = 0 
,
\end{equation}
where we are using the Whitney extension
of $\calM^{[n]}(\vomega\cdot\vnu;\eps)$ outside $\calE^{[n-1]}$.

Therefore one has to exclude from the set $\calE^{[n-1]}$
all the values $\eps$ around $\eps^{[n]}(\vnu)$ in
$I^{[n]}(\vnu)$, and this has to be done for
all $\vnu\in\Z^{d}_{*}$ satisfying
\begin{equation} \label{eq:32M}
\left| \vomega\cdot\vnu \right| \leq
\frac{3}{2} \left| \calM^{[n]}(\vomega\cdot\vnu;\eps) \right| ,
\end{equation}
because otherwise one can bound
$|i\vomega\cdot\vnu-\calM^{[n]}(\vomega\cdot\vnu;\eps)| \ge
|\vomega\cdot\vnu|/3 \geq C_{1} |\vnu|^{-\tau_{1}}$
as soon as $\tau_{1} \geq \tau$ and $C_{1} \leq C_{0}/3$.

For $\eps$ small enough and for all $n\geq n_{0}$ one can bound
\begin{equation*}
\left| \calM^{[n]}(0;\eps) - \calM^{[n_{0}]}(0;\eps) \right|
\leq 2D\left|\eps\right| e^{-D'2^{n_{0}/\tau_{1}}}
\leq \beta_{1} B_{1}
\left| \eps \right| e^{-\beta_{2} B_{2}2^{n_{0}/\tau_{1}}} ,
\end{equation*}
by the third inequality in (\ref{eq:boundsMderM})
of Lemma~\ref{lemma:lemma8.2}, applied repeatedly from
scale $n_{0}+1$ to scale $n$, and having used that
one has $\beta_{2} B_{2}< D'$ and $2D \leq \beta_{1} B_{1}$
(see Remark~\ref{rmk:betafixed}),
\begin{equation*}
\left| \calM^{[n_{0}]}(0;\eps) - \calM^{[n_{0}]}_{j_{0}}(0;\eps) 
\right|
= O \left( \left| \eps \right|^{j_{0}} \sqrt{|\eps|} \right)
\end{equation*}
if $p$ in (\ref{eq:epslogroot}) is chosen
according to Remark~\ref{rmk:pj0}, and
\begin{equation*}
\left| \calM^{[n_{0}]}_{j_{0}}(0;\eps) -
\ov{\calM}^{[n_{0}]}_{j_{0}}(0;\eps) \right| \leq
B_{1} \left| \eps \right| e^{-B_{2}2^{n_{0}/\tau_{1}}} ,
\end{equation*}
by Lemma~\ref{lemma:M-ovM}, so that one finds
\begin{equation*}
\left| \calM^{[n]}(0;\eps) - \ov{\calM}^{[n_{0}]}_{j_{0}}(0;\eps)
\right| < 2 G_{j_{0}} \left| \eps \right|^{j_{0}} ,
\end{equation*}
if $n_{0}$ is fixed as said in Remark~\ref{rmk:n0e1}, so that
$2\beta_{1} B_{1} \,|\eps\,| e^{-\beta_{2}B_{2}2^{n_{0}/\tau_{1}}}
\le |\eps|^{j_{0}} |G_{j_{0}}|$.

Therefore one can bound
\begin{eqnarray*}
\left| \calM^{[n]}(x;\eps) \right| & \leq &
\left| \calM^{[n]} (0;\eps) \right| +
\left| \calM^{[n]}(x;\eps)  - \calM^{[n]}(0;\eps) \right|
\nonumber \\
& \leq & \left| \ov{\calM}^{[n_{0}]}_{j_{0}}(0;\eps) \right| +
2 G_{j_{0}} \left| \eps \right|^{j_{0}} +
D \sqrt{|\eps|} \left| x \right| \nonumber \\
& \leq & 3 G_{j_{0}} \left| \eps \right|^{j_{0}} +
\frac{3}{2} D \sqrt{|\eps|} \left| \calM^{[n]}(x;\eps) \right| ,
\end{eqnarray*}%
with $x=\vomega\cdot\vnu$, for all $\vnu$ satisfying (\ref{eq:32M}).
We can conclude that there exists a constant
$\gotD$ such that one has
\begin{equation} \label{eq:boundMD}
|\calM^{[n]}(\vomega\cdot\vnu;\eps)|
\leq \eps_{m}^{j_{0}} \gotD 
\end{equation}
for all $\vnu$ satisfying (\ref{eq:32M}).

Hence we have to consider only the vectors
$\vnu\in\Z^{d}_{*}$ satisfying not only (\ref{eq:32M})
but also the inequality
$\left| \vomega\cdot\vnu \right| < 2 \eps_{m}^{j_{0}} \gotD$,
i.e. for all $\vnu\in\Z^{d}_{*}$ such that
\begin{equation} \label{eq:enne0}
\left| \vnu \right| \geq \left( \frac{C_{0}}{2\eps_{m}^{j_{0}}
\gotD } \right)^{1/\tau} := N_{0} .
\end{equation}
We call $\calN_{0}$ the set of $\vnu\in\Z^{d}_{*}$
which satisfy (\ref{eq:32M}) and (\ref{eq:enne0}).

For such $\vnu$, by setting $x=\vomega\cdot\vnu$, one has
\begin{eqnarray} \label{eq:boundderMn}
\left| \partial_{\eps} \calM^{[n]}(x;\eps) \right| & \geq &
\gotm_{1} \left| \eps \right|^{j_{0}-1} -
\frac{3}{2} \gotm_{2} \sqrt{|\eps|} \left( 2 \eps_{m}^{j_{0}}
\gotD \right) \nonumber \\
& \geq & \frac{\gotm_{1}}{2^{j_{0}-1}} \eps_{m}^{j_{0}-1}
\left( 1 - \frac{3\gotm_{2}}{2\gotm_{1}}
\eps_{m}^{3/2} 2^{j_{0}} \gotD \right)
\geq \frac{\gotm_{1}}{2^{j_{0}}} \eps_{m}^{j_{0}-1} ,
\end{eqnarray}
so that the measure of the corresponding excluded set,
which can be written as
\begin{equation}
\int_{I^{[n]}(\vnu)} \der \eps =
\int_{-1}^{1} \der t \, \frac{\der\eps(t)}{\der t} ,
\end{equation}
where $\eps(t)$ is defined by
\begin{equation}
i\vomega\cdot\vnu - \calM^{[n]}(\vomega\cdot\vnu;\eps(t)) =
t \frac{C_{1}}{|\vnu|^{\tau_{1}}} ,
\end{equation}
will be bounded by
\begin{equation}
\int_{I^{[n]}(\vnu)} \der \eps \leq \int_{-1}^{1} \der t \,
C_{1} |\vnu|^{-\tau_{1}} \frac{1}{|\partial_{\eps}
\calM^{[n]}(\vomega\cdot\vnu;\eps(t))| } 
\le \frac{2^{j_0+1}}{\gotm_{1} \eps_{m}^{j_{0}-1}}
\frac{C_{1}}{|\vnu|^{\tau_{1}}} ,
\end{equation}
by (\ref{eq:boundderMn}).

This yields that we have to exclude from $\calE^{[n-1]}$ a set
\begin{equation}
\calI^{[n]}=\bigcup_{\vnu \in \calN_{0}} I^{[n]}(\vnu)
\end{equation}
of measure bounded by
\begin{eqnarray}
{\rm meas}(\calI^{[n]}) & \leq & \sum_{\vnu \in \calN_{0}}
{\rm meas}(I^{[n]}(\vnu)) \leq {\rm const.}
\sum_{|\vnu|\geq N_{0}} \frac{C_{1}}{\eps_{m}^{j_{0}-1}}
 |\vnu|^{-\tau_{1}} \nonumber \\
& \leq & {\rm const.} \frac{C_{1}}{\eps_{m}^{j_{0}-1}}
 \left( \frac{\eps_{m}^{j_{0}}}{C_{1} }
\right)^{ (\tau_{1}-d)/\tau} = {\rm const.} \eps_{m}^{1+\xi'} ,
\end{eqnarray}
where $\xi' = j_{0}(\tau_{1}-\tau - d)/\tau$, so that $\xi'>0$ if
$\tau_{1} > \tau + d$, which fixes the value of $\tau_{1}$.

We can easily prove that there exist two positive constants
$E_{1}$ and $E_{2}$ such that one has
\begin{equation} \label{eq:diffcenter}
\left| \eps^{[n]}(\vnu)-\eps^{[n-1]}(\vnu) \right| \leq
\sqrt{\eps_{m}} E_{1} e^{-E_{2} 2^{n/\tau_{1}}} 
\end{equation}
for all $n\geq n_{0}+1$ and for all $\vnu\in\Z^{d}_{*}$.
By setting $\delta\eps = \eps^{[n]}(\vnu)-\eps^{[n-1]}(\vnu)$,
we obtain (again by using Whitney extensions)
\begin{eqnarray}
0 & = & i\vomega\cdot\vnu -
\calM^{[n]}(\vomega\cdot\vnu;\eps^{[n]}(\vnu)) \nonumber \\
& = & i\vomega\cdot\vnu -
\calM^{[n-1]}(\vomega\cdot\vnu;\eps^{[n-1]}(\vnu)+\delta\eps) \nonumber 
\\
& & \qquad - \calM^{[n]}(\vomega\cdot\vnu;\eps^{[n]}(\vnu)) +
\calM^{[n-1]}(\vomega\cdot\vnu;\eps^{[n]}(\vnu)) \nonumber \\
& = & - \partial_{\eps}\calM^{[n-1]}
(\vomega\cdot\vnu;\eps^{[n-1]}(\vnu))\,\delta\eps +
o(\delta\eps) \nonumber \\
& & \qquad - \left( \calM^{[n]}(\vomega\cdot\vnu;\eps^{[n]}(\vnu)) -
\calM^{[n-1]}(\vomega\cdot\vnu;\eps^{[n]}(\vnu)) \right) \nonumber ,
\end{eqnarray}
by (\ref{eq:Whitney}) in Lemma~\ref{lemma:lemma9.1};
hence one can use that
\begin{eqnarray} \label{eq:diffMnimproved}
& & \left| \calM^{[n]}(\vomega\cdot\vnu;\eps) -
\calM^{[n-1]}(\vomega\cdot\vnu;\eps) \right| \nonumber \\
& & \qquad \qquad
\leq D \, |\eps| \, e^{-D' 2^{n/\tau_{1}}} \nonumber \\
& & \qquad \qquad
\leq \frac{D}{\beta_{1}B_{1}} \, |\eps| \left( \beta_{1} B_{1}
e^{-\beta_{2} B_{2}  2^{n/\tau_{1}}} \right) 
\leq B' |\eps|^{j_{0}} e^{-\beta_{2} B_{2} 2^{n/\tau_{1}}} ,
\end{eqnarray}
with $B'$ a suitable constant,
by the third inequality of (\ref{eq:boundsMderM})
in Lemma~\ref{lemma:lemma8.2}, by (\ref{eq:Mn01bound}) and
by Remark~\ref{rmk:betafixed}.
Hence by (\ref{eq:boundderMn}) and (\ref{eq:diffMnimproved})
we obtain (\ref{eq:diffcenter}) with $E_{1}=4B'/\gotm_{1}$
and $E_{2}=\beta_{2} B_{2}$.

For all $|\vnu|\ge \calN_{0}$ fix $n_{*}=n_{*}(\vnu)$ such that
$| \eps^{[n_{*}+1]}(\vnu)-\eps^{[n_{*}]}(\vnu)|
\leq C_{1} |\vnu|^{-\tau_{1}}$. One can choose $n_{*}(\vnu)
\leq {\rm const.} \, \tau_{1} \log \log |\vnu|$.

Then for all $n_0+1 \le n\le n_{*}$ define $J^{[n]}(\vnu)$ as
\begin{equation} \label{eq:Jn}
J^{[n]}(\vnu) = \left\{
\eps \in \calE^{[n-1]} \;:\;
\left| i\vomega\cdot\vnu - \calM^{[n-1]}(\vomega\cdot\vnu;\eps) \right|
< 2 C_{1} |\vnu|^{-\tau_{1}} \right\} ;
\end{equation}
by construction all the sets $I^{[n]}(\vnu)$ fall inside
$J^{[n_*]}(\vnu)$ as soon as $n>n_{*}$.
Then we can bound ${\rm meas}(\calI)$ by the sum of the measures of
the sets $J^{[n_0+1]}(\vnu), \ldots,J^{[n_*]}(\vnu)$ for all
$\vnu\in\Z^{d}_{*}$ such that $|\vnu| \geq \calN_{0}$.
Such a measure will be bounded by
\begin{equation}
{\rm const.} \sum_{|\vnu|\geq \calN_{0}}
n_*(\vnu)\frac{C_1}{\eps_m^{j_0-1}}|\vnu|^{-\tau_1} 
\leq {\rm const.} \eps_{m}^{1+\xi} ,
\end{equation}
with a value $\xi$ smaller than $\xi'$ in order to take into account
the logarithmic corrections due to the factor $n_{*}(\vnu)$. $\EP$

%%%%%%%%%%%%%%%%%%%%%%%%%%%%%%%%%%%%%%%%%%%%%%%%%%%%%%%%%%%%%%%%%%%%%%%%%
\begin{prop} \label{prop:prop9}
Define the set of admissible values of $\calE_{*}$ as in
Definition \ref{def:epsadm} with $C_{1}=C_{0}/3$ and
$\tau_{1}>\tau+d$. Then one has
\begin{equation*}
\lim_{m\to\infty}
\frac{{\rm meas}(\calE_{m}\cap\calE_{*})}
{{\rm meas}(\calE_{m})} = 1 \, .
\end{equation*}
$\ES$
\end{prop}
%%%%%%%%%%%%%%%%%%%%%%%%%%%%%%%%%%%%%%%%%%%%%%%%%%%%%%%%%%%%%%%%%%%%%%%%%

\prova It is an immediate consequence of the definitions
and of Lemma~\ref{lemma:lemma9.2}.$\EP$

%%%%%%%%%%%%%%%%%%%%%%%%%%%%%%%%%%%%%%%%%%%%%%%%%%%%%%%%%%%%%%%%%%%%%%%%%
%%%%%%%%%%%%%%%%%%%%%%%%%%%%%%%%%%%%%%%%%%%%%%%%%%%%%%%%%%%%%%%%%%%%%%%%%
\zerarcounters
\section{Properties of the renormalized expansion}
\label{sec:prenormalized}
%%%%%%%%%%%%%%%%%%%%%%%%%%%%%%%%%%%%%%%%%%%%%%%%%%%%%%%%%%%%%%%%%%%%%%%%%
%%%%%%%%%%%%%%%%%%%%%%%%%%%%%%%%%%%%%%%%%%%%%%%%%%%%%%%%%%%%%%%%%%%%%%%%%

To complete the proof of existence of a quasi-periodic solution
of (\ref{eq:equ}) we have to show that the function defined
by the renormalized expansion (\ref{eq:uren})
solves the equation (\ref{eq:equ}). Set $\calE_{+}=\cup_{m=0}^{\infty}
\calE_{m} \cap \calE_{*}$: such a set contains the
admissible values of $\eps$ in $[0,\eps_{0}]$.
Define analogously $\calE_{-}$ for the interval $[-\eps_{0},0]$,
and set $\calE=\calE_{+} \cup \calE_{-}$.

%%%%%%%%%%%%%%%%%%%%%%%%%%%%%%%%%%%%%%%%%%%%%%%%%%%%%%%%%%%%%%%%%%%%%%%%%
\begin{lema} \label{lemma:lemma5}
For all $\eps\in\calE$ the function $\olu(t)$
defined through (\ref{eq:uren}) solves the equation
\begin{equation}
\olu = g \left( R + \eps Q \olu^{2} \right) ,
\end{equation}
where $g$ is the pseudo-differential operator with
kernel $g(\vomega\cdot\vnu)=1/i\vomega\cdot\vnu$.$\ES$
\end{lema}
%%%%%%%%%%%%%%%%%%%%%%%%%%%%%%%%%%%%%%%%%%%%%%%%%%%%%%%%%%%%%%%%%%%%%%%%%

\prova As in Section 8 of Ref.~\cite{Gentile}.$\EP$

\vspace{0.4cm}

So far we proved that there exists a function $\olu(t)= U(\vomega
t;\eps)$ which solves (\ref{eq:equ}) for $\eps$ in a suitable large
measure Cantor set $\calE$.  For $g$ given by $g(t)=i\eps Q(t)
\olu(t)$, Proposition~\ref{proposition:average} proves that $\phi(t)$
given in (\ref{eq:phi}) solves (\ref{eq:hill}) and is quasi-periodic.

In principle, if we set $\Omega_{\eps}=\Omega_{0}+\Me{g}$, $\phi$
could be of the form
\begin{equation*}
\phi(t) \; =\;  \Phi(\undomega_{1}t,\, \omega_{0}t,\,
\Omega_{0}t,\,\Omega_{\eps}t)
\;\equiv\; e^{i\Omega_{\eps}t}
\tilde\Phi (\undomega_{1}t,\,\omega_{0}t,\,\Omega_{0}t) ,
\end{equation*}
as it depends on $\olu(t)$, and an extra frequency arises from the
integral of the average of $g_{0}+g$ in the definition of $\Phi(t)$.
But this is not the case, because the function $\tilde\Phi$
is of the form $\tilde\Phi=(\undomega_{1}t,\omega_{0}t)$,
that is its dependence on $t$ is only through the variables
$\omega_{0}t$ and $\undomega_{1}t$. This follows from the
following property.

%%%%%%%%%%%%%%%%%%%%%%%%%%%%%%%%%%%%%%%%%%%%%%%%%%%%%%%%%%%%%%%%%%%%%%%%%
\begin{lema} \label{lemma:frequencies}
Let $\olu$ be the function defined through the renormalized
expansion (\ref{eq:uren}): then $u^{[k]}_{\vnu} \neq 0$
requires that in $\vnu=(\undm,n_{1},n_{2})$ one has $n_{2}=2$.$\ES$
\end{lema}
%%%%%%%%%%%%%%%%%%%%%%%%%%%%%%%%%%%%%%%%%%%%%%%%%%%%%%%%%%%%%%%%%%%%%%%%%

\prova The proof is by induction on $k$. For $k=0$ the result
is obvious from the relation $(i \vomega \cdot \vnu) u^{(0)}_{\vnu} =
R_{\vnu}$ in (\ref{eq:recursive}) and from the identity $R_{\vnu} =
P^{(1)}_{\undm} \calF_{n_1}^{(2)} \delta_{n_2, 2}$.
Let us assume that $u^{[k']}_{\vnu} \propto \delta_{n_{2},2}$ for
all $k'<k$. Then to order $k$ the second relation in
(\ref{eq:recursive}) yields that one can have
$u^{[k]}_{\vnu} \neq 0$ only if $\vnu=\vnu_{0}+\vnu_{1}+\vnu_{2}$:
for the last component $n_{2}$ of the vector $\vnu$ the identity
$Q_{\vnu}=\delta_{\undm,\undzero}\calF_{n_{1}}^{(-2)}\delta_{n_{2},-2}$
and the inductive assumption give $n_{2}=-2+2+2=2$.$\EP$

\vspace{0.4cm}

Hence $\olu(t)=e^{2i\Omega_{0} t} \tilde 
U(\undomega_{1}t,\omega_{0}t)$,
with $\tilde U$ analytic and periodic in its arguments.
By taking into account that one has
$Q(t)=e^{-2i\Omega_{0}t-2i\psi_{0}(t)}$, with $\psi_{0}(t)$
depending on $t$ only through the variable $\omega_{0}t$, one has
$Q(t)\olu(t) = e^{-2i\psi_{0}(t)} \tilde U(\undomega_{1}t,\omega_{0}t)$.
As a consequence $\phi(t)$ is a quasi-periodic function
with $d$ fundamental frequencies 
$\undomega_{1},\omega_{0},\Omega_{\eps}$,
and the dependence on the last frequency is only through the factor
$e^{i\Omega_{\eps} t}$, exactly as in the unperturbed case
(\ref{eq:bohr}). As anticipated in Remark~\ref{rmk:menoiomega0}
the same result can be obtained by starting from
the unperturbed solution given by the second function
in (\ref{eq:generalsolution}), and an analogous result is found,
so that we can conclude that the system is reducible for $\eps\in\calE$.

So the solution $\olu(t)$ describes the
motion on a $d$-dimensional maximal torus which is the
continuation in $\eps$ of an unperturbed $d$-dimensional torus.
The rotation vector of the latter is
$\vomega=(\undomega_{1},\omega_{0},\Omega_{0})$, while,
as an effect of the perturbation, only the last
component of the rotation vector is changed into a new
frequency $\Omega_{\eps}=\Omega_{0}+\Me{g}$: this provides a simple
physical interpretation of the the quantity $\Me{g}$.
It is likely that the new frequency $\Omega_{\eps}$ is
such that the vector $(\undomega_{1},\omega_{0},\Omega_{\eps})$
is still Diophantine. This does not follow directly from our
analysis, but we expect that this is the case.

%%%%%%%%%%%%%%%%%%%%%%%%%%%%%%%%%%%%%%%%%%%%%%%%%%%%%%%%%%%%%%%%%%%%%%%%%
%%%%%%%%%%%%%%%%%%%%%%%%%%%%%%%%%%%%%%%%%%%%%%%%%%%%%%%%%%%%%%%%%%%%%%%%%
\zerarcounters
\section{Null renormalization}
\label{sec:null}
%%%%%%%%%%%%%%%%%%%%%%%%%%%%%%%%%%%%%%%%%%%%%%%%%%%%%%%%%%%%%%%%%%%%%%%%%
%%%%%%%%%%%%%%%%%%%%%%%%%%%%%%%%%%%%%%%%%%%%%%%%%%%%%%%%%%%%%%%%%%%%%%%%%

We are left with the case in which one has $G_{j}=0$ for all $j\in\N$.
In such a case we need no resummations, as it will become clear
from the analysis. Hence we use the simpler multiscale
decomposition of the propagators given by (\ref{eq:escalas}),
with $C_{1}=C_{0}$. The following result holds.

%%%%%%%%%%%%%%%%%%%%%%%%%%%%%%%%%%%%%%%%%%%%%%%%%%%%%%%%%%%%%%%%%%%%%%%%%
\begin{lema} \label{lemma:sumpsi}
One has $\psi_{n-1}(x)\psi_{n}(x)=\psi_{n-1}(x)$ and
\begin{equation} \label{eq:sumpsi}
\psi_{0}(x) + \sum_{j=1}^{n} \chi_{j-1}(x) \psi_{j}(x) = \psi_{n}(x) ,
\end{equation}
for all $n\in\N$ and for all $x\in\R$.
\end{lema}
%%%%%%%%%%%%%%%%%%%%%%%%%%%%%%%%%%%%%%%%%%%%%%%%%%%%%%%%%%%%%%%%%%%%%%%%%

\prova Both relations follow immediately from the definitions.$\EP$

\vspace{0.4cm}

Then we consider the same tree expansion leading to (\ref{eq:exp2}),
where no resummation is performed.
The following result allows us to get rid of some trees.

%%%%%%%%%%%%%%%%%%%%%%%%%%%%%%%%%%%%%%%%%%%%%%%%%%%%%%%%%%%%%%%%%%%%%%%%%
\begin{lema} \label{lemma:discard}
Suppose that one has $G_{j}=0$ for all $j\in\N$.
Then in the tree expansion of $u^{(k)}_{\vnu}$ in (\ref{eq:exp2})
the sum over $\Theta_{k,\vnu}$ can be restricted only to trees
which do not contain any vertex $v$ such that one of the entering
lines carries the same momentum of the exiting line.$\ES$
\end{lema}
%%%%%%%%%%%%%%%%%%%%%%%%%%%%%%%%%%%%%%%%%%%%%%%%%%%%%%%%%%%%%%%%%%%%%%%%%

\prova If there were no the scale labels this would follow
from item (a) in Proposition~\ref{prop:MeQu}. The presence
of the scales could destroy in principle the compensation
mechanism responsible of the cancellation among the values of
the various trees. But it is sufficient to note that
the coefficient $u^{(k)}_{\vnu}$ is obtained by summing
over all the possible scale labels, and in this way
we reconstruct for each line $\ell$ the original propagator
$1/i\vomega\cdot\vnu_{\ell}$ (just use (\ref{eq:sumpsi})
for $n\to\infty$), hence we can apply the cited result.$\EP$

\begin{rmk} \label{rmk:alphanul}
If $G_{j}=0$ for all $j\in\N$ formal solubility of the equation (\ref{eq:u})
requires no condition on the coefficients $\alpha^{(k)}$,
which therefore can be arbitrarily fixed (cf. Lemma~\ref{lemma:Gj}).
For simplicity we can still fix $\alpha^{(k)}=0$ for all $k$,
even if this not strictly necessary. Of course one can ask what
happens for other choices of the coefficients $\alpha^{(k)}$,
but we do not investigate further such a problem
because the case in which all $G_{j}$ are
vanishing is rather special, and likely it can really arise
only in trivial situations (like $p_{1}\equiv0$).
\end{rmk}

We define the clusters according to the definition previously done,
whereas we slightly change the definition of self-energy graph,
to make it more suitable for our purposes in the present case
(cf. Ref.~\cite{Bartuccelli_Gentile}). An important feature
is that the propagators are not changed by any resummation
procedure, so that for any line $\ell$ the (two) scales
for which the corresponding propagator is not vanishing
are uniquely fixed by $\vnu_{\ell}$.

\begin{defi}[Self-Energy Graph]
We call {\rm self-energy graph} any cluster $T$ of a tree $\theta$
which satisfies
\begin{enumerate}
\item $T$ has only one entering line $\ell_T^{\rm in}$ and only one
exiting line $\ell_T^{\rm out}$;
\item The momentum of $T$ is zero, i.e. $\vnu_T = \sum_{v \in B(T)}
\vnu_v = \vz$.
\item The mode labels $\vnu_{v}$, $v\in B(T)$, satisfy the relation
$\sum_{v \in B(T)} |\vnu_{v}| < 2^{(n_{\rm ext}-4)/\tau}$, where
$n_{\rm ext}$ is the minimum
between the scales of the external lines of $T$.
\end{enumerate}
We call {\rm self-energy line} any line $\ell_T^{\rm out}$ which
exits from a self-energy graph $T$. We call {\rm normal line} any
line which is not a self-energy line.
\end{defi}

The self-energy value is then defined as before (see (\ref{eq:calVT})),
with the only difference that now the propagators
are $g^{(n_{\ell})}_{\ell}$ (because they are not renormalized).

The aim of the last item in the definition of self-energy graph
is that, given a self-energy graph, if we sum over all
the scales of the internal lines compatible with the
cluster structure, which yields that for each line
$\ell\in L(T)$ one has $n_{\ell}<n_{\rm ext}$, if $n_{\rm ext}=
\min\{n_{\ell_{T}^{\rm out}},n_{\ell_{T}^{\rm in}}\}$,
then we reconstruct for each line $\ell$ a propagator
$\psi_{n_{\rm 
ext}+1}(\vomega\cdot\vnu_{\ell})/i\vomega\cdot\vnu_{\ell}$,
with $\psi_{n_{\rm ext}+1}(\vomega\cdot\vnu_{\ell})=1$.
The last assertion is implied from the following result.

%%%%%%%%%%%%%%%%%%%%%%%%%%%%%%%%%%%%%%%%%%%%%%%%%%%%%%%%%%%%%%%%%%%%%%%%%
\begin{lema} \label{lemma:seg-scales}
For any self-energy graph $T$, by setting $n_{\rm ext}=
\min\{n_{\ell_{T}^{\rm out}},n_{\ell_{T}^{\rm in}}\}$, one can have
$\calV_{T}(\vomega\cdot\vnu) \neq0$ only
if $n_{\ell} \leq n_{\rm ext}-2$ for any line $\ell\in L(T)$.$\ES$
\end{lema}
%%%%%%%%%%%%%%%%%%%%%%%%%%%%%%%%%%%%%%%%%%%%%%%%%%%%%%%%%%%%%%%%%%%%%%%%%

\prova By definition of scales one has $C_{0}2^{-n_{\rm ext}-1}
\leq |\vomega\cdot\vnu| \leq C_{0} 2^{-n_{\rm ext}+1}$.
The third item in the definition of self-energy graph gives
$|\vomega\cdot\vnu_{\ell}^{0}|>C_{0} 2^{-(n_{\rm ext}-4)/\tau}$ (see
(\ref{eq:xell0}) for the definition of $\vnu_{\ell}^{0}$), hence by the
Diophantine condition (\ref{eq:Diophantine}) on $\vomega$ one obtains
\begin{equation*}
|\vomega\cdot\vnu_{\ell}| \geq
|\vomega\cdot\vnu_{\ell}^{0}| - |\vomega\cdot\vnu|
\geq C_{0} 2^{-(n_{\rm ext}-4)} - C_{0} 2^{-(n_{\rm ext}-1)}
\geq C_{0} 2^{-(n_{\rm ext}-3)} ,
\end{equation*}
so that $\chi_{n'-1}(\vomega\cdot\vnu_\ell)=0$ for 
$n'>n_{\rm ext}-2$.$\EP$

\begin{defi}[Localization]
For any self-energy graph $T$ we can define the {\rm localized part}
of the self-energy value $\calV_{T}(\vomega\cdot\vnu)$ as
\begin{equation} \label{eq:localization}
\calL \calV_{T}(\vomega\cdot\vnu) = \calV_{T}(0) ,
\end{equation}
and the {\rm regularized part} as
\begin{equation} \label{eq:regularization}
\calR \calV_{T}(\vomega\cdot\vnu) =
(\vomega\cdot\vnu) \int_{0}^{1} {\rm d}t \,
\partial \calV_{T}(t\vomega\cdot\vnu) ,
\end{equation}
where $\partial$ denotes derivative with respect to the argument,
so that $\partial \calV_{T}(t\vomega\cdot\vnu)=
\partial \calV_{T}(x)/\partial x |_{x=t\vomega\cdot\vnu}$.
We shall call $\calL$ and $\calR$ the {\rm localization}
and {\rm regularization operator}, respectively.
\end{defi}

\vspace{0.3cm}

By definition of self-energy value, one has
\begin{equation} \label{eq:dercalVT}
\partial \calV_{T}(t\vomega\cdot\vnu) 
= \eps^{k_T} \left(\prod_{v \in B(T)} F_v\right)
\sum_{\ell \in L(T)} \partial 
g^{(n_{\ell})}(\vomega\cdot\vnu_{\ell}(t))
\left(\prod_{\ell' \in L(T) \setminus \ell }
g^{(n_{\ell'})}(\vomega\cdot\vnu_{\ell'}(t)) \right) ,
\end{equation}
where $\vnu_{\ell}(t)=\vnu_{\ell}^{0}$ if $\ell$ is not
along the path connecting the external lines of $T$,
and $\vnu_{\ell}(t)=\vnu_{\ell}^{0}+t\vnu$ otherwise.

The definition above suggests a further splitting of the tree values
To each self-energy graph $T$ we associate
a localization label which can be either $\calL$ or $\calR$:
the first one means that we have to compute the self-energy value
for $\vomega\cdot\vnu=0$, while the second one tells us
that we have to replace $\calV_{T}(\vomega\cdot\vnu)$ with
$\calR \calV_{T}(\vomega\cdot\vnu)$
as given by (\ref{eq:regularization}). Since a
self-energy graph can contain other self-energy graphs,
the application of the localization and regularization operators
has to be performed iteratively by starting from the outermost
(or maximal) self-energy graphs to end up with the innermost ones.

%%%%%%%%%%%%%%%%%%%%%%%%%%%%%%%%%%%%%%%%%%%%%%%%%%%%%%%%%%%%%%%%%%%%%%%%%
\begin{lema} \label{lemma:nulllocalized}
Suppose that one has $G_{j}=0$ for all $j\in\N$.
Then in the tree expansion of $u^{(k)}_{\vnu}$ in (\ref{eq:exp2})
only trees with localization label $\calR$ have to been retained.$\ES$
\end{lema}
%%%%%%%%%%%%%%%%%%%%%%%%%%%%%%%%%%%%%%%%%%%%%%%%%%%%%%%%%%%%%%%%%%%%%%%%%

\prova Given a maximal self-energy graph $T$ consider the localized 
part of its self-energy graph. First of all note that the entering
line of $T$ cannot enters the same vertex $v$ which
the exiting line of $T$ comes out from (as a consequence
of Lemma~\ref{lemma:discard}). For the remaining trees
we can sum over all the scale labels compatible with the cluster
structure, and apply Lemma~\ref{lemma:seg-scales}
(which allows us to replace the support compact functions with 1).
Then we can apply the cancellation mechanism leading
to Lemma~\ref{lemma:Gjc}: indeed one immediately realizes that
the cancellation works for fixed mode labels
(see Remark~\ref{rmk:derivative}).

Then $\calV_{T}(0)=0$, so that we can replace
$\calV_{T}(\vomega\cdot\vnu)$ with
$\calR\calV_{T}(\vomega\cdot\vnu)$, as given by (\ref{eq:dercalVT}).
Here $\vnu$ is the momentum of the line entering $T$.

Next look at a self-energy graph $T'$ contained inside $T$
and which is maximal (that is the only self-energy graph
containing $T'$ is $T$ itself), and suppose we are considering
a contribution to $\calR\calV_{T}(\vomega\cdot\vnu)$
in which the derivative acts on some propagator external to $T'$.
The momentum $\vnu'(t)$ flowing through the entering
line $\ell_{T'}^{\rm in}$ of $T'$ is either
$\vnu'(t)=\vnu_{\ell_{T'}^{\rm in}}^{0}$ or
$\vnu'(t)=\vnu_{\ell_{T'}^{\rm in}}^{0}+t\vnu$,
so that for each line $\ell'\in L(T')$ one has
either $\vnu_{\ell'}=\vnu_{\ell'}^{0}$ or
$\vnu_{\ell'}=\vnu_{\ell'}^{0}+\vnu'(t)$.
So when we compute the localized part of the
self-energy value of $T'$, we have to put $\vnu'(t)=0$,
and we can reason exactly as before for $T$:
then the same cancellation mechanism applies.

If instead the derivative in (\ref{eq:dercalVT}) acts
on the self-energy value $\calV_{T}(\vomega\cdot\vnu'(t))$
than we can write $\calV_{T}(\vomega\cdot\vnu'(t))=
\calL \calV_{T}(\vomega\cdot\vnu'(t))+
\calR \calV_{T}(\vomega\cdot\vnu'(t))$, and of course
the first term gives no contribution as it is a constant.
Hence also in such a case we can get rid of the localized part
of the self-energy value.

We can iterate the argument until no further
self-energy graph is left, and the assertion follows.$\EP$

\vspace{0.4cm}

Hence we have to consider the tree expansion (\ref{eq:exp2}),
and retain only self-energy clusters with localization label $\calR$. 
The discussion then becomes standard (see for instance
Ref.~\cite{Gentile_Mastropietro}), and for each self-energy graph
$T$, if $\vnu$ is the momentum flowing through its external lines,
we obtain a gain factor $\vomega\cdot\vnu$, which
compensate exactly one of the propagators of the external lines
of $T$, say that of the exiting line (self-energy line).
Of course one has to control that no line
is differentiated more than once, but this is a standard
argument (again we refer to Ref.~\cite{Gentile_Mastropietro}
for details). At the end we obtain that $\Val(\theta)$
admits a bound like (\ref{eq:boundvalue}), with the
only difference that the propagators can be differentiated
so that they have to be bounded as they were quadratic
and not linear. On the other hand only normal lines
have to be considered, as the self-energy lines
are compensated by the mechanism described above,
and they are bounded through (\ref{eq:Nnnorm}). And the bound
(\ref{eq:Nnnorm}) still holds with the new definition of
self-energy graph, as shown in Ref.~\cite{Gentile_Mastropietro}.

The conclusion is that the series defining $u(t)$
is convergent, and it turns out to be analytic in $\eps$.
In particular this means that no value of $\eps$ has to
be discarded in such a case. Moreover $\Me{g}=0$,
because $\Me{g}=i\eps\Me{Qu}$, and
$\Me{Qu}=0$ by item (a) in Proposition~\ref{prop:MeQu}
and the hypothesis that one has $G_{j}=0$ for all $j\in\N$.
In particular one has $\Omega_{\eps}=\Omega_{0}$.

Therefore the case in which $G_{j}=0$ for all $j$
corresponds to have an integrable system. Note that
the condition $G_{j}=0$ for all $j\in\N$ is a condition
on the perturbation itself, so that it is not something
that has to be checked while carrying on
any iterative scheme to solve the problem.

%\newpage

%%%%%%%%%%%%%%%%%%%%%%%%%%%%%%%%%%%%%%%%%%%%%%%%%%%%%%%%%%%%%%%%%%%%%%%%%%%
%%%%%%%%%%%%%%%%%%%%%%%%%%%%%%%%%%%%%%%%%%%%%%%%%%%%%%%%%%%%%%%%%%%%%%%%%%%

\end{document}